\documentclass[lettersize,journal]{IEEEtran}
\IEEEoverridecommandlockouts

\usepackage{cite}
\usepackage{amsmath,amssymb,amsfonts}
\usepackage[ruled,linesnumbered]{algorithm2e}
\usepackage{graphicx}     
\usepackage{textcomp}
\usepackage{xcolor}
\usepackage{bbm}
\usepackage{dblfloatfix}
\usepackage{pifont}
\usepackage{bm}
\usepackage{booktabs}      
\usepackage{array}         
\usepackage{enumitem}
\usepackage{afterpage}
\usepackage[T1]{fontenc}
\usepackage{tabularx}      
\usepackage{float}
\usepackage{wrapfig}
\usepackage{pbox}
\usepackage{bbding}
\usepackage{balance}
\usepackage{url}

\usepackage{tabularx}  
\usepackage{ragged2e} 

\usepackage[caption=false]{subfig} 

\ifodd 0
\newcommand{\rev}[1]{{\color{blue}#1}}
\newcommand{\hbrev}[1]{{\color{blue}#1}}
\newcommand{\lxrev}[1]{{\color{purple}#1}}

\newcommand{\revhu}[1]{{\color{blue}#1}}

\else
\newcommand{\rev}[1]{#1}

\newcommand{\hbrev}[1]{#1}
\newcommand{\lxrev}[1]{#1}
\newcommand{\revhu}[1]{#1}
\fi

\renewcommand{\textit}[1]{#1}

\def\BibTeX{{\rm B\kern-.05em{\sc i\kern-.025em b}\kern-.08em
    T\kern-.1667em\lower.7ex\hbox{E}\kern-.125emX}}
\begin{document}

\title{Path to Diversity: A Primer on ISAC-izing Commodity Wi-Fi for Practical Deployments}

\author{
    Hongbo Wang,~\IEEEmembership{Graduate Student Member,~IEEE,}
    Xin Li,~\IEEEmembership{Member,~IEEE,}
    Yinghui He,~\IEEEmembership{Member,~IEEE,}\protect\\ 
    Jingzhi Hu,~\IEEEmembership{Member,~IEEE,}
    Mingming Xu,~\IEEEmembership{Member,~IEEE,}
    Zhe Chen,~\IEEEmembership{Member,~IEEE,} \protect\\ 
    Fu Xiao,~\IEEEmembership{Senior Member,~IEEE,}
    and~Jun~Luo,~\IEEEmembership{Fellow,~IEEE}


    \thanks{H. Wang is with the Collaborative Initiative, Interdisciplinary Graduate Programme, Nanyang Technological University (NTU), Singapore 639798, and also with the College of Computing and Data Science, Nanyang Technological University (NTU), Singapore 639798 (e-mail: hongbo001@ntu.edu.sg).}

    \thanks{X. Li, Y. He, and J. Luo are with the College of Computing and Data Science, Nanyang Technological University (NTU), Singapore 639798 (e-mail: l.xin@ntu.edu.sg; yinghui.he@ntu.edu.sg; junluo@ntu.edu.sg).}
    
    \thanks{J. Hu is with the School of Cyber Science and Engineering, Xi'an Jiaotong University, Xi'an 710049, China (e-mail:jingzhi.hu518@gmail.com).}
    
    \thanks{M. Xu and F. Xiao are with the School of Computer, Nanjing University of Posts and Telecommunications, Nanjing 210023, China (e-mail: 2020070137@njupt.edu.cn; xiaof@njupt.edu.cn).}

    \thanks{Z. Chen is with the Institute of Space Internet at Fudan University, Shanghai 200433, China (e-mail: zhechen@fudan.edu.cn).}

}



\maketitle

\begin{abstract}
\textit{Integrated Sensing and Communication (ISAC)} has emerged as a key paradigm in next-generation wireless networks. While the ubiquity and low deployment cost of commodity Wi-Fi render 
it 
an ideal platform for wide-scale sensing, it is the continuous evolution of Wi-Fi standards—towards higher frequency bands, wider bandwidths, and larger antenna arrays—that fundamentally unlocks the physical resources required for high-performance ISAC, thereby triggering a surge of related proposals. To structure this rapidly expanding field, a plethora of surveys have appeared. However, prevailing literature predominantly adopts a top-down perspective, emphasizing upper-layer applications or deep learning models while treating the physical layer as an opaque abstraction. Consequently, these works often fail to touch the ``bottom layer'' of signal formation and lack substantive technical guidance on how to overcome the fundamental physical barriers that constrain sensing performance.

To bridge this gap, this tutorial takes a bottom-up approach, systematically analyzing the sensing gains brought by Wi-Fi advancements through the lens of physical-layer diversity. We organize the technical framework around four orthogonal dimensions: i) \textit{Temporal Diversity} addresses synchronization gaps to enable absolute ranging, ii) \textit{Frequency Diversity} expands the effective bandwidth to sharpen range resolution, iii) \textit{Link Diversity} leverages distributed topologies and digital feedback
to achieve ubiquitous observability, and iv) \textit{Spatial Diversity} utilizes multi-antenna arrays to combine passive angular discrimination with active directional control. Collectively, these orthogonal dimensions resolve fundamental ambiguities in time, range, and space, thereby bridging physical capabilities with challenging sensing diversities such as general-purpose multi-person sensing. By synthesizing these dimensions, this tutorial provides a comprehensive guide for ``ISAC-izing'' commodity Wi-Fi, paving the way for future standardization and robust deployment.
\end{abstract}


\section{Introduction}     \label{sec:Introduction}

\subsection{The ISAC Paradigm and the Wi-Fi Potential}

The domain of wireless communication is undergoing a paradigm shift%
\revhu{, which is} transitioning from mere data pipes to intelligent, environment-aware platforms%
. As emerging applications like autonomous driving, smart manufacturing, and immersive XR demand simultaneous high-speed connectivity and environmental awareness, the traditional separation between communication and sensing is becoming increasingly inefficient~\cite{liu2022integrated}.
To address this, Integrated Sensing and Communication (ISAC) envisions wireless systems performing a dual function: concurrently exchanging data and perceiving the surrounding environment by leveraging intrinsic radio signal dynamics~\cite{liu2022survey}.
This unified design not only enhances spectral and infrastructure efficiency, but also lays the essential foundation for next-generation context-aware and adaptive wireless systems~\cite{tan2021integrated}.

Among the various technologies supporting ISAC, commodity Wi-Fi stands out as a practical and promising platform. This prominence is primarily driven by its ubiquity and low deployment cost—intrinsic strengths that render it an ideal foundation for scalable sensing without requiring dedicated infrastructure~\cite{ma2020wifi,liu2020wireless}%
. 
\revhu{Since Channel State Information  (CSI)~\cite{halperin2011tool, gringoli2019free} became explicitly accessible},
Wi-Fi has evolved into a versatile sensor supporting diverse device-free applications, ranging from localization~\cite{kotaru2015spotfi, he2016wifi, xie2019mdtrack, wang2025wical} and activity recognition~\cite{zhang2022wi, su2023realtime, chi2024xfall, meng2024graphar} to vital sign monitoring~\cite{liu2015tracking, zeng2020multisense}. Furthermore, distinct from vision-based modalities, Wi-Fi sensing inherently preserves user privacy and \hbrev{ensures robustness against poor lighting and severe occlusion}~\cite{sclrr, Xu2020low_light}. These properties establish Wi-Fi as the most accessible entry point for real-world ISAC deployments.

\subsection{\hbrev{Bridging Physical Capabilities to Sensing Diversity}}
While inherent strengths provide the foundation, the capability of Wi-Fi to perform high-performance sensing is fundamentally driven by the continuous expansion of its physical layer resources. The evolution of Wi-Fi standards has introduced substantial diversity in hardware and system configurations~\cite{ieee80211ax, garcia2021ieee}. Modern devices operate over higher frequency bands (e.g., 5 GHz, 6 GHz) with wider bandwidths (e.g., 80 MHz, 160 MHz), directly enabling finer temporal and spectral resolution. Simultaneously, MIMO techniques offer spatial diversity and directional control, while the rise of multi-device networking—via mesh and multi-link architectures—enhances connectivity and deployment flexibility. Furthermore, advanced communication protocols facilitate smarter coordination and signal steering. 
\revhu{Originally developed to enhance communication performance}, these developments \revhu{also potentially} enrich the diversity of physical-layer characteristics, enabling the extraction of rich sensing features across time, frequency, link, and spatial domains.

However, bridging these raw physical resources to reliable \textit{sensing diversity}—the capability to resolve specific target attributes like range, 
\revhu{doppler}, and angle to achieve general-purpose multi-person sensing—is non-trivial. The fundamental gap arises from a mismatch in design objectives: \textit{commodity Wi-Fi is engineered for bursty digital communication, not precision analog sensing}~\cite{jiang2021picoscenes}. While communication systems treat channel variations as impairments to be equalized, sensing systems view them as the primary source of information. This conflicting paradigm creates four systemic barriers that correspond to the four dimensions of diversity:

\begin{itemize}
    \item \textbf{Temporal Indeterminacy (The Synchronization Gap):} Unlike radar systems with unified clocks, Wi-Fi devices operate asynchronously with independent oscillators. The resulting hardware offsets disrupt the phase coherence required for Doppler tracking and the absolute timing reference needed for Time-of-Flight (ToF) ranging, restricting sensing to coarse relative measurements~\cite{vasisht2016decimeter, xie2019mdtrack}.
    
    \item \textbf{Spectral Insufficiency and Fragmentation (The Resolution Gap):} Standard Wi-Fi channels are inherently constrained in bandwidth (e.g., 20--160 MHz), limiting range resolution to the meter level which is insufficient for fine-grained sensing~\cite{dardari2009ranging, li2025uceiverfi}. 
    Wider spectrum resources are often fragmented into non-contiguous slices due to regulatory constraints. Standard protocols process these channels independently, preventing the synthesis of the effective ultra-wide bandwidth required to resolve centimeter-level multipath details.

    \item \textbf{Observational and Accessibility Barriers (The Ubiquity Gap):} Achieving ubiquitous sensing faces dual hurdles. First, 
    a single Wi-Fi link suffers from inherent blind spots (e.g., tangential motion defined by Fresnel zones) and 
    \revhu{coverage constraints}, limiting spatial observability~\cite{wang2016human}. 
    Second, 
    standard communication protocols 
    \revhu{conceal CSI from upper layers,}
    \revhu{with its access confined to only} specific legacy hardware or firmware hacks~\cite{halperin2011tool, gringoli2019free}. 
    This dependency creates a severe 
    \revhu{CSI accessibility}
    bottleneck for ubiquitous
    \revhu{sensing with} commodity devices.
    
    \item \textbf{Directional Coarseness and Passivity (The Directional Gap):} Commodity Wi-Fi arrays possess limited apertures, leading to coarse angular resolution 
    that blurs spatially adjacent targets~\cite{xiong2013arraytrack, kotaru2015spotfi}. 
    Moreover, the system exhibits 
    \revhu{directional passivity. S}tandard beamforming is designed solely to maximize SNR for the receiver, not to actively scan or illuminate specific areas of interest. This absence of programmable directional control fundamentally limits the capability to isolate sensing targets or suppress interference in complex environments.
\end{itemize}

Consequently, ``ISAC-izing'' commodity Wi-Fi requires a fundamental 
\rev{architectural redesign} of physical-layer strategies to bridge these gaps. To systematically address these challenges, we adopt a structured framework that organizes the transformation into four primary dimensions of diversity: \textbf{temporal}, \textbf{frequency}, \textbf{link}, and \textbf{spatial}.

\subsection{Overview of Key Diversity Domains}

\revhu{To fully harness the potential of Wi-Fi ISAC, physical-layer resources should be interpreted not only from a communication perspective but also as rich sources of environmental observation.}
\revhu{In the context of this tutorial, we identify four primary diversity domains in the Wi-Fi physical layer.}
Each domain is presented here as a targeted solution to the specific 
\revhu{systematic} barrier 
\revhu{listed}
above.

\subsubsection{Temporal Diversity (Synchronization for Absolute Ranging and Coherence)} 
To bridge the 
synchronization gap, temporal diversity 
necessitates a transition from asynchronous bistatic setups to \textit{monostatic architectures}, thereby establishing \textit{Synchronized ISAC}. By enabling the transmitter and receiver to share a common clock, this architecture eliminates timing indeterminacy, unlocking \textit{absolute ToF} for precision ranging and \textit{phase coherence} for fine-grained motion tracking~\cite{chen2024isac,song2024siwis}. While effective, this co-location introduces overwhelming self-interference~\cite{bharadia2013full}. As detailed in Sec.~\ref{sec:temporal}, harnessing this diversity hinges on mastering \textit{signal separation}—incorporating techniques such as \textit{active cancellation} and \textit{physical isolation} to recover faint environmental echoes from 
\revhu{strong} transmission leakage.

\subsubsection{Frequency Diversity (High Resolution via Spectrum Expansion)}
To overcome the Resolution Gap, Frequency Diversity governs the system's delay resolution by synthesizing a wide effective bandwidth to realize \textit{High-Resolution ISAC}. Instead of relying on contiguous channels which are prone to boundary effects \revhu{(elaborated in Sec.~\ref{sec:frequency})},
state-of-the-art designs shift towards \textit{discrete channel sampling}. By aggregating irregular spectral slices across the 2.4/5/6 GHz bands, this approach breaks the bandwidth limitations of standard protocols
to achieve 
\revhu{\textit{centimeter-level resolution comparable to UWB,}}
capable of distinguishing closely spaced targets~\cite{li2024uwb, li2025ccs}.

\subsubsection{Link Diversity (Ubiquitous Observability via Physical and Digital Paths)}
To resolve the Ubiquity Gap, Link Diversity aggregates observations from distributed physical paths and standardized digital feedback to establish \textit{Ubiquitous ISAC}. As detailed in Sec.~\ref{sec:link}, we categorize this domain into two complementary forms. \textit{Physical Link Diversity} exploits the geometric distribution of distributed pairs to eliminate blind spots and leverage \textit{near-field domination} for spatial disentanglement~\cite{hu2023muse}. Complementing this is \textit{Digital Link Diversity}, which utilizes standardized protocol feedback like \textit{Beamforming Feedback Information (BFI)}. Serving as a stable, low-pass filtered ``digital proxy'', BFI overcomes hardware fragmentation barriers, enabling robust sensing on unmodified commodity devices~\cite{wu2023beamsense, wang2024mukifi, yi2024bfmsense}.

\subsubsection{Spatial Diversity (Directional Discrimination and Control via Active and Passive MIMO)}
To address the Directional Gap, Spatial Diversity exploits Multiple-Input Multiple-Output (MIMO) techniques enabled by multi-antenna arrays to establish \textit{Directional ISAC}. We structure this domain into two complementary paradigms. On the receiver side, \textit{Passive Rx-MIMO} provides \textit{Directional Discrimination}, employing advanced signal processing algorithms to overcome the angular resolution limit~\cite{kotaru2015spotfi, xiong2013arraytrack}. On the transmitter side, \textit{Active Tx-MIMO} provides \textit{Directional Control}. It introduces a paradigm shift towards programmable directionality, leveraging feedback manipulation to actively steer electromagnetic energy for sensing. As examined in Sec.~\ref{sec:spatial}, this joint capability transforms Wi-Fi from a passive radiator into an active probe, enabling the system to both ``look'' (Rx) and ``illuminate'' (Tx) with high angular precision~\cite{he2025versabeam, xu2024beamforming}.

\subsection{Existing Survey and Tutorial Landscape}

\begin{table*}[t]
\centering
\caption{Comparison of Representative Wi-Fi Sensing Surveys}
\label{tab:survey_comparison}
\renewcommand{\arraystretch}{1.3}
\begin{tabular}{
    >{\raggedright\arraybackslash}p{2.5cm} 
    >{\raggedright\arraybackslash}p{2.2cm} 
    >{\raggedright\arraybackslash}p{5.3cm} 
    >{\raggedright\arraybackslash}p{5.5cm}
}
\toprule
Reference & Survey Category & Primary Focus and Contribution & Identified Gap or Limitation \\
\midrule

He et al. (2020) \cite{he2020wifi} & Application: \newline Sensing Tasks & Frames Wi-Fi sensing as “Wi-Fi vision”; categorizes perception-level tasks such as imaging, detection, and pose. & Uses vision-based analogies; lacks generalizable signal-layer principles; diversity implicit but not abstracted. \\

\addlinespace

Tan et al. (2022) \cite{tan2022commodity} & Application: \newline Sensing Tasks & Reviews a decade of commodity Wi-Fi sensing; taxonomizes systems by functional domains (activity, object, localization). & A retrospective on application evolution; lacks bottom-up analysis of how physical-layer upgrades drive these capabilities. \\

\addlinespace

Xiao et al. (2023) \cite{xiao2023surveya} & Application: \newline Scenarios & Proposes a dual-axis taxonomy (task type × motion granularity); reviews diverse human sensing scenarios. & High-level abstraction of application logic; lacks analysis of signal-layer capabilities or diversity mechanisms. \\

\addlinespace

Chen et al. (2023) \cite{chen2023crossdomain} & Algorithm: \newline Frameworks & Focuses on cross-domain generalization techniques (e.g., transfer learning) for CSI-based sensing. & Solves domain shift via algorithmic adaptation; overlooks how physical diversity can intrinsically reduce environmental dependency. \\

\addlinespace

Yang et al. (2023) \cite{yang2023sensefi} & Algorithm: \newline Frameworks & Builds SenseFi, a benchmarking library for comparing DL models across sensing tasks and datasets. & Empirical emphasis on model/dataset comparison; abstracted from signal-level or physical-layer sensing implications. \\

\addlinespace

Ahmad et al. (2024) \cite{ahmad2024wifibased} & Algorithm: \newline Learning Models & Surveys Transformers and CNNs for Wi-Fi sensing; discusses training strategies and deployment challenges. & Strong DL focus; limited discussion on CSI formation mechanism or physical modeling. \\

\addlinespace

Radwan et al. (2025) \cite{radwan2025tutorial} & Algorithm: \newline Learning Models & Surveys Self-Supervised Learning (SSL) paradigms to address data scarcity; discusses pretext tasks and contrastive learning. & Addresses label efficiency through learning paradigms; does not explore physical-layer enhancements to reduce data requirements. \\

\addlinespace

Ma et al. (2020) \cite{ma2020wifi} & Signal: \newline Signal Processing & Surveys CSI signal modeling, preprocessing, and feature extraction pipelines. & Covers signal chain depth but lacks unified diversity abstraction. \\

\addlinespace

Hernandez \& Bulut (2023) \cite{hernandez2023wifia} & Signal: \newline Signal Processing & Surveys signal processing for resource-constrained edge devices; evaluates computational efficiency. & Focuses on implementation bottlenecks (computation); does not explore active physical-layer diversity. \\

\addlinespace

Du et al. (2024) \cite{du2024overview} & Signal: \newline Standards & Comprehensive overview of the IEEE 802.11bf standard; details PHY/MAC specifications and protocols. & Focuses on standard compliance and protocol definitions; does not analyze the physics regarding diversity exploitation. \\

\bottomrule
\end{tabular}
\end{table*}

A plethora of surveys have emerged to structure the rapidly growing field of Wi-Fi sensing, which serves as the practical precursor to Wi-Fi-based ISAC. To delineate our tutorial's unique scope, we categorize these representative works into three broad groups: application-centric, algorithm-centric, and 
\revhu{signal-level, based} 
on their primary abstraction levels. While these works provide extensive summary of existing literature, they largely adopt a top-down perspective that obscures the fundamental physical mechanisms—specifically the diversity dimensions—that are essential for transforming Wi-Fi into a 
ISAC platform \revhu{for diverse realistic scenarios}.

\subsubsection{Application-Centric---Sensing Tasks and Deployment Scenarios}
These surveys primarily organize the literature by sensing tasks or deployment scenarios, treating the physical layer as an abstracted interface. 
He et al. (2020)~\cite{he2020wifi} frame sensing through the lens of “Wi-Fi vision,” categorizing tasks such as imaging and pose estimation to highlight RF perception capabilities. 
Tan et al. (2022)~\cite{tan2022commodity} review a decade of commodity Wi-Fi sensing, classifying systems into three functional domains: activity recognition, object sensing, and localization. 
Xiao et al. (2023)~\cite{xiao2023surveya} propose a dual-axis taxonomy intersecting sensing task types with motion granularity, mapping human behaviors to system requirements. 
While valuable for high-level classification, these works offer limited insight into the underlying signal structure, abstracting away \textit{how} physical signal diversity fundamentally constrains or enables these capabilities.

\subsubsection{Algorithm-Centric---Learning Models and Frameworks}
These surveys focus on the computational backend, reviewing learning models while often treating Channel State Information (CSI) as a given input. 
Chen et al. (2023)~\cite{chen2023crossdomain} address sensing robustness, surveying cross-domain generalization techniques like transfer learning. 
Yang et al. (2023)~\cite{yang2023sensefi} introduce SenseFi, a comprehensive benchmark library for evaluating deep learning models across diverse datasets. 
Ahmad et al. (2024)~\cite{ahmad2024wifibased} provide a unified overview of modern architectures, including Transformers and CNNs, specifically for human sensing. 
Most recently, Radwan et al. (2025)~\cite{radwan2025tutorial} survey Self-Supervised Learning (SSL) paradigms, discussing how pretext tasks can mitigate data scarcity. 
However, these works generally decouple learning from signal formation. They tend to solve sensing challenges solely through model complexity, overlooking how physical-layer diversity dictates the intrinsic information content of the input data.
\subsubsection{Signal-Level---Signal Processing and Standards}
A third line of surveys engages directly with physical layer implementation, spanning signal modeling to standardization. 
Ma et al. (2020)~\cite{ma2020wifi} systematically review the foundational signal processing pipeline, covering CSI channel modeling, noise reduction, and feature extraction. 
Hernandez and Bulut (2023)~\cite{hernandez2023wifia} shift the focus to edge implementation, surveying low-complexity processing techniques tailored for resource-constrained IoT devices. 
Du et al. (2024)~\cite{du2024overview} provide a comprehensive overview of the IEEE 802.11bf amendment, detailing the official PHY/MAC specifications for native WLAN sensing. 
While these works offer a holistic view of the physical layer, they predominately view the physical layer as a static set of constraints or specifications, lacking a unified abstraction on how to \textit{actively} exploit physical-layer diversity to break performance limits.

\vspace{0.5em}
\noindent \textbf{Summary of the Gap.} 
To facilitate comparison and synthesize the landscape, Table~\ref{tab:survey_comparison} summarizes these representative surveys by category, core focus, and identified limitations. As the table illustrates, the prevailing literature predominantly adopts a \textit{top-down perspective}, prioritizing high-level application logic or post-processing algorithms while often treating the physical layer as an opaque abstraction. In contrast, there is a scarcity of work that systematically explores how sensing capability itself is fundamentally governed by the orthogonal dimensions of physical-layer diversity. This theoretical gap—\textit{the disconnect between evolving hardware resources and their intrinsic sensing limits}—motivates the central perspective of this tutorial.

\subsection{Tutorial Motivation, Contribution, and Structure}

\begin{table*}[t]
\caption{Organization of this Tutorial: Mapping Diversity Dimensions to Sensing Gaps}
\label{tab:tutorial_organization}
\centering
\small
\renewcommand{\arraystretch}{1.3}
\newcolumntype{L}{>{\raggedright\arraybackslash}X}
\begin{tabularx}{\textwidth}{|c|l| >{\hsize=0.8\hsize}L |l| >{\hsize=1.2\hsize}L |}
\hline
\textbf{Sec.} & \textbf{Focus Area} & \textbf{Targeted Physical Gap} & \textbf{Diversity Dimension} & \textbf{Key Solution / Paradigm} \\ \hline
\hline

\textbf{II} & \textbf{Foundations} & 
Signal Abstraction \& Imperfections & 
N/A (Signal Basis) & 
CSI/BFI Models, Hardware Offset Compensation, Feature Extraction \\ \hline

\textbf{III} & \textbf{Synchronization} & 
\textbf{The Synchronization Gap:} \newline Lack of common clock prevents absolute ranging. & 
\textbf{Temporal Diversity} & 
\textbf{Synchronized ISAC:} \newline Monostatic architecture via Self-Interference Cancellation (SIC). \\ \hline

\textbf{IV} & \textbf{Resolution} & 
\textbf{The Resolution Gap:} \newline Limited bandwidth limits range resolution. & 
\textbf{Frequency Diversity} & 
\textbf{High-Resolution ISAC:} \newline Synthesizing ultra-wide bandwidth via Discrete Channel Sampling. \\ \hline

\textbf{V} & \textbf{Ubiquity} & 
\textbf{The Ubiquity Gap:} \newline Single-link blind spots \& Accessibility barriers. & 
\textbf{Link Diversity} & 
\textbf{Ubiquitous ISAC:} \newline Synergizing Physical Multi-links (Coverage) \& Digital Feedback (BFI). \\ \hline

\textbf{VI} & \textbf{Directionality} & 
\textbf{The Directional Gap:} \newline Coarse angular resolution \& Passive illumination. & 
\textbf{Spatial Diversity} & 
\textbf{Directional ISAC:} \newline From Passive Rx-MIMO to Active Tx-MIMO. \\ \hline

\textbf{VII} & \textbf{Future Directions} & 
Systemic Integration \& Deployment Barriers & 
\textbf{Cross-Diversity Fusion} & 
Native Standardization, Physics-Aware AI, and Privacy Preservation. \\ \hline

\end{tabularx}
\end{table*}

Motivated by the lack of theoretical abstraction in existing literature (Table~\ref{tab:survey_comparison}), this tutorial presents the first \textit{bottom-up framework} for Wi-Fi ISAC. 
In contrast to prior surveys that focus on high-level applications, we center our perspective on \textbf{physical-layer diversity}—a set of orthogonal signal dimensions that fundamentally determine the sensing upper bound. We argue that transforming a communication device into a robust sensor relies on the rigorous manipulation of diversity across four key dimensions: temporal, frequency, link, and spatial. 
These diversity dimensions are not only inherent in Wi-Fi hardware, but are increasingly enriched by modern standards such as 802.11ax/be, 6 GHz operation, and multi-link support. This tutorial provides a structured guide to harnessing these evolving resources, transforming raw physical capabilities into reliable sensing diversity and ability.

\vspace{0.5em}
\noindent
\textbf{Tutorial Structure.} 
The organization of this tutorial is visually summarized in \textbf{Table~\ref{tab:tutorial_organization}}, which maps each chapter to the specific physical gap it addresses and the corresponding solution paradigm.

\begin{itemize}
    \item \textbf{Sec.~\ref{sec:background}} establishes the \textbf{Foundations}, introducing the fundamental signal models (CSI/BFI) and addressing the critical hardware imperfections that distort them.
    
    \item \textbf{Sec.~\ref{sec:temporal} to Sec.~\ref{sec:spatial}} constitute the core, where each section leverages a specific diversity dimension to achieve a distinct ISAC capability:
    \begin{itemize}
        \item \textbf{Sec.~\ref{sec:temporal} (Temporal Diversity)} establishes \textbf{\textit{Synchronized ISAC}} to enable absolute ranging and phase coherence.
        \item \textbf{Sec.~\ref{sec:frequency} (Frequency Diversity)} realizes \textbf{\textit{High-Resolution ISAC}} by synthesizing effective ultra-wide bandwidths.
        \item \textbf{Sec.~\ref{sec:link} (Link Diversity)} constructs \textbf{\textit{Ubiquitous ISAC}} through the synergy of distributed physical and digital links.
        \item \textbf{Sec.~\ref{sec:spatial} (Spatial Diversity)} enables \textbf{\textit{Directional ISAC}} via active beam steering and shaping.
    \end{itemize}
    
    \item \textbf{Sec.~\ref{sec:future_work}} and \textbf{Sec.~\ref{sec:conclusion}} discuss future directions and conclude the tutorial.
\end{itemize}

Through this structure, the tutorial aims to provide both conceptual clarity and practical design guidance for transforming commodity Wi-Fi systems into deployable ISAC platforms.

\section{Foundations of Wi-Fi Sensing: Signals and Imperfections}     \label{sec:background}
The ability of Wi-Fi systems to perform sensing tasks is fundamentally reliant on their capacity to extract and interpret information about how radio signals interact with the surrounding environment. This section delves into the primary signal modalities used for Wi-Fi sensing—Channel State Information (CSI) and Beamforming Feedback Information (BFI)—and critically examines the inherent imperfections in these signals due to hardware limitations, along with the techniques developed to mitigate these issues. Building upon these signal foundations, this section further introduces representative sensing features derived from CSI, as well as multipath parameter estimation methods.

\subsection{Channel State Information (CSI): The Primary Sensing Modality}

\subsubsection{Physical Principles and Mathematical Representation of CSI}

Channel State Information (CSI) provides a fine-grained characterization of the physical layer of a wireless channel. Conceptually, it is a discrete, sampled measurement of the underlying \textit{Channel Frequency Response (CFR)}, which is the continuous function that describes how a signal's amplitude and phase change across all frequencies. In modern Wi-Fi systems based on Orthogonal Frequency-Division Multiplexing (OFDM), CSI is estimated for each subcarrier from known training sequences, such as the Long Training Field (LTF), embedded in each data packet's preamble.

\vspace{0.5em} \noindent \textbf{Signal Model.}
Mathematically, for a specific subcarrier $k$, the received signal vector $\mathbf{y}_k$ relates to the transmitted signal vector $\mathbf{x}_k$ via the linear system model~\cite{van1995channel}:
\begin{equation}
    \mathbf{y}_k = \mathbf{H}_k \mathbf{x}_k + \mathbf{n}_k,
\label{eq:signal_model}
\end{equation}
where $\mathbf{n}_k$ represents the Additive White Gaussian Noise (AWGN), and $\mathbf{H}_k$ is the CSI matrix we seek to estimate. By removing the known pilot $\mathbf{x}_k$, the receiver obtains $\mathbf{H}_k$, which encapsulates the combined effect of the environment on the signal.

\vspace{0.5em} \noindent \textbf{Multipath Propagation Model.}
To understand how the estimated matrix $\mathbf{H}_k$ encodes environmental dynamics, we can model the wireless channel as a superposition of multiple propagation paths. For a full Multiple-Input Multiple-Output (MIMO) system, the time-variant CSI between transmit antenna $i$ and receive antenna $j$ on subcarrier $k$ at time $t$ can be expressed as:
\begin{equation}
H_{i,j,k}(t) =  \sum_{l=1}^{L} a_l(t) e^{-j 2\pi d_{i,j,l}(t) f_k / c}, 
\label{eq:csiijk_dynamic}
\end{equation}
where $L$ is the number of paths, and for each path $l$, $a_l(t)$ is its time-variant complex attenuation, $d_{i,j,l}(t)$ is its time-variant propagation distance, $f_k$ is the frequency of subcarrier $k$, and $c$ is the speed of light.



\begin{figure}[t]
\vspace{-1em}
\setlength{\abovecaptionskip}{6pt}  
    \centerline{
    \includegraphics
    [width=0.6 \columnwidth]{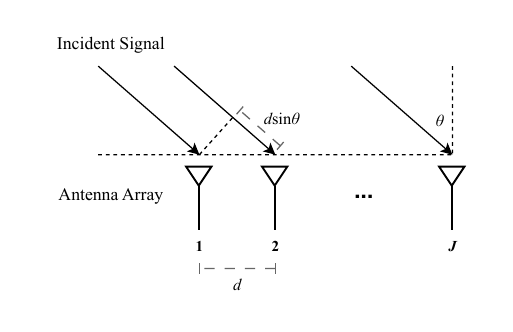}
    }
     \caption{Uniform linear array consisting of $J$ receive antennas }
    \label{fig:spotfi}
\end{figure}

To make this model more concrete for sensing, we can simplify it to a common scenario with a single-antenna transmitter and a multi-antenna receiver. Specifically, we consider a receiver equipped with a uniform linear array (ULA), as depicted in Fig.~\ref{fig:spotfi}. Furthermore, assuming the channel coherence time exceeds the packet duration, we can analyze the CSI from a single packet as a quasi-static snapshot, temporarily dropping the time variable $(t)$. The total propagation distance of the $l$-th path to the $j$-th antenna, denoted by $d_{j,l}$, can be broken down as:
\begin{equation}
d_{j,l} = d_l + (j-1) d \sin \theta_l,
\label{eq:distance_static}
\end{equation}
where $d_l = c \tau_l$ is the distance of the path to the first reference antenna, with $\tau_l$ being its Time of Flight (ToF), and $\theta_l$ is its Angle of Arrival (AoA). Substituting this into the general channel model gives the complete CSI for subcarrier $k$ at antenna $j$:
\begin{align}
H_{j,k} 
&= \sum_{l=1}^{L} a_l \cdot e^{-j 2\pi f_k \tau_l} \cdot e^{-j \frac{2\pi f_k}{c} (j-1) d \sin \theta_l}.
\label{eq:CSI_joint_static}
\end{align}
This expression is the foundation of model-based Wi-Fi sensing. It explicitly reveals how the fundamental physical parameters of the environment—such as range and bearing—are encoded into a single, multi-dimensional CSI frame.

\subsubsection{Fundamental Sensing Parameters Encoded in CSI}
\label{ssec:sensing_parameters}

The value of CSI in Wi-Fi sensing stems from its high sensitivity to environmental dynamics. Any physical change—such as human motion or object displacement—alters the multipath structure of the wireless channel. These changes are encoded into specific signal parameters governed by the available physical-layer diversity. Below, we dissect these parameters and explicitly map them to the diversity dimensions defined in our framework.

\paragraph{\textbf{Time-of-Flight (ToF)}}
Time-of-Flight, denoted as $\tau_l$, represents the signal propagation delay along path $l$. Physically, ToF is embedded within the phase slope of the CSI across the frequency domain. From Eqn.~\eqref{eq:CSI_joint_static}, for a single path, the phase $\psi_{k}^{(l)}$ at subcarrier frequency $f_k$ is:
\begin{equation}
\psi_{k}^{(l)} = \angle H_{k}^{(l)} \approx C_l - 2\pi f_k \tau_l.
\label{eq:tof_relation}
\end{equation}
Here, $C_l$ is a constant phase offset. This equation reveals a dual dependency on physical-layer diversity. First, the absolute accuracy of $\tau_l$ relies on \textbf{Temporal Diversity} in the form of strict synchronization. Without a common clock reference, random timing offsets introduce unknown biases to $\tau_l$, rendering absolute ranging infeasible. Second, the resolution of $\tau_l$ estimation (i.e., the capacity to distinguish closely spaced multipath components) is fundamentally determined by \textbf{Frequency Diversity}—specifically, the signal bandwidth. A wider spectral aperture sharpens the channel impulse response, reducing peak overlap. Therefore, the standard method to extract ToF in a multi-path environment is to apply an \textit{Inverse Fast Fourier Transform (IFFT)} to the CSI vector. This transforms the signal to the time domain, yielding a \textit{Power Delay Profile (PDP)} where each peak corresponds to the ToF of a distinct multipath component.

\paragraph{\textbf{Angle-of-Arrival (AoA)}}
Angle-of-Arrival, $\theta_l$, is the spatial direction from which a signal arrives at the receiver array. It is encoded in the phase differences between spatially separated antennas, captured by the term $e^{-j \frac{2\pi f_k}{c} (j-1) d \sin \theta_l}$. This parameter is the direct product of \textbf{Spatial Diversity}. The number of antennas and the array aperture size constitute the spatial degrees of freedom that dictate the system's angular resolution and its ability to discriminate spatially distinct targets. High-resolution algorithms like MUSIC are typically employed to invert this spatial structure and estimate $\theta_l$.

\paragraph{\textbf{Doppler Frequency Shift (DFS)}}
When a target is in motion, the path length $d_l(t)$ changes over time, introducing a frequency shift known as the Doppler Effect. This manifests as a phase rotation along the time dimension of the CSI packet stream. The Doppler shift $f_{D,l}$ is defined as~\cite{soumekh1999synthetic}:
\begin{equation}
    f_{D,l}(t) = -\frac{1}{2\pi} \frac{d\phi_l(t)}{dt} = -\frac{v_l(t)}{\lambda_k},
\end{equation}
where $v_l(t)$ is the radial velocity and $\lambda_k$ is the wavelength. This parameter is governed by \textbf{Temporal Diversity}—specifically, the packet sampling rate (frame rate) and the continuous observation duration. A higher sampling rate expands the detectable velocity range (preventing aliasing), while a longer observation window improves the velocity resolution. DFS is commonly extracted using Short-Time Fourier Transform (STFT) to generate dynamic spectrograms.


\paragraph{\textbf{Complex Attenuation (Amplitude)}}
Finally, the complex attenuation $a_l$ (or the channel magnitude $\|\mathbf{H}\|$) captures the signal strength scaling induced by path loss and shadowing. While phase-based parameters (ToF, AoA, DFS) describe the \textit{geometry} of the environment, amplitude describes the \textit{quality} of the link. This amplitude is directly linked to the system's fundamental metric: the \textit{Signal-to-Noise Ratio (SNR)}.
Based on the signal model in Eqn.~\eqref{eq:signal_model}, SNR is proportional to the squared norm of the channel matrix:
\begin{equation}
    \eta = \frac{\|\mathbf{H}_k\|^2 P_{tx}}{\sigma_n^2},
\end{equation}
where $P_{tx}$ denotes the transmit power and $\sigma_n^2$ represents the noise power (variance of the AWGN). In ISAC systems, this metric plays a dual role:
\begin{itemize}
    \item \textbf{Communication SNR ($\eta^C$):} Determines the data rate via the discrete Modulation and Coding Scheme (MCS). The relationship between $\eta^C$ and throughput follows a \textbf{staircase function}.
    \item \textbf{Sensing SNR ($\eta^S$):} Quantifies the strength of target reflections relative to noise. In contrast to the discrete nature of MCS, sensing precision improves \textbf{continuously} with increasing $\eta^S$, serving as the fundamental limit for estimation variance.
\end{itemize}

\subsubsection{Practical CSI Acquisition Tools}
\label{ssec:csi_tools}

Accessing fine-grained CSI from commodity hardware requires bypassing standard drivers. The research community has developed open-source tools to unlock this capability on specific Network Interface Cards (NICs), enabling the vast majority of academic research:

\begin{itemize}
    \item \textbf{Intel CSI Tool:} The pioneering tool for the Intel 5300 NIC~\cite{halperin2011tool}. It spurred early foundational research but is limited to older 802.11n hardware.
    \item \textbf{Atheros CSI Tool:} Enables CSI extraction on Qualcomm Atheros chipsets~\cite{xie2015precise}, offering an alternative to Intel with different subcarrier granularities.
    \item \textbf{Nexmon CSI:} A C-based firmware patching framework for Broadcom chipsets~\cite{schulz2017nexmon}. It crucially enables CSI extraction on smartphones (e.g., Nexus, Raspberry Pi), expanding ISAC to mobile and embedded IoT scenarios.
    \item \textbf{PicoScenes:} A modern, unified platform supporting diverse hardware (Intel, Qualcomm, AX200/210)~\cite{jiang2021picoscenes}. It supports the latest Wi-Fi standards (802.11ac/ax) and provides extra-large bandwidth data required for high-resolution sensing.
    \item \textbf{ESP32 CSI Tool:} A lightweight solution for IoT microcontrollers~\cite{espressif_esp_csi}. Officially supported by Espressif, it facilitates large-scale, low-cost distributed sensing deployments.
\end{itemize}

\subsection{Hardware Imperfections and Compensation}
\label{ssec:offset}

While CSI provides a rich source of information for sensing, the raw measurements obtained from commodity Wi-Fi hardware are invariably affected by hardware-induced artifacts and impairments. These imperfections cause the measured CSI, denoted as $\hat{H}_{i,j,k}(t)$, to deviate from the true physical channel response $H_{i,j,k}(t)$. Among all CSI components, the phase is particularly sensitive to such errors. If these offsets are not properly understood and corrected, they can severely distort the CSI, degrading the accuracy, stability, and robustness of ISAC systems.

\subsubsection{The Reality of CSI: Hardware-Induced Offsets}
\label{ssec:offset}
Modeling frameworks typically decompose these hardware-induced offsets into four major categories, each originating from specific components of the RF and baseband processing chain~\cite{xie2015precise, ratnam2024optimal}. These include Carrier Frequency Offset (CFO), Carrier Phase Offset (CPO), Sampling Frequency Offset (SFO), and Packet Detection Delay (PDD). While all four manifest as phase distortions, they differ fundamentally in their temporal characteristics and impact on sensing.

\paragraph{Carrier Frequency Offset (CFO)}
CFO arises from the inherent frequency mismatch between the local oscillators of the transmitter and receiver. Due to manufacturing variability and temperature drift, there exists a frequency difference $\gamma_c$ (Hz). This results in a cumulative phase shift that increases linearly over time:
\begin{equation}
H_{i,j,k}^{\text{CFO}}(t) = H_{i,j,k}(t) \cdot e^{-j 2\pi \gamma_c t}.
\end{equation}
This time-varying rotation disrupts inter-packet phase coherence. For sensing tasks relying on phase continuity—such as Doppler-based gesture recognition or micro-motion tracking—CFO introduces a drifting baseline that can completely mask the true motion signals.

\paragraph{Carrier Phase Offset (CPO)}
CPO, denoted by $\phi_c$, is a static but random phase offset introduced by the Phase-Locked Loop (PLL) during the synchronization process. Although the PLL ensures frequency stability once locked, it locks to a random initial phase for each packet. This adds a constant phase term uniformly across all subcarriers:
\begin{equation}
H_{i,j,k}^{\text{CPO}}(t) = H_{i,j,k}(t) \cdot e^{-j 2\pi \phi_c}.
\end{equation}
Unlike CFO, CPO is random per packet. This stochastic behavior destroys the absolute phase reference between consecutive packets, making it particularly challenging for long-duration tasks like gait recognition where a stable phase baseline is essential.

\paragraph{Sampling Frequency Offset (SFO)}
SFO arises when the sampling clocks at the transmitter and receiver are not perfectly synchronized. This mismatch, denoted by $\beta$, results in a progressive timing error as the packet is sampled, manifesting as a phase slope across subcarriers in the frequency domain:
\begin{equation}
H_{i,j,k}^{\text{SFO}}(t) = H_{i,j,k}(t) \cdot e^{-j 2\pi \beta k / N_{\text{FFT}}}.
\label{eq:SFO}
\end{equation}
Since this distortion is frequency-dependent, it alters the measured Angle of Arrival (AoA) and blurs the range resolution in ToF estimation, effectively ``stretching''or ``compressing'' the estimated distance.

\paragraph{Packet Detection Delay (PDD)}
PDD is caused by the jitter in the receiver's packet detection algorithm, which aligns the FFT window with the start of the OFDM symbol. A misalignment of $\varepsilon$ samples introduces a linear phase shift similar to SFO:
\begin{equation}
H_{i,j,k}^{\text{PDD}}(t) = H_{i,j,k}(t) \cdot e^{-j 2\pi \varepsilon k / N_{\text{FFT}}}.
\label{eq:PDD}
\end{equation}
However, unlike the systematic SFO, PDD varies randomly from packet to packet. This introduces abrupt ``jumps'' in the phase slope, complicating the extraction of smooth phase trends over time.

\subsubsection{Mitigating Imperfections: CSI Offset Compensation Technologies}
\label{ssec:offset_compensation}

To counteract these effects, researchers have developed various compensation methods summarized in Table~\ref{tab:csi_offsets}. These techniques aim to ``clean'' the raw CSI by algorithmically removing distortions.

\begin{table*}[t]
\centering
\caption{Summary of CSI Offsets and Compensation Methods}
\label{tab:csi_offsets}
\renewcommand{\arraystretch}{1.3}
\begin{tabular}{
    >{\raggedright\arraybackslash}p{1.0cm} 
    >{\raggedright\arraybackslash}p{3.2cm} 
    >{\raggedright\arraybackslash}p{2.5cm} 
    >{\raggedright\arraybackslash}p{4.2cm} 
    >{\raggedright\arraybackslash}p{5.1cm}
}
\toprule
\textbf{Offset} & \textbf{Primary Cause} & \textbf{Mathematical Effect} & \textbf{Common Compensation Methods} & \textbf{Features / Limitations} \\
\midrule

\textbf{CFO} & Frequency mismatch between Tx/Rx oscillators & $e^{-j 2\pi \gamma_c t}$ & Uplink-Downlink Multiplication, Conjugate Multiplication, CSI Ratio & Requires bidirectional exchange or multiple antennas; absolute phase lost after conjugation or ratio \\
\addlinespace

\textbf{CPO} & Random phase due to Rx PLL initialization & $e^{-j 2\pi \phi_c}$ & Statistical Averaging, Uplink-Downlink Multiplication, Conjugate Multiplication, CSI Ratio & Averaging is simple but requires a static channel; other methods have stricter hardware requirements. \\
\addlinespace

\textbf{SFO} & Sampling Clock Mismatch (Tx/Rx) & $e^{-j 2\pi \beta k / N_{\text{FFT}}}$ & Linear Regression, Conjugate Multiplication, CSI Ratio & Linear regression is lightweight; conjugate/ratio methods lose absolute phase; multi-antenna required. \\
\addlinespace

\textbf{PDD} & FFT window misalignment with OFDM symbol start & $e^{-j 2\pi \varepsilon k / N_{\text{FFT}}}$ & Statistical Averaging, Linear Regression, Conjugate Multiplication, CSI Ratio & Averaging removes the random component but requires a static channel; linear regression corrects the slope. \\

\bottomrule
\end{tabular}
\end{table*}

\paragraph{Statistical Averaging}
This method relies on the law of large numbers to mitigate random, zero-mean errors like CPO and PDD. By collecting $N$ CSI snapshots over a short window where the physical channel is assumed static, the random phase errors average out:
\begin{equation}
\bar{H}_{i,j,k} = \frac{1}{N} \sum_{n=1}^{N} \hat{H}_{i,j,k}(t_n).
\end{equation}
While simple and hardware-agnostic, this approach significantly increases latency and fails in dynamic scenarios where the target moves faster than the averaging window allows.

\paragraph{Linear Regression}
Targeting the linear-in-frequency distortions (SFO and PDD), this technique models the unwrapped phase $\psi_{i,j,k}$ as a linear function of the subcarrier index $k$. The observed phase is approximated as:
\begin{equation}
\psi_{i,j,k}(t) \approx \angle H_{i,j,k}(t) - 2\pi \left(\frac{\beta + \varepsilon}{N_{FFT}}\right) k.
\end{equation}
A least-squares regression is applied to estimate the slope $\hat{\alpha}_{i,j}$ that minimizes the phase error. The compensated phase is then obtained by removing this linear trend:
\begin{equation}
\psi'_{i,j,k}(t) = \psi_{i,j,k}(t) + 2\pi \hat{\alpha}_{i,j} k.
\end{equation}
This method effectively removes the random phase slopes, as seen in systems like SpotFi~\cite{kotaru2015spotfi}. However, it inadvertently removes the physical ToF information, which also manifests as a linear slope, thus rendering the method unsuitable for absolute ranging.

\paragraph{Uplink-Downlink Multiplication}
This active compensation strategy leverages channel reciprocity to cancel time-varying offsets (CFO/CPO). By exchanging packets bidirectionally, the offsets appear with opposite signs in the Uplink (UL) and Downlink (DL) channels:
\begin{equation}
\hat{H}^{\text{DL}}_{i,j,k} \propto e^{-j(\gamma_c t + \phi_c)}, \quad \hat{H}^{\text{UL}}_{j,i,k} \propto e^{+j(\gamma_c t + \phi_c)}.
\end{equation}
Multiplying these two measurements cancels the exponential terms:
\begin{equation}
\hat{H}^{\text{comp}}_{i,j,k} = \hat{H}^{\text{DL}}_{i,j,k} \cdot \hat{H}^{\text{UL}}_{j,i,k} \approx (H_{i,j,k})^2.
\end{equation}
Used in systems like Chronos~\cite{vasisht2016decimeter}, this method recovers a stable signal magnitude but requires strict synchronization of bidirectional traffic, which introduces protocol overhead.

\paragraph{Conjugate Multiplication Across Antennas}
For multi-antenna receivers, all RF chains are driven by a single local oscillator, sharing identical phase errors $\Phi_{\text{err}}$. We can exploit this by multiplying the CSI of one antenna $j_1$ by the conjugate of another antenna $j_2$:
\begin{align}
\hat{H}^{\text{comp}}_{i,j_1,j_2,k} &= \hat{H}_{i,j_1,k} \cdot (\hat{H}_{i,j_2,k})^* \nonumber\\
&= (H_{i,j_1,k} e^{-j2\pi \Phi_{\text{err}}}) \cdot (H_{i,j_2,k} e^{-j2\pi \Phi_{\text{err}}})^* \nonumber\\
&= H_{i,j_1,k} \cdot H^*_{i,j_2,k}.
\end{align}
The error terms cancel completely. This operation preserves the \textit{relative} phase difference between antennas, which is sufficient for Angle of Arrival (AoA) estimation (e.g., in Widar2.0~\cite{qian2018widar2}), though absolute channel information is lost.

\paragraph{CSI Ratio}
The CSI Ratio method extends the cancellation principle to both amplitude and phase. By computing the complex division of CSI from two antennas:
\begin{equation}
R_{j_1,j_2,k} = \frac{\hat{H}_{i,j_1,k}}{\hat{H}_{i,j_2,k}} = \frac{H_{i,j_1,k}}{H_{i,j_2,k}}.
\end{equation}
All common-mode multiplicative distortions (gain variations, phase noise) are eliminated. This yields a highly stable metric for detecting subtle changes (e.g., respiration in FarSense~\cite{zeng2019farsense}), at the cost of losing absolute signal strength and phase reference.

\subsubsection*{Summary and Design Implications}
The selection of a compensation technique represents a fundamental design trade-off. Simple methods like Linear Regression can sanitize signal structure but strip away critical absolute delay information. More robust hardware-based methods, such as CSI Ratio, effectively eliminate noise but reduce the sensing dimensionality to relative measurements. Consequently, while these algorithmic sanitization techniques are sufficient for classification or relative tracking, they fall short for tasks requiring precise absolute ranging. As we will discuss in Sec.~\ref{sec:temporal}, achieving absolute ranging necessitates a move beyond post-processing—specifically, the transition to Monostatic Sensing configurations that inherently eliminate synchronization errors by design.

\subsection{Derived Sensing Features from CSI}
\label{ssec:derived_features}

\begin{figure}[t]
\centering
\includegraphics[width=\columnwidth]{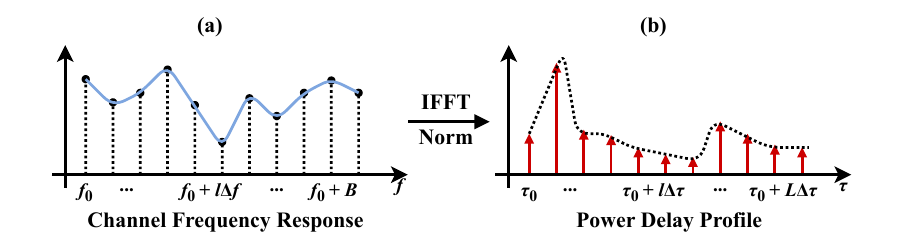} 
\caption{Transformation from frequency-domain CSI to time-domain PDP via IFFT and norm operation~\cite{xie2015precise}.}
\label{fig:cfr_to_pdp}
\end{figure}

Raw CSI parameters are often transformed into higher-level representations to provide intuitive ``images'' of channel characteristics. This section introduces three critical derived features used in advanced Wi-Fi sensing.

\subsubsection{Power Delay Profile (PDP)}
The PDP visualizes channel structure in the time-delay domain. As illustrated in Fig.~\ref{fig:cfr_to_pdp}, the process begins by applying an Inverse Fast Fourier Transform (IFFT) to the frequency-domain CSI samples. This transformation yields the Channel Impulse Response (CIR), denoted as $h(\tau)$, which characterizes the channel as a superposition of delayed impulses. The PDP is then computed as the squared magnitude of the CIR: $P(\tau) = |h(\tau)|^2$.
The resulting peaks correspond to the ToF of dominant multipath components, making PDP a fundamental tool for ranging and localization.

\subsubsection{Body-coordinate Velocity Profile (BVP)}
\label{para:bvp_def}
Introduced by Widar3.0, the BVP addresses the location-dependency of traditional device-centric features (e.g., Doppler spectrograms). BVP is a domain-invariant representation that projects signal power onto a coordinate system aligned with the human body. By aggregating Doppler Frequency Shifts (DFS) from multiple observations and applying sparse recovery, it reconstructs the velocity distribution of specific body parts (e.g., torso, limbs). This isolates intrinsic kinetic signatures from environmental geometry, enabling robust activity recognition across different locations and orientations without retraining.

\begin{figure}[t]
\centering
\includegraphics[width=0.5\columnwidth]{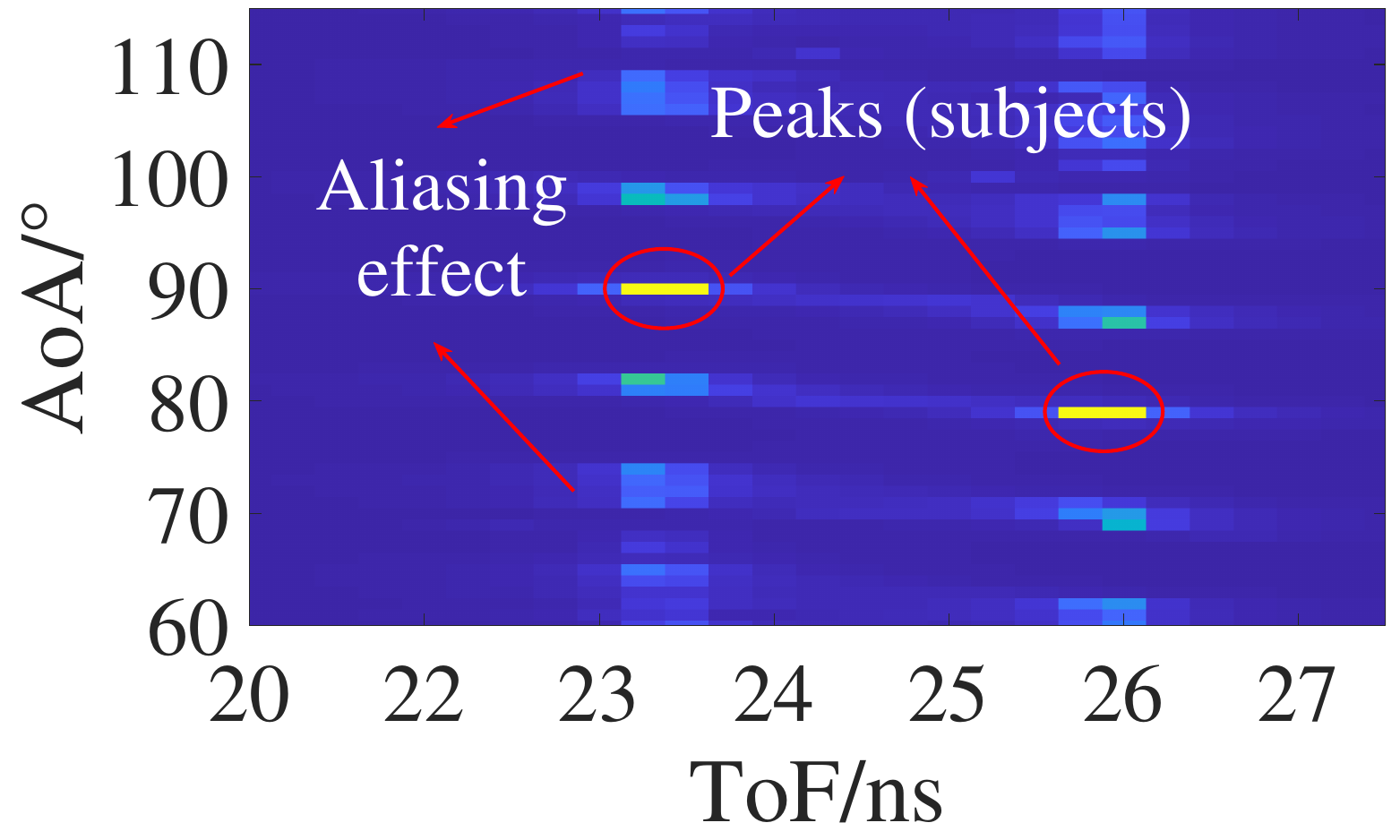} 
\caption{A ToF-AoA spectrum generated from CSI, where peaks indicate the ToF and AoA of distinct signal paths~\cite{li2024uwb}.}
\label{fig:tof_aoa_spectrum}
\vspace{-1em}
\end{figure}

\subsubsection{ToF-AoA Spectrum}
Multi-antenna systems benefit from visualizing channel structure jointly in time and space via the ToF-AoA Spectrum. This 2D ``image'' maps signal intensity against distance (ToF) and angle (AoA). Unlike PDP, this spectrum requires 2D super-resolution algorithms (e.g., 2D-MUSIC used in SpotFi) applied to spatio-temporal CSI data rather than a simple IFFT. As shown in Fig.~\ref{fig:tof_aoa_spectrum}, distinct peaks indicate different subjects or reflectors. While artifacts like periodic aliasing may appear due to limited antenna counts, they can typically be filtered to isolate targets of interest.

\subsection{Multipath Parameter Estimation: Super-Resolution Algorithms}
\label{ssec:super_resolution_algos}

Estimating multipath parameters (e.g., ToF, AoA, Doppler) is restricted by the limited bandwidth and aperture of commodity Wi-Fi. Basic transformation methods, such as IFFT, are fundamentally bound by these physical constraints, resulting in coarse meter-level uncertainties. To overcome these resolution barriers, super-resolution algorithms exploit signal structures—subspace orthogonality, sparsity, or statistical likelihood—to achieve fine-grained accuracy. This section reviews three primary classes of these algorithms.

\paragraph{\textbf{Subspace-Based Methods (MUSIC)}}
MUSIC separates signal and noise subspaces via eigen-decomposition of the CSI covariance matrix. It exploits the orthogonality between the true signal steering vector $\mathbf{s}(\theta)$ and the noise eigenvectors $\mathbf{E}_N$ to identify parameters by searching for peaks in the pseudospectrum:
\begin{equation}
    P_{\text{MUSIC}}(\theta) = \frac{1}{\mathbf{s}(\theta)^H \mathbf{E}_N \mathbf{E}_N^H \mathbf{s}(\theta)}.
\end{equation}
While MUSIC offers asymptotically infinite resolution in high-SNR regimes, exceeding physical array limits, its performance degrades significantly with coherent multipath signals, often necessitating spatial smoothing preprocessing. Furthermore, the requirement for an exhaustive grid search imposes a heavy computational burden, challenging its deployment in real-time embedded systems.

\paragraph{\textbf{Sparse Recovery Methods (OMP and IHT)}}
Leveraging the inherent sparsity of indoor multipath, these methods formulate estimation as finding a sparse vector $\mathbf{x}$ in the linear system $\mathbf{h} = \mathbf{\Psi}\mathbf{x}$, where $\mathbf{\Psi}$ is an overcomplete dictionary. Algorithms like \textit{Orthogonal Matching Pursuit (OMP)} solve this greedily by iteratively selecting the dictionary atom most correlated with the residual signal, offering high computational efficiency. Alternatively, \textit{Iterative Hard Thresholding (IHT)} combines gradient descent with a hard sparsity constraint, minimizing reconstruction error while forcing weak components to zero. Both approaches strike a balance between resolution and speed to facilitate real-time tracking, though stability can be compromised if dictionary columns are highly coherent.

\paragraph{\textbf{Iterative Refinement Methods (SAGE)}}
The Space-Alternating Generalized Expectation-maximization (SAGE) algorithm avoids the dimensionality curse by decomposing the joint multi-path search into a sequence of single-path estimations. As an extension of the Expectation-Maximization (EM) algorithm, it operates in cycles, estimating parameters for one path while ``subtracting'' the interference of all others ($L-1$) from the received signal. This interference cancellation capability allows SAGE to handle highly correlated signals robustly—a key advantage over MUSIC—and converges to optimal Maximum Likelihood (ML) estimates without requiring the computation of large covariance matrices.

\subsection{Beamforming Feedback Information (BFI): A Digital Proxy for Physical Beams}
\label{ssec:bfi_intro}

While CSI represents the raw channel state matrix ($\mathbf{H}$) estimated at the receiver, Beamforming Feedback Information (BFI) is its compressed spatial derivative fed back to the transmitter. Mathematically, BFI is obtained by performing Singular Value Decomposition (SVD) on the CSI, discarding amplitude and absolute phase to retain only the optimal spatial steering directions (the $\mathbf{V}$ matrix). Consequently, BFI serves as a digital ``instruction manual'' that dictates how the Access Point (AP) shapes its radiated energy. To understand its sensing capability, we must first examine how it is acquired via protocol, dissect its physical role in shaping spatial beams, and finally evaluate its unique signal characteristics.


\subsubsection{Protocol Mechanism: The Channel Sounding Process}
\label{ssec:bfi_protocol}
The acquisition of BFI is governed by the standardized channel sounding protocol (e.g., IEEE 802.11ac/ax/be). The process initiates when the AP broadcasts a Null Data Packet (NDP). Upon reception, a specific user $u$ estimates its downlink channel matrix $\mathbf{H}_u$. To generate the feedback, the user decomposes this channel using SVD:
\begin{equation}
    \mathbf{H}_u = \mathbf{U}_u \mathbf{\Sigma}_u \mathbf{V}_u^H,
\end{equation}
where $\mathbf{U}_u$ and $\mathbf{V}_u$ are the unitary matrices representing the receive and transmit spatial directions for user $u$, respectively, and $\mathbf{\Sigma}_u$ contains the singular values. The right singular matrix $\mathbf{V}_u$, which encodes the optimal spatial transmission directions, is compressed into quantized angles and fed back to the AP as the \textit{BFI}.  Since $\mathbf{V}_u$ is a direct mathematical derivative of $\mathbf{H}_u$, it inherently mirrors the physical channel state. Consequently, any environmental variation captured by the raw CSI is propagated into this feedback, establishing the fundamental validity of BFI as a sensing signal.

\subsubsection{Physical Essence (Mapping Feedback to Spatial Beams)}
\label{ssec:bfi_physics}
To comprehend the physical meaning of BFI, we must analyze how the AP utilizes this feedback to construct the precoding (beamforming) matrix $\mathbf{W}_u$. In commodity Wi-Fi systems, to maximize transmission efficiency, the AP typically derives the precoder directly from the feedback matrix, i.e., $\mathbf{W}_u = \mathbf{V}_u$~\cite{he2025beamfi}.
This engineering reality implies that BFI is a direct digital representation of the physical beam. We analyze this representation in two distinct operational modes:

\paragraph{\textbf{SU-MIMO}}
In the simplest Single-User Multiple-Input Multiple-Output (SU-MIMO) scenario, the AP serves a single client. The system aims to maximize the received signal-to-noise ratio (SNR) by aligning the transmitted beam with the dominant channel paths. The received signal model is expressed as:
\begin{equation}
\label{eq:su_mimo_model}
    \mathbf{y}_u = \mathbf{H}_u \mathbf{W}_u \mathbf{x}_u + \mathbf{n}_u,
\end{equation}
where $\mathbf{x}_u$ is the data stream and $\mathbf{n}_u$ is noise. From this equation, it is evident that the effective channel observed by the receiver is no longer the raw physical channel $\mathbf{H}_u$, but the composite response $\mathbf{H}_u \mathbf{W}_u$.

\paragraph{\textbf{MU-MIMO}}
The scenario becomes more complex in Multi-User Multiple-Input Multiple-Output (MU-MIMO) groups, where the AP simultaneously serves a set of users $\mathcal{G}$. The received signal for a specific user $u$ is:
\begin{equation}
\label{eq:mu_mimo_model}
    \mathbf{y}_u = \mathbf{H}_u \mathbf{W}_u \mathbf{x}_u + \underbrace{\sum_{u' \in \mathcal{G} \setminus \{u\}} \mathbf{H}_u \mathbf{W}_{u'} \mathbf{x}_{u'}}_{\text{Multi-User Interference}} + \mathbf{n}_u.
\end{equation}
Here, the summation term represents the signal leakage from beams intended for other users ($u'$). 

\subsubsection{Characteristics: Stability vs. Fidelity}
\label{sssec:bfi_characteristics}
The distinction between CSI and BFI reflects a fundamental trade-off between signal stability and information granularity.

\begin{figure}[t]
    \centering
    \setlength\abovecaptionskip{3pt} 
    
    \subfloat[Respiration.]{
        \includegraphics[height=2.1cm]{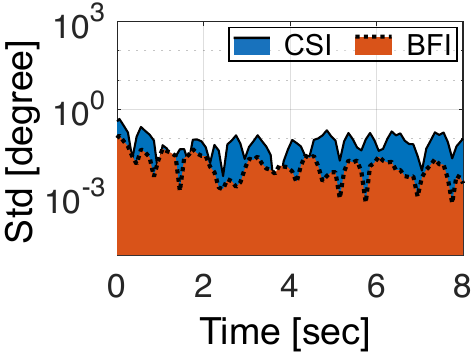}
        \label{fig:bfi_csi_compare_box_re}
    }
    \hfill
    \subfloat[Gesture.]{
        \includegraphics[height=2.1cm]{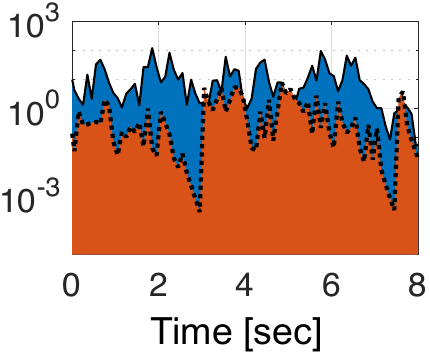}
        \label{fig:bfi_csi_compare_box_ge}
    }
    \hfill
    \subfloat[Activity.]{
        \includegraphics[height=2.1cm]{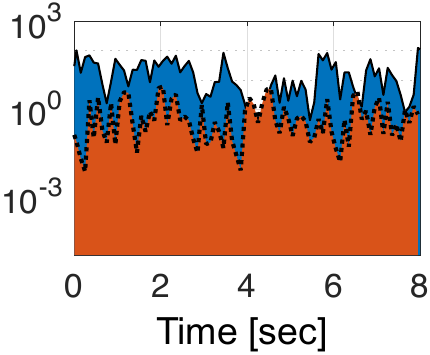}
        \label{fig:bfi_csi_compare_box_ac}
    }
    
    \vspace{1pt} 
    
    \subfloat[Respiration.]{
        \includegraphics[height=2.2cm]{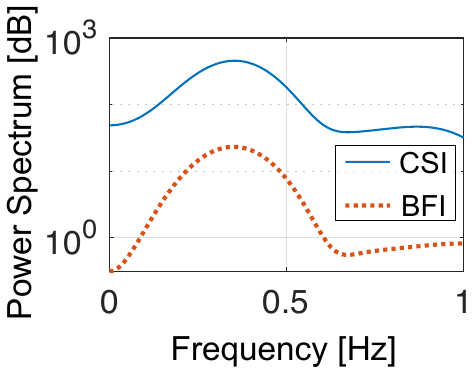}
        \label{fig:bfi_csi_compare_time_re}
    }
    \hfill
    \subfloat[Gesture.]{
        \includegraphics[height=2.2cm]{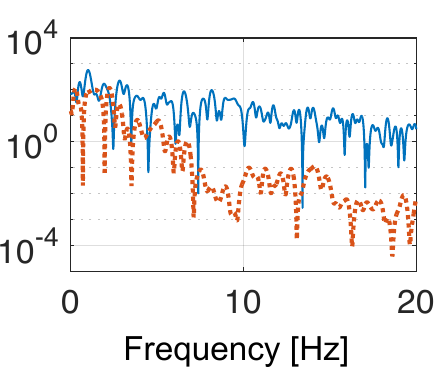}
        \label{fig:bfi_csi_compare_time_ge}
    }
    \hfill
    \subfloat[Activity.]{
        \includegraphics[height=2.2cm]{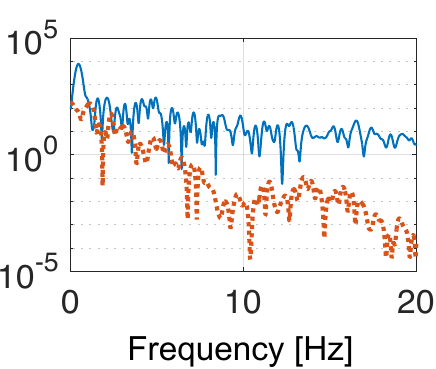}
        \label{fig:bfi_csi_compare_time_ac}
    }
    
    \caption{Comparison of signal characteristics between CSI (blue) and BFI (orange). Top Row (a)-(c): Standard Deviation (Std) analysis showing BFI's reduced jitter. Bottom Row (d)-(f): Power Spectrum demonstrating the low-pass filtering effect of BFI generation~\cite{hu2023muse}.}
    \label{fig:bfi_csi_compare}
\end{figure}

\paragraph{\textbf{Merit: Inherent Stability}}
BFI exhibits superior stability compared to raw CSI. As analyzed in MUSE-Fi~\cite{hu2023muse}, raw CSI is sensitive to time-variant hardware offsets (e.g., CFO and PDD), resulting in significant signal jitter (blue areas in Fig.~\ref{fig:bfi_csi_compare}).
In contrast, BFI effectively mitigates these fluctuations. This is primarily because the hardware-induced offsets and random noise are partially decomposed and compressed during SVD. Furthermore, the standard quantization process imposes a discrete resolution limit, where minor noise fluctuations fall within the quantization intervals and are ignored, leaving only significant structural changes in the feedback.

\paragraph{\textbf{Limitation: Information Loss and Sparsity}}
However, stability comes at the cost of fidelity. Spatially, BFI discards amplitude information (\(\boldsymbol{\Sigma}\)) and absolute phase, rendering it ill-suited for attenuation-based ranging~\cite{jiang2023design}. Temporally, unlike continuous CSI streams, real-world BFI is inherently sparse and bursty. It is generated only during traffic-dictated sounding intervals, leading to irregular sampling rates that challenge continuous tracking---a bottleneck addressed by advanced recovery algorithms in Sec.~\ref{ssec:musefi}.

\vspace{1em}
\noindent
\textbf{Transition to Diversity Domains.} 
Equipped with these fundamental signal models and sanitization techniques, we have established the ``physical toolbox'' for Wi-Fi sensing. However, realizing high-performance ISAC demands more than clean signals; it requires strategically exploiting physical dimensions to resolve sensing ambiguities. In the following sections, we dissect this exploitation across four orthogonal domains, beginning with \textbf{Temporal Diversity} (Sec.~\ref{sec:temporal}), where we address the challenge of synchronization to unlock absolute ranging.

\section{Temporal Diversity: Achieving Synchronized ISAC via Monostatic Signal Separation}
\label{sec:temporal}

This chapter explores the realization of \textit{Synchronized ISAC} through \textbf{Temporal Diversity}. We begin by analyzing the fundamental limitations of standard bistatic Wi-Fi and demonstrating why a transition to a monostatic architecture is essential for absolute ranging. The discussion then centers on the primary engineering barrier of this transition—self-interference—and introduces the two dominant signal separation methodologies: \textit{Active Cancellation} and \textit{Physical Separation}. Furthermore, we present a detailed case study of ISAC-Fi to exemplify how multi-stage cancellation transforms commodity Wi-Fi into a fully synchronized, radar-like sensor. Finally, we conclude with a comparative analysis of these paradigms.

\subsection{Architectural Foundations: The Shift from Bistatic to Monostatic Sensing}
\label{ssec:temporal_concept}

\begin{figure}[t]
  \centering
  \setlength{\abovecaptionskip}{6pt}
  \subfloat[Conventional bistatic sensing.]{
    \includegraphics[width=0.9\linewidth]{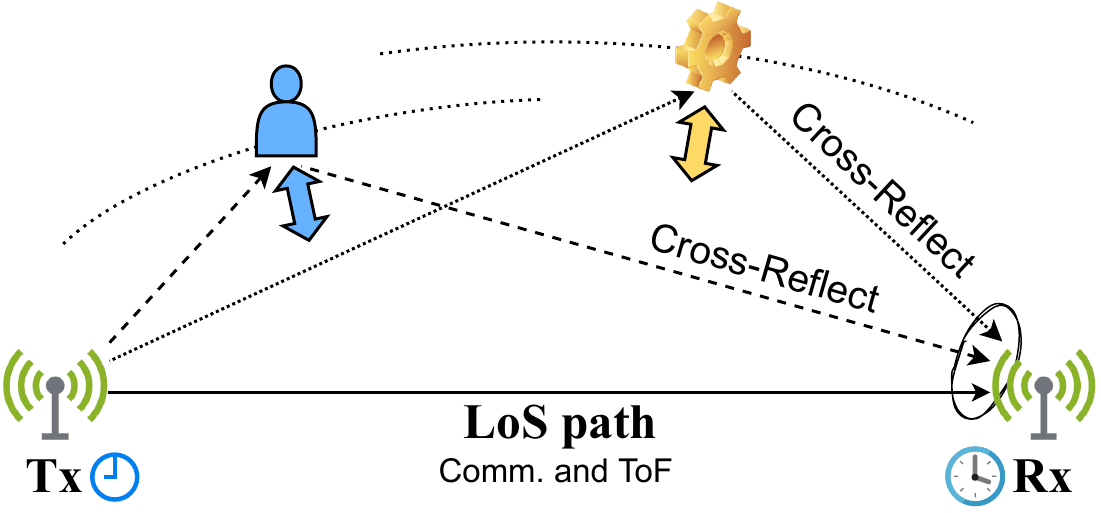}
    \label{sfig:bistatic}
  }
  \vspace{1ex} 
  \subfloat[Monostatic sensing.]{
    \includegraphics[width=0.9\linewidth]{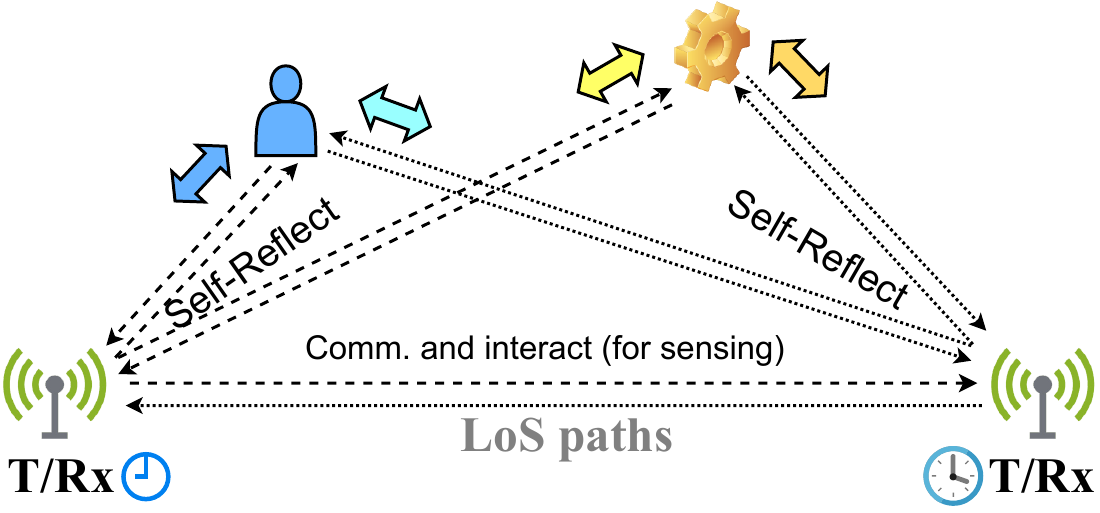}
    \label{sfig:monostatic}
  }
  \caption{Bistatic (a) vs Monostatic (b) sensing. The thin arrows represent RF propagation, while thick arrows denote sensed subject motions~\cite{chen2024isac}.}
  \label{fig:AoA spectrum}
  \vspace{-1em}
\end{figure}

Temporal Diversity empowers the system to utilize a unified time reference to acquire precise time-domain measurement capabilities. Unlike bistatic systems where the lack of a common clock introduces random timing jitters, a unified reference ensures that the received signal's delay and phase are deterministically locked to the physical propagation path. This synchronization serves as the cornerstone of precision sensing, enabling absolute ranging and sub-centimeter motion tracking with perfect phase coherence. In this section, we examine how achieving this diversity necessitates a fundamental architectural shift from asynchronous bistatic topologies to synchronized monostatic designs, while exposing the critical self-interference barrier that emerges from this transition.

\subsubsection{The Synchronization Barrier and Geometric Ambiguity in Bistatic Sensing}
\label{ssec:geo_ambiguity}
Traditional Wi-Fi sensing has predominantly operated in a bistatic configuration (Fig.~\ref{sfig:bistatic}), where physically separated Tx and Rx rely on independent local oscillators. While sufficient for communication, this topology suffers from two fundamental limitations:

\begin{itemize}
    \item \textbf{Clock Asynchrony (The Temporal Barrier):} The independent oscillators introduce random hardware offsets (i.e., CFO, CPO, SFO, and PDD). Because the receiver lacks knowledge of the exact transmission instant, it cannot distinguish between the phase slope caused by physical propagation and that caused by hardware timing errors. Consequently, bistatic systems are restricted to measuring \emph{relative} changes, rendering absolute ranging mathematically infeasible without complex compensation.

    \item \textbf{Geometric Ambiguity (The Spatial Barrier):} Beyond the temporal barrier, bistatic setups face inherent limitations due to the Fresnel-zone model~\cite{wang2016human}. A single ToF measurement defines an elliptical locus rather than a specific point, creating \emph{location ambiguity}. Furthermore, motion along the tangential direction of these ellipses does not alter path length, resulting in \emph{motion blindness} where significant movements remain undetectable due to the spatially varying sensitivity~\cite{chen2024isac}.
\end{itemize}

\subsubsection{The Monostatic Solution and the Interference Paradox}
To recover Temporal Diversity, the optimal strategy is the monostatic architecture (Fig.~\ref{sfig:monostatic}), where Tx and Rx share a unified hardware clock. This co-location theoretically eliminates synchronization errors, enabling absolute ToF measurement while removing the geometric blind spots inherent to bistatic Fresnel models.
However, this architectural shift introduces a fundamental paradox: the signal separation problem. Given Wi-Fi's half-duplex nature and continuous OFDM transmission, the powerful transmit signal leaks directly into the receiver. This results in overwhelming self-interference—often $80\sim100$\,dB stronger than the weak environmental reflections. Consequently, realizing Temporal Diversity becomes an engineering battle to suppress this leakage. Only by effectively isolating the informative echoes from the strong self-interference can the synchronized clock be exploited for absolute ranging.

\subsection{Methodologies for Self-Interference Management}
With signal separation established as the key enabler for Temporal Diversity, we categorize state-of-the-art solutions into two distinct philosophies: \textbf{Active Cancellation} (subtracting interference) and \textbf{Physical Separation} (avoiding interference).

\subsubsection{Active Self-Interference Cancellation (SIC)}
The first philosophy follows a ``receive first, cancel late'' principle. It allows the Tx leakage and Rx reflections to mix at the receiver front-end but employs a multi-stage pipeline—combining analog RF suppression and digital filtering—to reconstruct and subtract the interference signal. This approach aims to recover the complete channel response, including both static and dynamic components. By actively cancelling the known transmission signal from the received mixture, the system can expose the weak environmental echoes. The ISAC-Fi~\cite{chen2024isac} system is the prime exemplar of this philosophy. Given its significance as a comprehensive framework that fully unlocks absolute ranging capabilities, we dedicate Sec.~\ref{ssec:monostatic_sic} to a detailed case study of this approach.

\subsubsection{Physical Signal Path Separation}

\begin{figure}[t]
\vspace{-1em}
\setlength{\abovecaptionskip}{6pt}  
    \centerline{
    \includegraphics
    [width=0.8 \columnwidth]{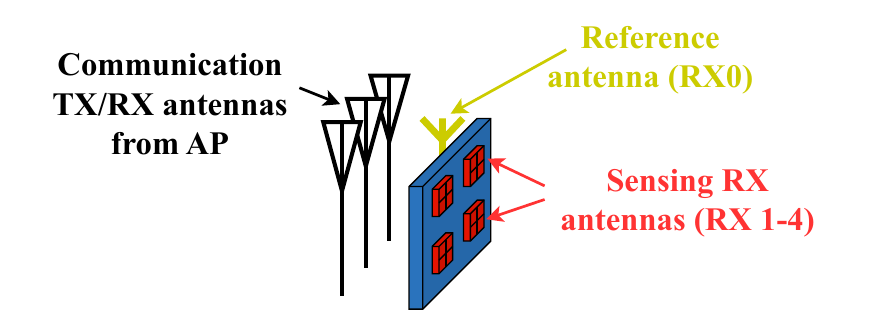}
    }
    \caption{Physical separation design of SiWiS using a dual-antenna system~\cite{song2024siwis}.}
    \label{fig:siwis}
\end{figure}

The second philosophy, ``prevention at the source,'' uses hardware design to physically isolate the sensing path from the leakage path at the RF front-end. A notable example is SiWiS~\cite{song2024siwis}, which introduces a novel self-mixing architecture. By utilizing a dedicated reference antenna to capture the clean transmission signal and mixing it with the environmental reflections, SiWiS achieves synchronization via analog self-mixing, bypassing the elaborate reconstruction pipeline required for full channel recovery.

\begin{itemize}
    \item \textbf{Hardware Architecture:} As shown in Fig.~\ref{fig:siwis}, SiWiS employs a specialized dual-antenna system configuration. A dedicated reference antenna (RX0) is positioned to capture the clean, high-power signal directly from the device's own transmitter. Simultaneously, a separate sensing antenna array (RX1--4) is oriented towards the environment to capture weak reflections while minimizing direct leakage~\cite{song2024siwis}.
    
    \item \textbf{Self-Mixing Mechanism:} The core innovation lies in using the signal from RX0 as the Local Oscillator (LO) to down-convert the reflection signals captured by the sensing antennas. Since both the LO (Tx signal) and the RF input (Reflection) originate from the same hardware source, they share identical phase noise and frequency offsets. Mixing them effectively cancels out these common-mode errors (CFO/CPO/STO), producing a phase-coherent baseband signal where phase variations depend \emph{exclusively} on the target's propagation delay~\cite{song2024siwis}.
    
    \item \textbf{Processing and Limitation:} To further extract stable features from the time-varying OFDM data, SiWiS integrates the mixed signal over the Long Training Field (L-LTF) of the packet preamble and removes the DC component. This effectively filters out the static leakage (and static objects). Crucially, this implies a trade-off: while SiWiS achieves perfect synchronization for motion tracking (leveraging Temporal Diversity for phase coherence), it discards the absolute delay information of static objects, making it specialized for fine-grained dynamic sensing rather than absolute ranging~\cite{song2024siwis}.
\end{itemize}

\subsection{Case Study: Realizing Absolute Ranging via Active Self-Interference Cancellation}
\label{ssec:monostatic_sic}

To demonstrate how \textbf{Temporal Diversity} can be fully realized to support absolute ranging, we discuss the methodology of ISAC-Fi system~\cite{chen2024isac} as an illustrative case. ISAC-Fi adopts a hybrid, multi-stage cancellation pipeline that follows a ``coarse-to-fine'' principle. The rationale lies in the extreme dynamic range between the transmitted leakage and the reflected echoes, which often exceeds 80\,dB~\cite{chen2024isac}. No single technique can bridge this gap: purely digital cancellation fails because the ADC would be saturated by the strong leakage, irreversibly clipping the weak reflections; conversely, purely analog cancellation lacks the precision to suppress interference down to the thermal noise floor. Therefore, ISAC-Fi orchestrates three sequential stages—supplemented by a critical calibration mechanism—to systematically suppress self-interference.

\subsubsection{System Design}
The ISAC-Fi pipeline is composed of three sequential cancellation stages—RF front-end isolation, analog cancellation, and digital cancellation—supplemented by a self-adapted calibration mechanism. Together, these components form a hybrid ``coarse-to-fine'' architecture that progressively suppresses self-interference while safeguarding the weak environmental reflections essential for sensing.

\begin{figure}[t]
    \setlength\abovecaptionskip{8pt}
    \vspace{-.5ex}
    \centering
    \includegraphics[width=.96\linewidth]{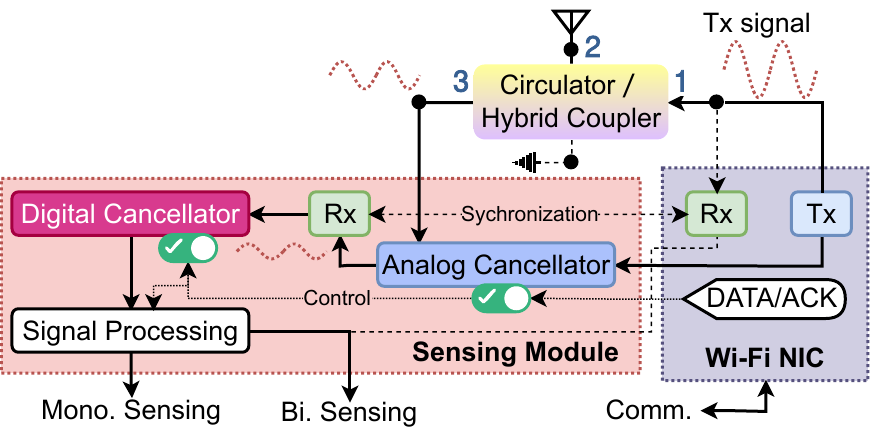}
    \caption{Tx-Rx separation for ISAC-Fi: The Tx-Rx separator orchestrates three levels of suppression: physical isolation via a Circulator/Hybrid Coupler, active analog cancellation via DQM, and fine-grained digital cancellation~\cite{chen2024isac}.}
    \label{fig:isac-fi}
    \vspace{-1.5ex}
\end{figure}

\paragraph{Stage 1---RF Front-End Isolation}
The first line of defense against self-interference is to create physical isolation between the transmit (Tx) and receive (Rx) paths at the RF front-end. ISAC-Fi achieves this by replacing the conventional Wi-Fi Tx/Rx switch—which only allows for half-duplex operation—with a passive component that enables concurrency. Depending on the specific hardware design, this component is typically either a Circulator or a Hybrid Coupler. These components function as RF traffic routers. A circulator, for example, is a three-port device that directs energy in a fixed rotation. As illustrated in Fig.~\ref{fig:isac-fi}, this enables a precise signal flow: the high-power transmitted signal enters at Port 1 and is directed to the antenna at Port 2. The weak, reflected signal then returns from the antenna to Port 2 and is routed to the receiver's sensing chain at Port 3. This provides an initial, passive level of isolation that, while insufficient on its own, is critical for reducing the burden on the subsequent active cancellation stages.

\paragraph{Stage 2---Analog Cancellation (Coarse Suppression)}
Since front-end isolation is imperfect, strong leakage still enters the receiver. The analog cancellator aims to suppress this signal before it saturates the Low Noise Amplifier (LNA) and ADC. ISAC-Fi employs a Direct Quadrature Modulator (DQM) to generate a cancellation signal. The DQM takes a copy of the baseband Tx signal and adjusts its amplitude and phase to create an inverted replica of the leakage ($G_A \approx -G_H$). The signal at this stage can be modeled as:
\begin{equation}
z(t) = [G_A \cdot x(t - \tau_0) + G_H \cdot x(t - \tau_0)] + G_C \cdot x(t - \tau_{l > 0})
\end{equation}
where $G_A$, $G_H$, and $G_C$ denote the channel gains of the analog cancellator, the RF hardware, and the circulator/hybrid coupler, respectively~\cite{chen2024isac}. $x(t - \tau_0)$ represents the self-interference signal with delay $\tau_0$, and $x(t - \tau_{l > 0})$ denotes the desired environmental reflections. Ideally, the system tunes $G_A$ to approach $-G_H$, thereby minimizing the interference term in the analog domain.


\paragraph{Stage 3---Digital Cancellation (Fine Suppression)}
Residual interference remains after the analog stage due to hardware imperfections. The digital cancellator removes this residue in the baseband domain. It leverages the known Wi-Fi packet preamble (L-STF/L-LTF) as a clean reference. An adaptive filter, trained using the Least Mean Squares (LMS) algorithm, estimates the channel of the residual leakage. The signal model at this stage is:
\begin{equation}
z[n] = [G_D \cdot s[n] + \omega'_A[n]] + x[n - \tau_{l > 0}],
\end{equation}
where $G_D$ denotes the adaptive filter coefficients obtained via the LMS algorithm, and $s[n]$ represents the baseband of the known Wi-Fi preamble (e.g., L-STF/L-LTF). $\omega'_A[n]$ represents the residue analog Tx interference that has passed through RF downconversion and ADC sampling~\cite{chen2024isac}, while $x[n - \tau_{l > 0}]$ is the desired target reflection signal. The estimated interference (the term $G_D \cdot s[n]$) is subtracted from the received signal, typically achieving a total suppression of $\approx 77$\,dB~\cite{chen2024isac}. This digital cleanup is the final step in suppressing the interference to a level sufficient for extracting reliable absolute ToF.

\begin{table*}[b] 
\centering
\small
\caption{Comparative Analysis of Temporal Diversity Methodologies}
\label{tab:temporal_comparison}
\renewcommand{\arraystretch}{1.3}
\begin{tabularx}{\textwidth}{ 
    >{\raggedright\arraybackslash\bfseries}p{4cm} 
    >{\raggedright\arraybackslash}X 
    >{\raggedright\arraybackslash}X 
}
\toprule
\textbf{Methodology} & \textbf{Active Self-Interference Cancellation (SIC)} & \textbf{Physical Signal Separation} \\
\midrule
\textbf{Representative System} & ISAC-Fi~\cite{chen2024isac} & SiWiS~\cite{song2024siwis} \\
\addlinespace
\textbf{Targeted Imperfection} & Self-Interference Leakage & Synchronization Errors (CFO/STO/CPO) \\
\addlinespace
\textbf{Primary Sensing Gain} & \textbf{Absolute ToF} (Round-Trip Time) & \textbf{Phase Coherence} (Deterministic Phase-Distance mapping) \\
\addlinespace
\textbf{Resulting Capability} & Absolute Ranging \& Single-Device Localization (Static + Dynamic) & Fine-grained Motion Tracking \& Zero-shot Generalization (Dynamic Only) \\
\addlinespace
\textbf{Key Limitation} & High hardware complexity (requires circulator/DQM/calibration). & Blind to static objects (cannot estimate absolute range $d$, only $\Delta d$). \\
\bottomrule
\end{tabularx}
\end{table*}

\paragraph{Self-Adapted Calibration---Preserving the Sensing Signal}
A critical risk in SIC systems is ``over-cancellation,'' where the system mistakenly identifies environmental reflections as interference and removes them. ISAC-Fi introduces a self-adapted calibration mechanism to solve this. It utilizes an RF switch to toggle the transmitter output between the actual antenna and a dummy load~\cite{chen2024isac}.
During calibration, the signal is sent to the dummy load, ensuring that no signal is radiated into the environment. The receiver thus captures only the internal hardware leakage. The cancellators update their parameters ($G_A$ and $G_D$) to minimize this specific signal. Once calibrated, the system switches back to the antenna. This ensures that the cancellation targets only the internal leakage profile, strictly preserving the external environmental reflections. This mechanism guarantees that the pursuit of interference suppression does not inadvertently sacrifice the very temporal diversity gain (i.e., the target reflection) we seek to harvest.

\subsubsection{Performance Evaluation and Capabilities}
To validate the transformative impact of \textbf{Temporal Diversity}, the ranging accuracy of ISAC-Fi is examined against traditional bistatic baselines.

\begin{figure}[t]
    \setlength\abovecaptionskip{6pt}
    \centering
    \includegraphics[width=0.5\linewidth]{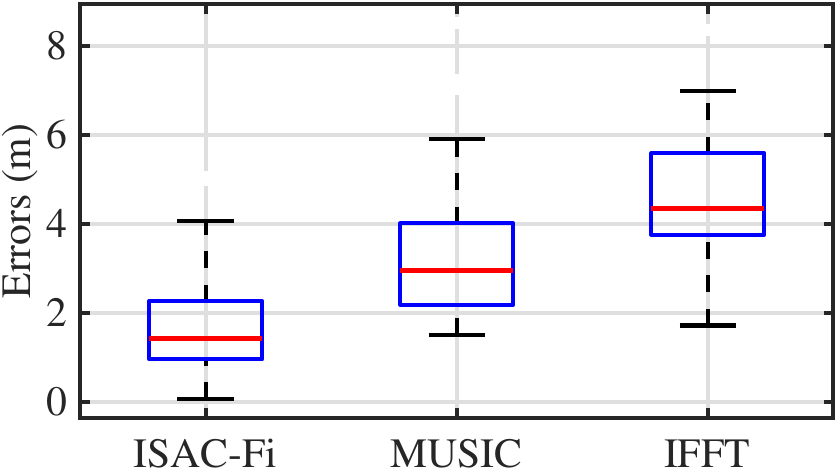}
    \caption{Ranging error comparison between ISAC-Fi (Monostatic) and traditional baselines (MUSIC/IFFT)~\cite{chen2024isac}.}
    \label{fig:ranging_errors}
\end{figure}

\paragraph{\textbf{Evaluation (Ranging Accuracy)}}
The core hypothesis of this chapter is that a unified time reference is the prerequisite for accurate absolute ranging. Fig.~\ref{fig:ranging_errors} provides empirical validation.
Standard algorithms (e.g., MUSIC and IFFT), when applied to asynchronous bistatic links, suffer from large median errors ($>3$\,m) and high variance. This is due to the stochastic nature of hardware offsets, which cannot be fully compensated. In contrast, by operating in the monostatic mode with intrinsic synchronization, ISAC-Fi achieves a median error of approximately 1.4\,m. This drastic reduction confirms that the architectural shift to a shared-clock topology effectively removes the temporal barriers preventing precise distance measurement.

\paragraph{\textbf{Unlocking Fundamental Sensing Capabilities}}
By establishing a reliable absolute ranging capability, the system unlocks a new class of sensing functions previously restricted to specialized radar:
\begin{itemize}
    \item \textbf{True Absolute Ranging:} It transitions Wi-Fi sensing from indirect or relative ranging (where distance is inferred or ambiguous) to quantitative absolute distance measurement. The receiver can now directly calculate the precise physical distance to a target based on the round-trip delay, independent of synchronization errors.
    \item \textbf{Single-Device Localization:} It breaks the reliance on multi-device triangulation. By explicitly resolving the absolute range dimension via Temporal Diversity, combined with the angle estimation from multi-antenna arrays, a single ISAC-Fi device can independently pinpoint a target's 2D coordinates~\cite{chen2024isac}.
    \item \textbf{Potential for Multi-subject Discrimination:} The transition to monostatic sensing eliminates the geometric ambiguity of bistatic models, ensuring that Doppler shifts uniquely map to absolute radial velocities. This deterministic feature provides a clear signal baseline that, when augmented with other diversities such as \textbf{Spatial Diversity} (multi-antenna arrays), enables more robust separation of multiple concurrent subjects.
\end{itemize}

\subsection{Summary: Towards Synchronized ISAC}
\label{ssec:temporal_summary}

This chapter demonstrated how \textbf{Temporal Diversity} resolves the synchronization-interference dilemma. While bistatic setups suffer from clock asynchrony (losing absolute accuracy), monostatic setups face self-interference (losing visibility). To address these challenges, we categorized the engineering solutions into two paradigms:

\begin{itemize}
    \item \textbf{Active Self-Interference Cancellation:} Employs a multi-stage pipeline (analog and digital) to actively subtract transmission leakage. By recovering the complete channel response, it enables true absolute ranging for both static and dynamic targets (e.g., ISAC-Fi~\cite{chen2024isac}), albeit with higher hardware complexity.
    
    \item \textbf{Physical Signal Separation:} Prevents interference at the source via hardware isolation and self-mixing. It achieves high-precision \textit{Phase Coherence} by eliminating clock errors, enabling robust motion tracking (e.g., SiWiS~\cite{song2024siwis}). However, its reliance on DC removal generally restricts it to dynamic sensing.
\end{itemize}

Table~\ref{tab:temporal_comparison} compares these approaches. Ultimately, these technologies transform the Wi-Fi physical layer into a precision timing instrument, achieving \textbf{Synchronized ISAC}. With accurate ranging and motion tracking established, the challenge shifts to improving measurement \emph{resolution}, a task governed by \textbf{Frequency Diversity} in the following chapter.
\section{Frequency Diversity: Achieving High-Resolution ISAC via Spectrum Expansion}
\label{sec:frequency}

This chapter explores the realization of \textit{High-Resolution ISAC} through \textbf{Frequency Diversity}. While Temporal Diversity guarantees measurement \textit{accuracy} via synchronization, Frequency Diversity fundamentally governs \textit{temporal resolution}—the capacity to distinguish closely spaced targets. We begin by quantifying the theoretical resolution limit that confines legacy narrowband Wi-Fi to coarse sensing. To dismantle this barrier, we trace the technological evolution through three distinct paradigms: from early \textit{Continuous Channel Stitching} and \textit{Multi-Band Frequency Sweeping}, to state-of-the-art \textit{Discrete Channel Sampling}. We then present a detailed case study of UWB-Fi, demonstrating how model-driven reconstruction synthesizes an effective ultra-wide bandwidth to unlock centimeter-level resolution. Finally, we conclude with a comparative analysis of these paradigms.
\subsection{Theoretical Foundations: The Bandwidth-Resolution Limit}
\label{ssec:bandwidth_resolution}

In the realm of Wi-Fi sensing, the utility of Frequency Diversity is
\lxrev{physically}
equivalent to the extent of the effective bandwidth used to probe the channel. Its primary role is to determine the temporal resolution—the minimum time delay difference required to distinguish two separate signal paths.

\subsubsection{The Physics of Resolution---From Bandwidth to Impulse Response}

To rigorously understand why bandwidth governs resolution, we must look at the time-domain representation of the channel. The Power Delay Profile (PDP) introduced in Sec.~\ref{ssec:derived_features} is essentially the squared magnitude of the Channel Impulse Response (CIR), $h(\tau)$. In any physical system with a finite bandwidth $B$, the estimated CIR is not a set of perfect Dirac delta functions, but rather the convolution of the true multipath channel with the system's point spread function.
Mathematically, the time-domain envelope of a single path with delay $\tau_0$ observed over a bandwidth $B$ is proportional to a Sinc function:
\begin{equation}
    |h(\tau)| \propto \left| \text{sinc}\left( B (\tau - \tau_0) \right) \right| = \left| \frac{\sin(\pi B (\tau - \tau_0))}{\pi B (\tau - \tau_0)} \right|.
    \label{eq:sinc_resolution}
\end{equation}
This equation reveals the physical constraint: the ``sharpness'' of a peak in the PDP is determined by the width of the Sinc function's main lobe. The first nulls of the Sinc function occur at $\pm 1/B$ from the center. According to the Rayleigh criterion, two multipath components with delays $\tau_1$ and $\tau_2$ can be resolved only if their peak separation exceeds half the main lobe width:
\begin{equation}
    \Delta \tau = |\tau_1 - \tau_2| \ge \frac{1}{B}.
\end{equation}
This $\Delta \tau$ is the fundamental temporal resolution limit. A wider effective bandwidth $B$ compresses the Sinc main lobe, reducing peak overlap and allowing the system to disentangle closely spaced reflections (e.g., distinguishing a hand from a torso).


\subsubsection{The Challenge of Narrowband Wi-Fi}

\begin{figure}[t]
  \setlength\abovecaptionskip{6pt}
  \centering
  
  \subfloat[20\,MHz, 1.2\,m.]{
    \includegraphics[width=0.3\linewidth]{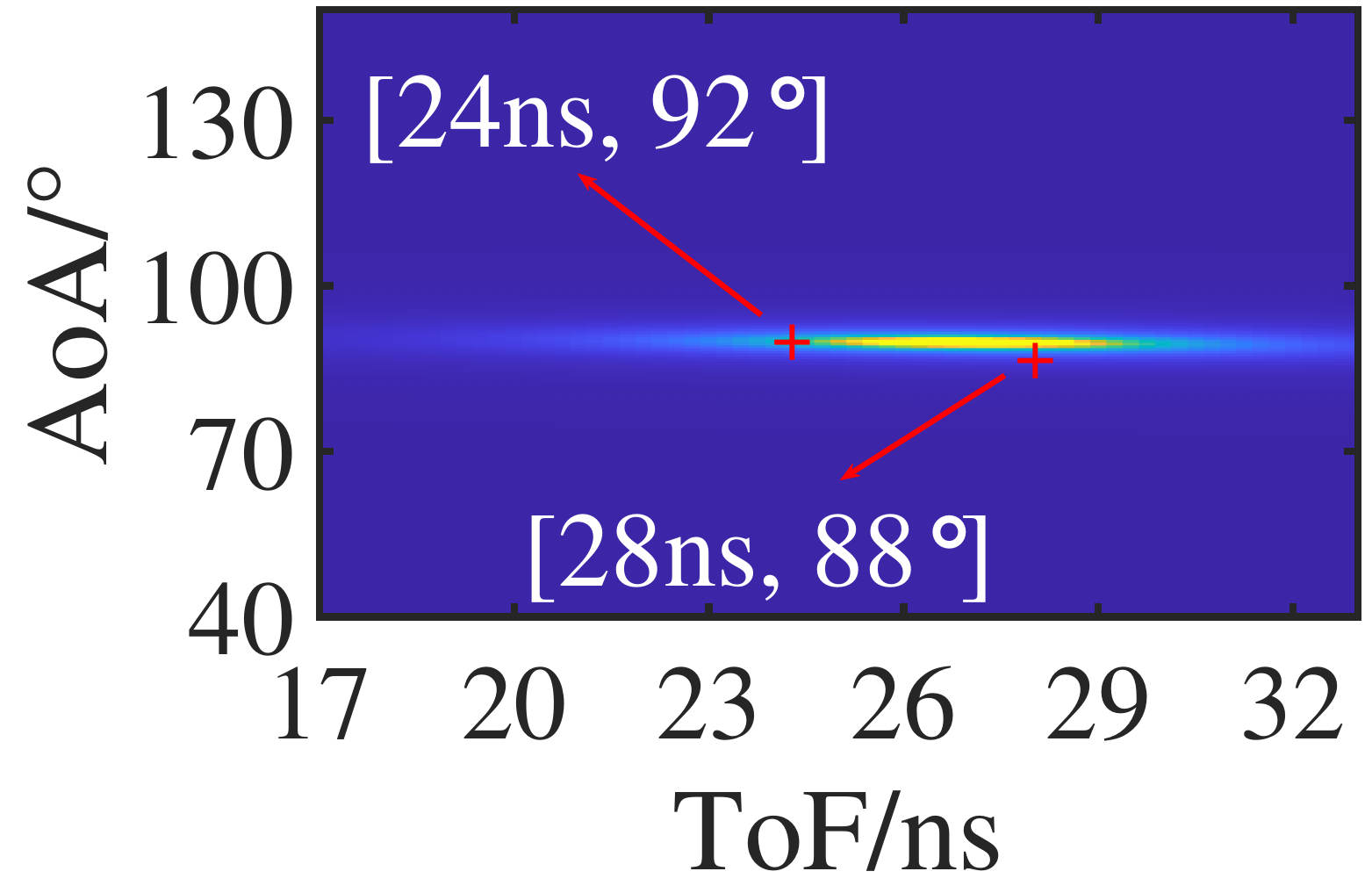}
    \label{sfig:20MHz_120cm}
  }
  \hfill
  \subfloat[320\,MHz, 1.2\,m.]{
    \includegraphics[width=0.3\linewidth]{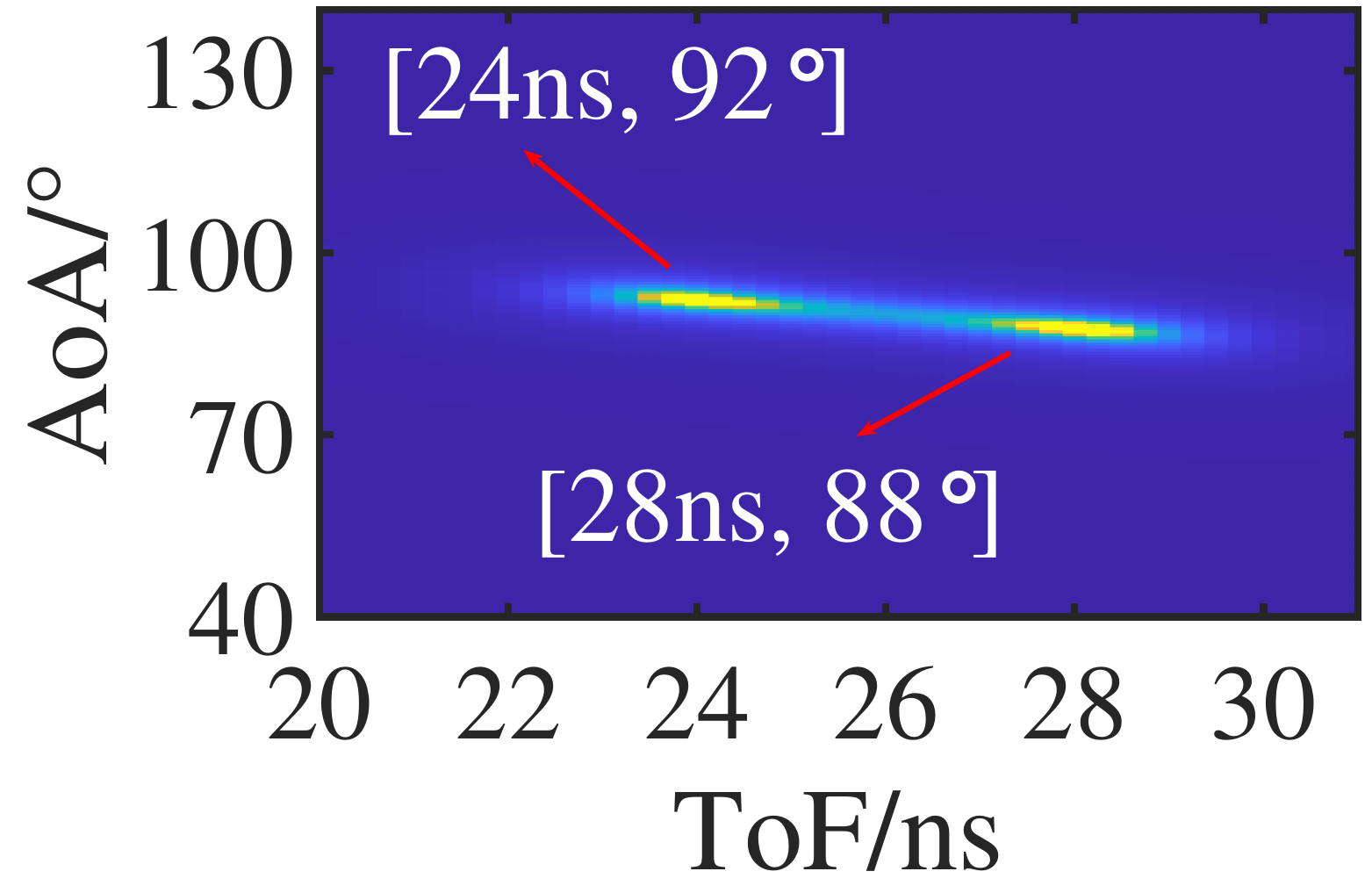}
    \label{sfig:320MHz_120cm}
  }
  \hfill
  \subfloat[320\,MHz, 0.8\,m.]{
    \includegraphics[width=0.3\linewidth]{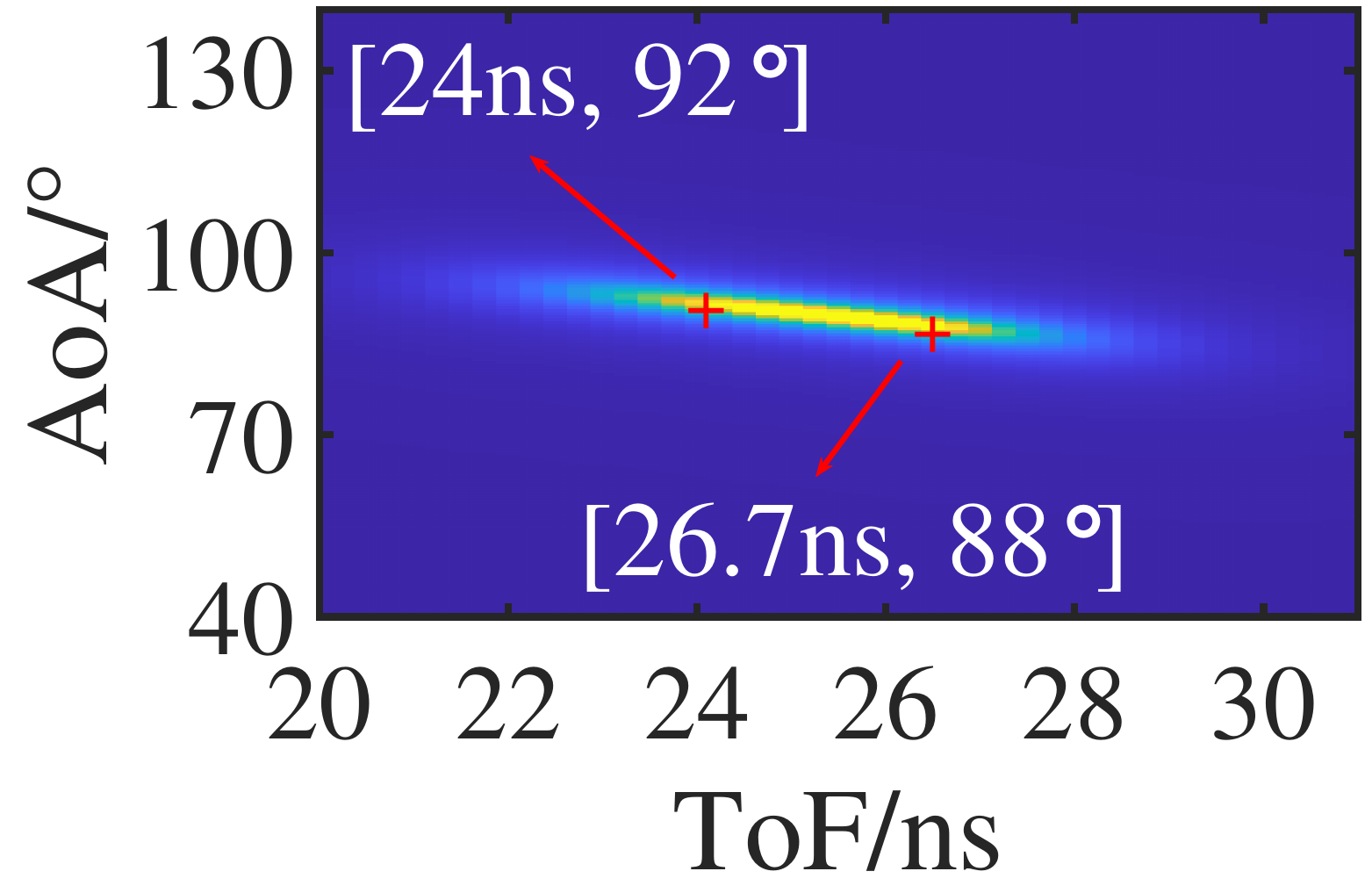}
    \label{sfig:320MHz_80cm}
  }
  
  \caption{Range resolution performance for two subjects at different distances under various Wi-Fi bandwidths~\cite{li2024uwb}.}
  \label{fig:Resolutions}
\end{figure}

Applying this \hbrev{theoretical resolution limit} to legacy Wi-Fi standards exposes their inadequacy for fine-grained sensing. A standard 802.11n channel with $B = 20$\,MHz yields a coarse resolution of $\Delta \tau = 50$\,ns, corresponding to a path length difference of $\approx 15$\,m. In typical indoor environments, multipath reflections from human limbs or nearby furniture often arrive within nanoseconds of each other. 
Consequently, as visually demonstrated in Fig.~\ref{sfig:20MHz_120cm}, narrowband sensing sees a ``blurred'' version of reality: reflections from two subjects separated by 1.2\,m merge into a single, unresolvable blob~\cite{li2024uwb}. Even with the wider 320\,MHz bandwidth introduced in Wi-Fi 7, which successfully resolves these targets as distinct peaks in Fig.~\ref{sfig:320MHz_120cm}, the resolution is still fundamentally limited to approximately 1\,m ($c/320\,\text{MHz}$). As revealed in Fig.~\ref{sfig:320MHz_80cm}, when targets move into closer proximity (e.g., 0.8\,m), the peaks merge again, failing to provide the centimeter-level granularity required for detailed human sensing or crowded multi-person scenarios~\cite{li2024uwb}. To break this persistent limit, modern ISAC systems must synthesize a much wider effective $B$ beyond standard channel limits.

\subsection{Evolutionary Path I: Continuous Channel Stitching}
\label{ssec:stitching}

The immediate engineering response to the resolution limit was to synthesize a wider aperture by aggregating adjacent channels. This approach, known as \emph{continuous channel stitching}, combines multiple narrowband CSI measurements (e.g., contiguous 20\,MHz channels) into a single, high-resolution wideband response~\cite{xie2015precise}.

\subsubsection{The Mathematical Challenge of Stitching}

Conceptually, stitching involves concatenating CSI vectors from a sequence of adjacent frequency bands. However, standard Wi-Fi transceivers are not designed for this coherent assembly. Each time the transceiver retunes, its hardware states---specifically the Phase-Locked Loop (PLL) and Automatic Gain Control (AGC)---are reset, introducing random discontinuities.
Based on the \hbrev{hardware-induced offset formulations} in Sec.~\ref{ssec:offset}, the concatenated channel $H_{\text{stitch}}(f)$ behaves as a disjoint piecewise function:
\begin{equation}
    \!\!\!H_{\text{stitch}}(f) \!=\!\!
    \begin{cases} 
      \!H_{\text{true}}(f) + n_1(f), & \!\!\! f \!\in \!\mathcal{B}_1 \\
      \!\alpha e^{-j(\Delta \phi_c + 2\pi f \delta_{\Delta})} H_{\text{true}}(f) + n_2(f), & \!\!\! f \!\in \!\mathcal{B}_2
   \end{cases}
\end{equation}
where $\mathcal{B}_1$ and $\mathcal{B}_2$ denote the frequency ranges. The parameters represent the differential errors:
\begin{itemize}
    \item $\alpha$: The amplitude scaling factor caused by inconsistent AGC settling.
    \item $\Delta \phi_c$: The jump in CPO arising because the PLL locks to a random initial phase.
    \item $\delta_{\Delta}$: The \hbrev{timing offset discontinuity, which manifests as a phase slope difference. It aggregates the sampling timing errors defined in Eqn.~\eqref{eq:SFO} and Eqn.~\eqref{eq:PDD} (i.e., SFO and PDD).}
\end{itemize}
Unless perfectly compensated, these offsets cause severe spectral leakage at the boundary, manifesting as spurious peaks in the PDP that destroy the resolution gain.

\subsubsection{Algorithmic Solutions---Splicer and ToneTrack}
To overcome these discontinuities between channels, pioneering systems developed algorithmic calibration.

\paragraph{Splicer (Stitching via PDP Invariance)}
Splicer~\cite{xie2015precise} exploits the physical invariant that the PDP represents the environment's multipath structure, which should remain consistent regardless of the carrier frequency. It iterates through potential phase and timing offsets to stitch the bands. The optimal parameters are those that minimize the entropy (sharpness) of the resulting PDP, thereby aligning the incoherent bands into a unified response.

\paragraph{ToneTrack (Resolution for ToA)}
While Splicer focuses on the profile, ToneTrack~\cite{xiong2015tonetrack} leverages stitching specifically to resolve Time-of-Arrival (ToA) ambiguity. It combines transmissions from frequency-agile radios to form a virtual wideband signal. By aligning separate transmissions in both time and frequency domains, ToneTrack achieves sub-meter localization accuracy even when the direct path is blocked or attenuated.

\subsubsection{The Scalability Bottleneck}
While continuous stitching demonstrated the feasibility of bandwidth synthesis, it faces physical limitations that prevent scaling to GHz-level \hbrev{resolution}:
\begin{itemize}
    \item \textbf{Hardware Boundary Effects:} Hardware band-pass filters exhibit non-ideal attenuation (roll-off) at channel edges. Simply concatenating adjacent channels results in distortions at the boundaries. Mitigating this requires capturing overlapping channels, which significantly reduces spectral efficiency~\cite{li2024uwb}.
    \item \textbf{Spectrum Availability:} Stitching relies on continuous blocks of clean spectrum. In crowded ISM bands, finding wide, contiguous idle channels (e.g., $>80$\,MHz) is statistically rare due to interference from other devices.
    \item \textbf{Sequential Scanning Latency:} Constructing a wide aperture requires sequentially scanning numerous channels. If the total scan time exceeds the channel coherence time
    \lxrev{($\sim$100\,ms~\cite{li2025enabling})}
    the physical channel state $H_{\text{true}}$ changes during the scan, rendering the result invalid for dynamic sensing.
\end{itemize}

These bottlenecks suggest that strictly adhering to continuity is not practical, motivating the search for alternative strategies that synthesize bandwidth across non-contiguous frequencies.

\subsection{Evolutionary Path II: Multi-Band Frequency Sweeping}
\label{ssec:sweeping}

The transition to the next evolutionary stage is \textit{Multi-Band Frequency Sweeping}. This paradigm effectively circumvents the hardware boundary effects and the strict spectrum contiguity constraints inherent to continuous stitching. By resolving range ambiguity through phase consistency across discrete bands to realize Device Localization, it eliminates the dependency on a single, uninterrupted wideband channel. Instead, it allows the system to aggregate fragmented spectrum resources—hopping across available subbands within the 2.4\,GHz and 5\,GHz range.



\subsubsection{Resolving Range Ambiguity}
\begin{figure}[t]
    \centering
    \setlength{\abovecaptionskip}{6pt}
    \includegraphics[width=0.7\columnwidth]{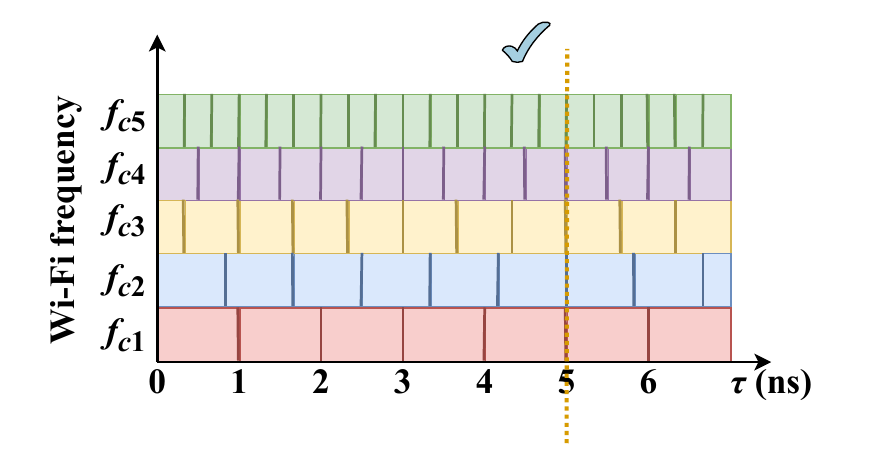}
    \caption{Visualizing range ambiguity resolution via frequency hopping. Due to phase wrapping, single-band measurements yield infinite candidate ranges (visualized as vertical lines). By combining measurements across multiple discrete bands, the system identifies the unique true ToF where all candidate solutions align~\cite{vasisht2016decimeter}.}
    \label{fig:chronos}
    \vspace{-1em}
\end{figure}

The primary technical challenge addressed by Chronos is range ambiguity. In a single narrowband channel, the measured phase wraps every $2\pi$, yielding an infinite set of possible ranges. As illustrated in Fig.~\ref{fig:chronos}, the system overcomes this by obtaining measurements at multiple discrete frequencies. While individual bands produce ambiguous candidates, the true physical ToF must be consistent across all observations. The system thus identifies the unique time instance where candidate solutions \emph{align}. This logic is mathematically analogous to the Chinese Remainder Theorem (CRT)~\cite{weisstein_crt}, where combining ambiguous remainders from different moduli yields a unique solution, effectively disambiguating the range.

\subsubsection{Active Phase Cancellation}
To address the random phase offsets introduced by hardware retuning during hopping, Chronos employs an active, bidirectional protocol. By exchanging packets between the AP and client at each hop and multiplying the uplink and downlink channel measurements (detailed in Sec.~\ref{ssec:offset_compensation}), the random hardware phases cancel out due to reciprocity. This yields a clean phase measurement corresponding to the round-trip distance.

\subsubsection{Limitations (the Barrier to General Sensing)}
While multi-Band frequency sweeping partially solves the problem of hardware boundary effects and spectrum availability, it retains specific limitations that restrict its utility for general-purpose ISAC:

\begin{itemize}
    \item \textbf{Resolution Limit due to Sweeping Speed:} The finite hopping speed restricts the number of frequency steps measurable within the channel's coherence time. This strictly limits the total synthesized bandwidth, thereby placing a physical ceiling on the achievable range resolution.

    \item \textbf{Device-Based Constraints:} The system relies on multiplying uplink and downlink channel measurements to cancel phase offsets. This operation necessitates an active transceiver at the target end to generate the reverse link signal, thereby strictly limiting the system to device-based localization and precluding passive device-free sensing.

    \item \textbf{Algorithmic Restriction to Localization:} The core algorithm is specifically designed to extract a single scalar value: the ToF of the direct path. It treats the rich dynamic multipath components primarily as interference. This algorithmic design inherently limits the system to localization, rendering it incapable of general sensing tasks (e.g., imaging or activity recognition).
\end{itemize}

These limitations motivates the final evolutionary step: discrete channel sampling with deep learning, exemplified by UWB-Fi.

\subsection{Case Study: Realizing High-Resolution Sensing via Discrete Channel Sampling}
\label{ssec:uwbfi}

To overcome the latency and active dependency limitations of sequential sweeping, a new paradigm has emerged: Discrete Channel Sampling. Instead of attempting to fill the entire spectrum, it aggregates a small set of discrete, irregular spectral slices across the Wi-Fi spectrum to synthesize a massive effective aperture. We discuss the methodology of UWB-Fi~\cite{li2024uwb} as the representative instance.

\subsubsection{The Discrete Sampling Paradigm}
UWB-Fi leverages Wi-Fi 6E/7 NICs to perform discrete channel sampling across the 2.4, 5, and 6\,GHz bands, spanning a massive 4.7\,GHz bandwidth~\cite{li2024uwb}. Unlike the full sequential sweep of Chronos, UWB-Fi treats the spectrum as a sparse observation space, forming a discrete and irregular sampling pattern. The theoretical feasibility of this approach is grounded in Compressive Sensing (CS), as formalized in the companion works~\cite{li2025ccs, li2025enabling}. Since indoor multipath components are inherently sparse in the spatial-temporal domain, CS theory dictates that the \hbrev{underlying physical multipath profile (e.g., ToF and AoA) can be faithfully} recovered from a small set of random discrete frequency samples, provided the effective bandwidth is sufficiently wide~\cite{li2025ccs, li2025enabling}.

Crucially, to ensure this discrete channel sampling is valid for dynamic sensing, UWB-Fi implements a fast channel hopping scheme. By optimizing driver-kernel interactions to reduce switching time to $\approx 1$\,ms, the system captures 20 discrete channels within 20\,ms~\cite{li2024uwb}. This speed ensures the discrete samples represent a consistent, ``frozen'' environment well within the channel coherence time.

\subsubsection{The Reconstruction Challenge}
Reconstructing a clean response from such discrete, irregular data poses fundamental signal processing challenges:
\begin{itemize}
    \item\textbf{Inapplicability of Conventional Algorithms:}  Conventional signal processing approaches (e.g., IFFT or MUSIC) are incapable of deriving reliable estimation outcomes from such discrete samples, often producing severe artifacts that mask true targets.
    \item \textbf{Hardware Incoherence:} The primary barrier is the CPO, which randomly varies with every channel hop. With tens of samples, this \textit{accumulated randomness} totally overwhelms the information embedded in the signal~\cite{li2024uwb}. 
    
\end{itemize}

\begin{table*}[b]
\centering
\small
\caption{Comparative Analysis of Frequency Diversity Methodologies}
\label{tab:frequency_comparison}
\renewcommand{\arraystretch}{1.2}

\newcolumntype{L}{>{\raggedright\arraybackslash}X}

\begin{tabularx}{\textwidth}{@{} >{\raggedright\arraybackslash}p{3.2cm} L L L @{}}
\toprule
\textbf{Methodology} & \textbf{Continuous Channel Stitching} & \textbf{Multi-Band Frequency Sweeping} & \textbf{Discrete Channel Sampling} \\
\midrule

\textbf{Representative Systems} & 
Splicer~\cite{xie2015precise}, ToneTrack~\cite{xiong2015tonetrack} & 
Chronos~\cite{vasisht2016decimeter} & 
UWB-Fi~\cite{li2024uwb} \\
\addlinespace

\textbf{Core Goal} & 
High-Resolution Profiling (PDP) \& Ranging (ToA) & 
Single-AP Ranging \& Localization & 
General-Purpose, Multi-Subject Fine-Grained Sensing \\
\addlinespace

\textbf{Bandwidth Strategy} & 
Physical Adjacency & 
Sequential Full-Spectrum Sweep & 
Randomized Discrete Sampling \\
\addlinespace

\textbf{Key Enabling \newline Technology} & 
Optimization-based Alignment via PDP/ToA Consistency & 
Multi-band Phase Analysis & 
Model-Driven Deep Learning (\textit{SpecTrans} Network) \\
\addlinespace

\textbf{Sensing Output} & 
PDP or ToA & 
ToF of the Direct Path & 
ToF-AoA Spectrum \\
\addlinespace

\textbf{Key Limitation} & 
Constrained by spectrum availability and boundary effects. & 
Device-based only; Limited to localization. & 
Higher computational complexity for neural inference. \\

\bottomrule
\end{tabularx}
\end{table*}

\subsubsection{Model-Driven Learning Framework}

\begin{figure}[t]
    \centering
    \setlength{\abovecaptionskip}{6pt}  
    \includegraphics[width=0.9\columnwidth]{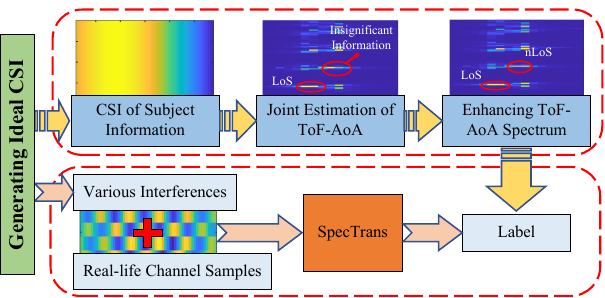}
    \caption{UWB-Fi's model-driven training strategy. The top branch generates ideal spectral labels using physical ray-tracing, while the bottom branch trains the network to reconstruct these spectra from discrete, noisy real-world inputs~\cite{li2024uwb}.}
    \label{fig:uwbfi_training}
\end{figure}

To solve this inverse problem without active calibration, UWB-Fi introduces a \emph{model-driven deep learning framework}. As illustrated in Fig.~\ref{fig:uwbfi_training}, this framework operates as a dual-branch pipeline designed to bridge the gap between noisy discrete samples and high-resolution spectral ground truth.

\paragraph{\textbf{Physics-Guided Label Synthesis (Top Branch)}}
The upper branch of Fig.~\ref{fig:uwbfi_training} is dedicated to generating high-fidelity training targets. Rather than simply using low-dimensional position tuples as labels, UWB-Fi reformulates the learning objective by projecting target coordinates into the high-dimensional signal domain. It uses the physical ray-tracing model (referenced in Eqn.~\eqref{eq:CSI_joint_static}) to mathematically synthesize \textit{Ideal CSI} based on simulated target locations. These synthetic signals are processed via MUSIC and refined using spectrum enhancement techniques (e.g., Laplacian filtering) to produce high-contrast, artifact-free ToF-AoA Spectra. This constructs a physics-compliant representation that explicitly guides the network to recover fine-grained multipath structures.

\paragraph{\textbf{The SpecTrans Network (Bottom Branch)}}
The lower branch of Fig.~\ref{fig:uwbfi_training} represents the inference pipeline. The specialized neural network, \textit{SpecTrans}, employs an Encoder-Decoder architecture enhanced with attention-based modules. Its design is explicitly inspired by two model-based algorithms: it emulates a trainable matched filter to remove hardware-related offsets, and performs a cross-domain transformation analogous to the MUSIC algorithm to convert discrete channel samples into ToF-AoA spectra. By minimizing the deviation from the physics-based labels, the network learns to reverse the complex distortions caused by discrete sampling and hardware-induced noise.

\paragraph{\textbf{Strategic Output (Spectrum vs. Tuples)}}
A critical design choice in UWB-Fi is the target output representation. As explained in the study, this choice is fundamental to the system's success and is rooted in the \emph{bias-variance tradeoff} in estimation. Directly training a network to regress a few real-valued tuples (e.g., coordinates) forces the model into a difficult compromise between bias and variance. By using the higher-dimensional spectrum as the training label and output, the network can offload the large estimation variance to irrelevant parts of the spectrum (i.e., background noise pixels), enabling an unbiased and low-variance estimation of the true ToF-AoA peaks~\cite{li2024uwb}.

\subsubsection{Performance Evaluation and Capabilities}
To validate the resolution gain achieved by the 4.7\,GHz synthesized bandwidth, UWB-Fi is tested in an extreme scenario designed to probe the physical limits of separability.

\begin{figure}[t]
    \centering
    \setlength\abovecaptionskip{3pt}
    \subfloat[ToF-AoA Spectrum.]{
        \includegraphics[width=0.4\linewidth]{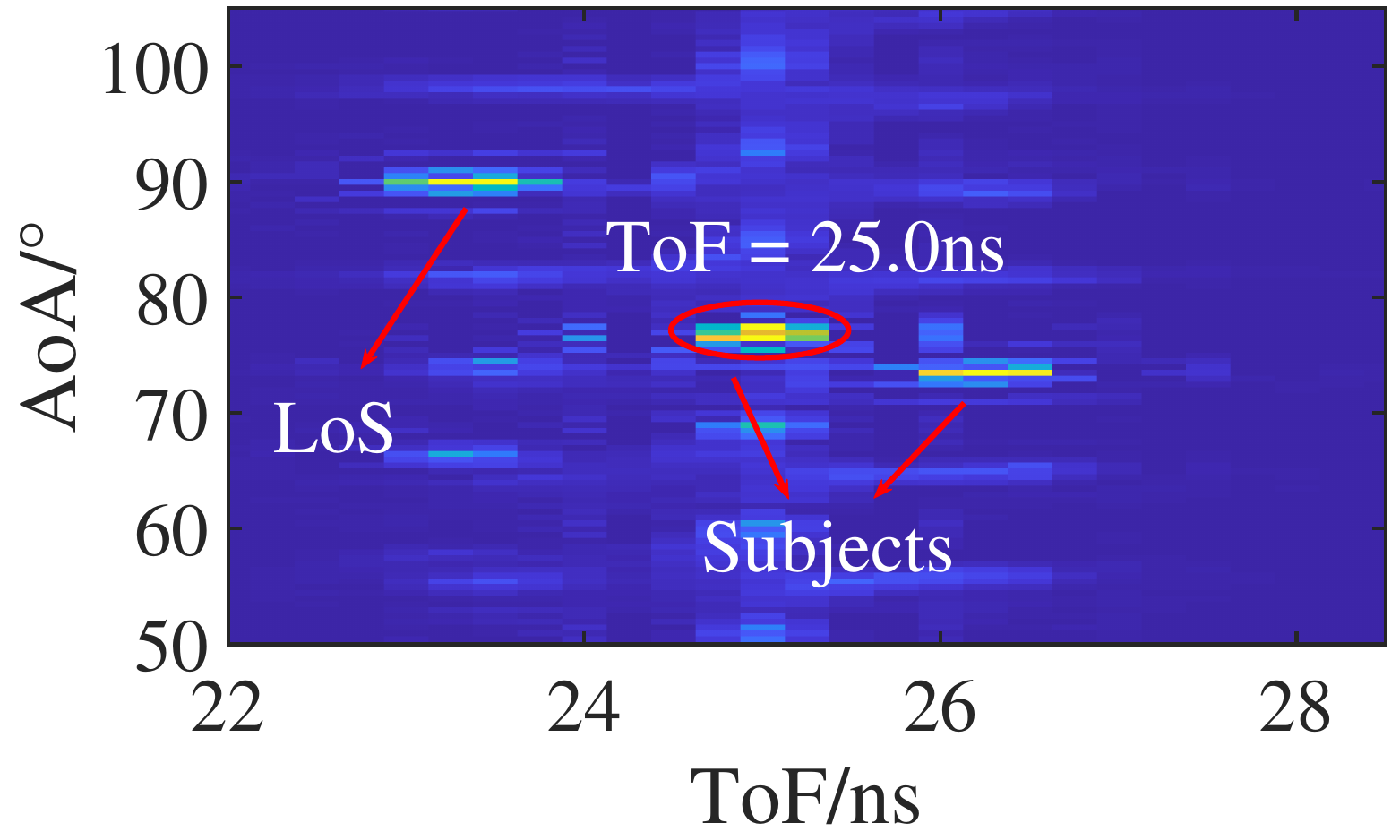}
        \label{sfig:mutil3_spe}
    }
    \subfloat[AoA-Amplitude Curve.]{
        \includegraphics[width=0.4\linewidth]{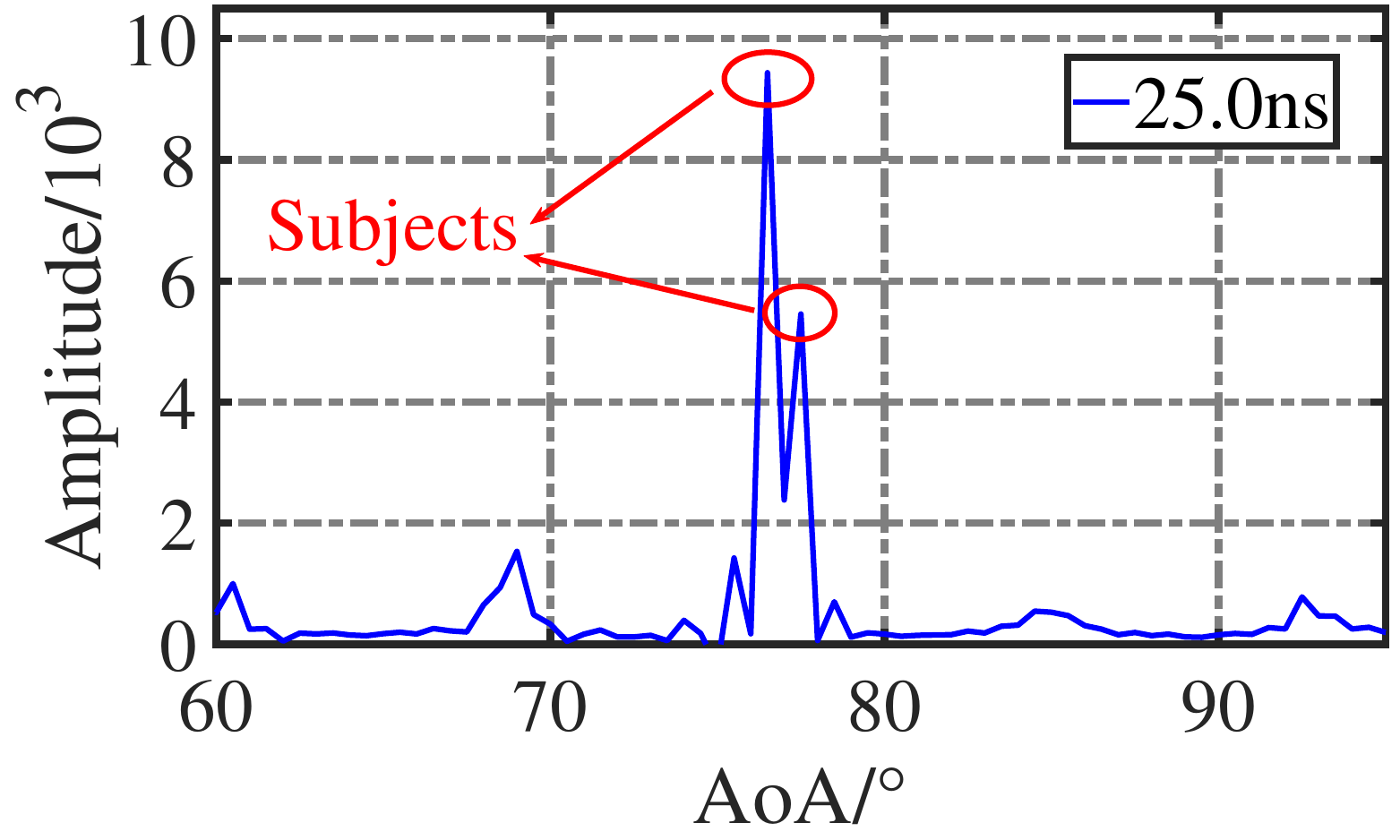}
        \label{sfig:mutil3_zoom}
    }
    \caption{Experimental validation of fine-grained resolution. The system successfully resolves three subjects standing in close proximity (within 0.4\,m). (a) The ToF-AoA spectrum reveals distinct clusters suggesting at least two reflection paths. (b) A sectional view at $\text{ToF} \approx 25$\,ns reveals two separate AoA peaks, allowing the complete distinction of all three subjects~\cite{li2024uwb}.}
    \label{fig:ThreeR}
\end{figure}

\paragraph{\textbf{Evaluation (Ability of Resolution)}}
The ultimate measure of Frequency Diversity is the minimum distance at which targets can be distinguished. To demonstrate this, we position three subjects in close proximity (within 0.4\,m). UWB-Fi successfully disentangles three distinct subjects, as shown in Fig.~\ref{fig:ThreeR}. This result confirms that the synthesized 4.7\,GHz bandwidth effectively breaks the theoretical resolution limit inherent to legacy narrowband hardware.

\paragraph{\textbf{Unlocking Fine-Grained Sensing Capabilities}}
This methodology effectively transforms a commodity Wi-Fi device into a UWB radar, unlocking two critical capabilities:
\begin{itemize}
    \item \textbf{Multi-Subject Separability:} The most direct benefit of the synthesized 4.7\,GHz bandwidth is the dramatic enhancement in range resolution. As evaluated in the theoretical framework of UWB-Fi (CCS-Fi), this discrete channel sampling achieves a spatial resolution of approximately 20\,cm (with 20 sampled channels), reaching centimeter levels with more samples~\cite{li2025ccs}. This capacity to resolve closely spaced subjects confirms that Frequency Diversity is the fundamental determinant of the system's physical resolving power.
    
    \item \textbf{General Sensing Capabilities:} Beyond static localization, the high-resolution spectrum serves as a universal interface for general sensing. The phase information of target motion is implicitly encoded in the spectral amplitudes surrounding the target's peak~\cite{li2024uwb}. By extracting and tracking these spectral bins across consecutive snapshots, the system can recover fine-grained micro-motion patterns (such as respiration waveforms and gesture signatures) directly from the synthesized UWB spectrum, supporting a wide range of downstream applications~\cite{li2024uwb}.
\end{itemize}

\subsection{Summary: Towards High-Resolution ISAC}
\label{ssec:frequency_summary}

This chapter examined exploiting \textbf{Frequency Diversity} to realize \textbf{High-Resolution ISAC}. As summarized in Table~\ref{tab:frequency_comparison}, the quest to synthesize wide effective bandwidth ($\Delta \tau = 1/B$) has evolved from hardware constraints to algorithmic innovation:

\begin{itemize}
    \item \textbf{Continuous Channel Stitching:} Aggregates contiguous channels but is severely constrained by spectrum scarcity and boundary effects~\cite{xie2015precise}.
    \item \textbf{Multi-Band Frequency Sweeping:} Resolves ambiguity via sequential sweeping, yet algorithmic design constraints limit its general applicability~\cite{vasisht2016decimeter}.
    \item \textbf{Discrete Channel Sampling:} Synthesizes a 4.7\,GHz aperture via randomized hopping and deep learning, unlocking centimeter-level resolution on commodity hardware~\cite{li2024uwb}.
\end{itemize}

With resolution addressed, the focus shifts to spatial coverage. The next chapter explores \textbf{Link Diversity} to overcome single-viewpoint occlusion through multi-perspective observations.

\section{Link Diversity: Realizing Ubiquitous ISAC via Physical Multi-Links and Digital Feedback}
\label{sec:link}

This chapter explores the realization of \textit{Ubiquitous ISAC} through \textbf{Link Diversity}. Unlike Temporal and Frequency diversities that optimize single-link fidelity, Link Diversity exploits the multiplicity of network paths and protocols to overcome the inherent limitations of a solitary viewpoint. We begin by analyzing the geometric blind spots and deployment barriers that constrain single-link sensing. To dismantle these barriers, we investigate two complementary paradigms: \textit{Physical Link Diversity}, which leverages distributed architectures to secure robust multi-perspective coverage, and \textit{Digital Link Diversity}, which harnesses \textit{Beamforming Feedback Information (BFI)} as a standardized proxy for commodity accessibility. We then present a detailed case study of MUSE-Fi, demonstrating how the synergy of physical near-field isolation and digital feedback resolves the multi-user sensing challenge. Finally, we conclude with a comparative analysis of these paradigms.

\subsection{Theoretical Foundations: Geometric Constraints and Accessibility Barriers}
\label{ssec:link_theory}

To understand the necessity of Link Diversity, we define it as the capability to aggregate observations from independent physical paths and distinct digital protocols. This two-fold definition addresses two fundamental bottlenecks that prevent the realization of ubiquitous sensing: the geometric constraints that limit observability, and the accessibility barriers that restrict device deployment.

\subsubsection{The Geometric Bottleneck (A Barrier to Geometric Ubiquity)}
From a physical perspective, a single Tx-Rx link projects the complex 3D environment onto a 1D time-series signal. This projection suffers from inherent information loss, preventing true \textbf{Geometric Ubiquity}—the ability to sense any target from any angle.
\begin{itemize}
    \item \textbf{Persisting Geometric Ambiguity:} Signal propagation is governed by the Fresnel zone geometry. As shown in Sec.~\ref{ssec:geo_ambiguity}, a single link is inherently blind to motion along the tangential direction of the Fresnel ellipsoid. Furthermore, static occlusion blocks the LoS, leaving targets in the shadow undetected. \textbf{Physical Link Diversity} resolves this by providing complementary viewing angles, ensuring that a target in the ``blind spot'' of one link is in the ``sweet spot'' of another.
    
    \item \textbf{The Under-determined Superposition Problem:} In multi-subject scenarios, the received signal is a linear superposition of reflections from all targets. With a single observation link, recovering the individual states of multiple targets is a mathematically ill-posed, under-determined problem (i.e., fewer equations than unknowns). Physical Link Diversity addresses this by constructing a system of independent equations from distributed links, rendering the inverse problem solvable.
\end{itemize}

\subsubsection{The Accessibility Bottleneck (A Barrier to Device Ubiquity)}
From a protocol perspective, relying solely on raw CSI creates deployment barriers that prevent \textbf{Device Ubiquity}—the ability to enable sensing on any standard Wi-Fi device.
\begin{itemize}
    \item \textbf{Hardware \revhu{Nondeployability}:} 
    Accessing high-fidelity CSI typically requires specific, often 
    \revhu{out-of-date}
    NICs (e.g., Intel 5300) and root-level firmware hacks. This dependency creates a ``closed garden,'' making it impossible to deploy sensing applications on the 
    \revhu{majority of mainstream}
    commercial IoT devices.
    \item \textbf{Signal Instability:} 
    \revhu{Raw CSI indiscriminately captures all channel variations, including high-frequency hardware phase noise and motion-induced perturbations. In dynamic settings (e.g., handheld devices), these unfiltered fluctuations can overwhelm subtle signals of interest, such as vital-sign signals.}
\end{itemize}

Consequently, overcoming these dual bottlenecks requires a complementary strategy: \textbf{Physical Link Diversity} is essential to dismantle geometric limits through multi-perspective observation, while \textbf{Digital Link Diversity}---leveraging standardized mechanisms like \textit{BFI}---is critical to break accessibility constraints and enable sensing across the commodity ecosystem.

\subsection{Methodology I: Physical Link Diversity}
\label{ssec:physical_link}

The most direct pathway to realizing \textbf{Geometric Ubiquity} is to physically populate the environment with multiple distributed transmitter-receiver pairs. Such Physical Link Diversity leverages macroscopic separation to gain complementary viewing angles, systematically dismantling the geometric limits of single-link sensing. We categorize these techniques into four functional paradigms.

\subsubsection{Omnidirectional Coverage and Geometric Triangulation}
The primary objective is to eliminate inherent blind spots by aggregating views. This philosophy originates from ~\cite{youssef2007challenges}, which utilized distributed links to probabilistically infer user presence. Building on this coverage capability, WiSee~\cite{pu2013whole} leveraged Doppler shifts for whole-home gesture recognition, while E-eyes~\cite{wang2014eyes} used fine-grained WiFi signatures from home appliances to identify activities.
Beyond detection, researchers aimed for finer geometric solving. WiDraw~\cite{sun2015widraw} enables hands-free drawing by exploiting the AoA and hand occlusions to track fine-grained trajectories. In parallel, LiFS~\cite{LiFS-MobiCom16} targets room-scale localization with low human effort, modeling CSI measurements as power fading equations to solve for position. To achieve higher fidelity, WiDeo~\cite{joshi2015wideo} mines backscatter reflections to realize high-resolution motion tracing. Most recently, Choi~\cite{choi2022sensor} advances this by fusing multi-link beacon CSI with inertial traces to enable unsupervised positioning without manual site surveys.

\subsubsection{View-Invariant Feature Extraction}
Observed signal patterns often change drastically with user location and device placement. To mitigate these geometric distortions, early works like CARM~\cite{wang2017device} pioneered the use of multi-link likelihood fusion to enhance recognition robustness against environmental variations. Then, systems like IndoTrack~\cite{li2017indotrack} and WiTraj~\cite{wu2021witraj} combine multi-link Doppler constraints to reconstruct absolute velocity vectors, correcting for view-dependent distortions. Extending this multi-link principle, RIM~\cite{fan2019rf} enables RF-based inertial measurement, while WiDance~\cite{qian2017inferring} utilizes orthogonal links for interactive motion direction inference. Pushing further towards true domain independence, Widar3.0~\cite{zhang2021widar3} fuses radial velocities to derive the \textit{Body-coordinate Velocity Profile (BVP)} (detailed in Sec.~\ref{para:bvp_def}), enabling location-agnostic recognition. Similarly, DeepMV~\cite{xue2020deepmv} employs a multi-view fusion mechanism to derive robust representations that are invariant to environmental variations.

\subsubsection{Distributed Imaging (Reconstructing 3D Structure)}
Treating the distributed network as a sparse imaging aperture allows for fine-grained 3D reconstruction. WiPose~\cite{jiang2020wipose} pioneered the fusion of multi-view CSI to reconstruct full 3D skeletons by encoding skeletal structural priors. To handle complex, non-standard movements, Winect~\cite{ren2021winect} leverages distributed links to track free-form 3D poses beyond predefined gesture sets, while GoPose~\cite{ren2022gopose} further refines estimation accuracy in dynamic scenarios through multi-view consistency.
Recent research has pushed beyond skeletons to visualize volumetric shape. Systems like Wi-Mesh~\cite{wang2022wimesh} and MultiMesh~\cite{wang2024multi} utilize spatial information from multiple distributed links to reconstruct the full 3D surface mesh of human bodies. These approaches effectively function as ``RF cameras,'' recovering detailed body shape and volume even under occlusion.

\subsubsection{Spatial Disentanglement (Multi-User Separation)}
A critical challenge involves disentangling signals from multiple concurrent targets. MultiTrack~\cite{tan2019multitrack} addresses this by leveraging distributed links to differentiate users based on distinct propagation paths. Advancing to 3D perception, Person-in-WiFi 3D~\cite{yan2024person} employs an end-to-end Transformer framework that fuses multi-view CSI. By learning spatial consistency across receivers, it effectively regresses independent 3D poses even under occlusion. In contrast, MUSE-Fi~\cite{hu2023muse} introduces a physics-based solution via \textit{Near-Field Domination}. It exploits rapid near-field attenuation to naturally isolate users, transforming the multi-user interference problem into distributed single-user tasks—a concept detailed in the subsequent Case Study.

\begin{table*}[h]
\caption{Taxonomy and Comparison of Physical Link Diversity Paradigms}
\label{tab:physical_link_taxonomy}
\centering
\small 

\newcolumntype{Y}{>{\RaggedRight\arraybackslash}X}

\begin{tabularx}{\textwidth}{@{} 
    >{\hsize=0.75\hsize}Y 
    >{\hsize=1.55\hsize}Y 
    >{\hsize=1.0\hsize}Y 
    >{\hsize=0.7\hsize}Y 
@{}}
\toprule
\textbf{Paradigm} & \textbf{Core Mechanism} & \textbf{Representative Works} & \textbf{Key Capability} \\ 
\midrule

\textbf{Coverage \& Triangulation} & 
Aggregates views to eliminate blind spots; exploits AoA, shadowing, and backscatter for geometric solving. & 
WiSee~\cite{pu2013whole}, E-eyes~\cite{wang2014eyes}, WiDraw~\cite{sun2015widraw}, LiFS~\cite{LiFS-MobiCom16}, WiDeo~\cite{joshi2015wideo}, \cite{choi2022sensor} & 
\textbf{Robustness:} \newline No blind spots. \\ 
\addlinespace

\textbf{View-Invariant Feature Extraction} & 
Combines multi-link Doppler to reconstruct absolute velocity vectors, or extracts BVP for domain independence. & 
CARM~\cite{wang2017device},
IndoTrack~\cite{li2017indotrack}, WiTraj~\cite{wu2021witraj}, RIM~\cite{fan2019rf}, WiDance~\cite{qian2017inferring}, Widar3.0~\cite{zhang2021widar3}, DeepMV~\cite{xue2020deepmv} & 
\textbf{Generalization:} \newline View/Location invariance. \\ 
\addlinespace

\textbf{Distributed Imaging} & 
Treats the distributed network as a sparse aperture to reconstruct 3D skeletons or surface meshes. & 
WiPose~\cite{jiang2020wipose}, Winect~\cite{ren2021winect}, GoPose~\cite{ren2022gopose}, Wi-Mesh~\cite{wang2022wimesh}, MultiMesh~\cite{wang2024multi} & 
\textbf{Fidelity:} \newline 3D Structure recovery. \\ 
\addlinespace

\textbf{Spatial Disentanglement} & 
Leverages distributed spatial diversity, multi-view consistency (Transformer), or near-field domination to separate users. & 
MultiTrack~\cite{tan2019multitrack}, Person-in-WiFi 3D~\cite{yan2024person}, MUSE-Fi~\cite{hu2023muse} & 
\textbf{Scalability:} \newline Multi-user separation. \\ 

\bottomrule
\end{tabularx}
\end{table*}

\textbf{Summary.} Table~\ref{tab:physical_link_taxonomy} summarizes these paradigms. While  \textbf{Physical Link Diversity} effectively overcomes geometric ambiguity, the associated deployment complexity motivates the exploration of lightweight \textbf{Digital Link Diversity} in the next section.

\subsection{Methodology II: Digital Link Diversity}
\label{ssec:bfi_evolution}

While Physical Link Diversity achieves \textit{Geometric Ubiquity}, it fails to address \textbf{Device Ubiquity}---the ability to sense on standard Wi-Fi devices. Acquiring raw CSI typically relies on legacy chipsets (e.g., Intel 5300) and firmware hacks, creating a barrier to large-scale adoption. To break this, research has shifted towards \textbf{Digital Link Diversity}, utilizing protocol-standardized feedback on modern commodity devices. The prominent realization of this philosophy is Beamforming Feedback Information (BFI), as illustrated in Sec.~\ref{ssec:bfi_intro}.

\subsubsection{Theoretical Foundations and Algorithmic Enablers}
BFI is a compressed, quantized representation of the channel matrix ($\mathbf{V}$ matrix from SVD), standardized in IEEE 802.11ac/ax/be. Recent research establishes BFI as a precision modality by addressing three challenges: mathematical invertibility, dimensional flexibility, and data efficiency.

\begin{itemize}
    \item \textbf{Mathematical Reconstruction \& Precision:} BeamSense~\cite{wu2023beamsense} formulated the BFI compression as an optimization problem, successfully reconstructing geometric parameters (e.g., AoA, ToF) despite quantization loss. Building on this, BFMSense~\cite{yi2024bfmsense} enhanced micro-motion sensitivity using a ratio model that cancels phase offsets to isolate and amplify subtle variations.
    
    \item \textbf{Hardware Agnosticism \& Dimensional Flexibility:} Unlike raw CSI which requires firmware hacks, BFI is protocol-standardized. BFMSense~\cite{yi2024bfmsense} validated that BFI can be extracted from heterogeneous chipsets (e.g., Qualcomm, Broadcom) without modification. Furthermore, FreeBFI~\cite{wang2025freebfi} achieved dimensional decoupling, enabling fine-grained sensing with an \textit{arbitrary} number of antennas, breaking the constraints of previous models that relied on specific antenna configurations.
    
    \item \textbf{Data Efficiency via Feature Selection:} Raw BFI streams often contain redundancy. Li et al.~\cite{li2024efficient} propose an efficient framework based on Cramer-Rao bound (CRB) analysis to select the most informative BFI features. This approach significantly reduces computational overhead while maintaining positioning accuracy, paving the way for resource-constrained edge/IoT devices.
\end{itemize}

\subsubsection{Benign Sensing: Stability and Capability Expansion}
For legitimate sensing, BFI offers a unique advantage over raw CSI: \textbf{Signal Stability}. The inherent quantization acts as a natural low-pass filter, suppressing hardware noise and jitter while preserving structural motion variations.

\paragraph{\textbf{Robustness in Dynamic Scenarios}}
While raw CSI is often overwhelmed by high-frequency hand tremors, BFI acts as a stable filter, enabling clear waveform extraction. M2-Fi~\cite{hu2024m2fi} exploits this unique property to enable robust respiration monitoring directly on handheld devices. This stability also underpins MUSE-Fi~\cite{hu2023muse}, which leverages BFI to capture reliable human activity signals in complex multi-user environments---a hybrid mechanism we examine in detail in the subsequent Case Study.

\paragraph{\textbf{From Crowd Analysis to Geometric Metrology}}
The scope has expanded beyond motion tracking. BeamCount~\cite{chen2024beamcount} utilizes BFI stability for robust indoor crowd counting without extensive calibration. Wi2DMeasure~\cite{wang2024wi2dmeasure} extends this to static metrology, modeling BFI variations to achieve accurate 2D size measurement, proving the compressed digital link captures sufficient geometric information.

\subsubsection{The Security Paradox---Sensitivity as a Vulnerability}
Ironically, the properties making BFI excellent for sensing---High Sensitivity and Plaintext Accessibility---transform it into a potent vector for side-channel attacks. Unlike CSI, BFI is broadcast in cleartext, allowing devices to ``eavesdrop'' on the environment.

\paragraph{\textbf{Micro-Motion Side Channels}}
BFI is exceptionally sensitive to near-field micro-motions. WiKI-Eve~\cite{hu2023wikieve} first revealed that BFI captures keystrokes without hacking, a capability advanced by MuKI-Fi~\cite{wang2024mukifi} to disentangle inputs from simultaneous users. Furthermore, BeamThief~\cite{chen2024beamthief} demonstrates that BFI acts as a microscopic probe to eavesdrop on sensitive passwords entered into POS terminals.

\paragraph{\textbf{Ubiquitous Surveillance}}
The threat extends to wide-area surveillance. LeakyBeam~\cite{xiao2025leakybeam} demonstrates passive sniffing from outside buildings to infer indoor occupancy. Furthermore, BFId~\cite{todt2025bfid} enables gait-based identification. These works highlight a trade-off: democratizing sensing via Digital Link Diversity comes at the cost of exposing physical-layer privacy.
\subsection{Case Study: Realizing Multi-User Separability via Hybrid Physical-Digital Diversity}
\label{ssec:musefi}

\subsubsection{Introduction---The Practicality Gap}
The methodologies reviewed in the preceding sections represent two distinct evolutionary paths. Physical Link Diversity 
pushes the limit of sensing resolution but often necessitates dense, dedicated infrastructure. 
Digital Link Diversity
achieves hardware ubiquity through standardized feedback, \revhu{such as BFI,} 
but struggles with robustness in complex environments.
Bridging these two paths to realize \textit{Practical Ubiquity} faces two 
\revhu{major} barriers in real-world deployments:

\begin{itemize}
    \item \textbf{\revhu{Multi-Subject Signal Superposition}:} 
    In realistic multi-subject scenarios, signals from different users superimpose at the receiver. Traditional blind source separation (e.g., ICA) often fails when interference is strong or users are closely 
    \revhu{located} rendering single-link sensing unreliable.
    \item \textbf{Traffic Sparsity and Burstiness:} Unlike lab experiments with continuous high-rate sounding (e.g., 1000~Hz), real-world Wi-Fi traffic is 
    \revhu{driven} by user applications. It is inherently bursty and sparse (e.g., intermittent packets during web browsing or video buffering). This temporal discontinuity makes continuous tracking mathematically ill-posed for conventional algorithms.
\end{itemize}

To bridge this gap, we discuss the methodology of MUSE-Fi~\cite{hu2023muse}, a representative system that exemplifies the synergy of physical and digital link diversity. Instead of treating these challenges separately, MUSE-Fi introduces a hybrid architecture: it leverages near-field domination (a physical link property) to spatially isolate users, and a sparse recovery algorithm (a digital link innovation) to reconstruct continuous motion from fragmentary BFI streams. This design transforms the intractable multi-user problem into a scalable, distributed sensing task executable on commodity smartphones.

\subsubsection{Physical Pillar---Near-Field Disentanglement}

\begin{figure}[t]
    \centering
    \includegraphics[width=0.75\columnwidth]{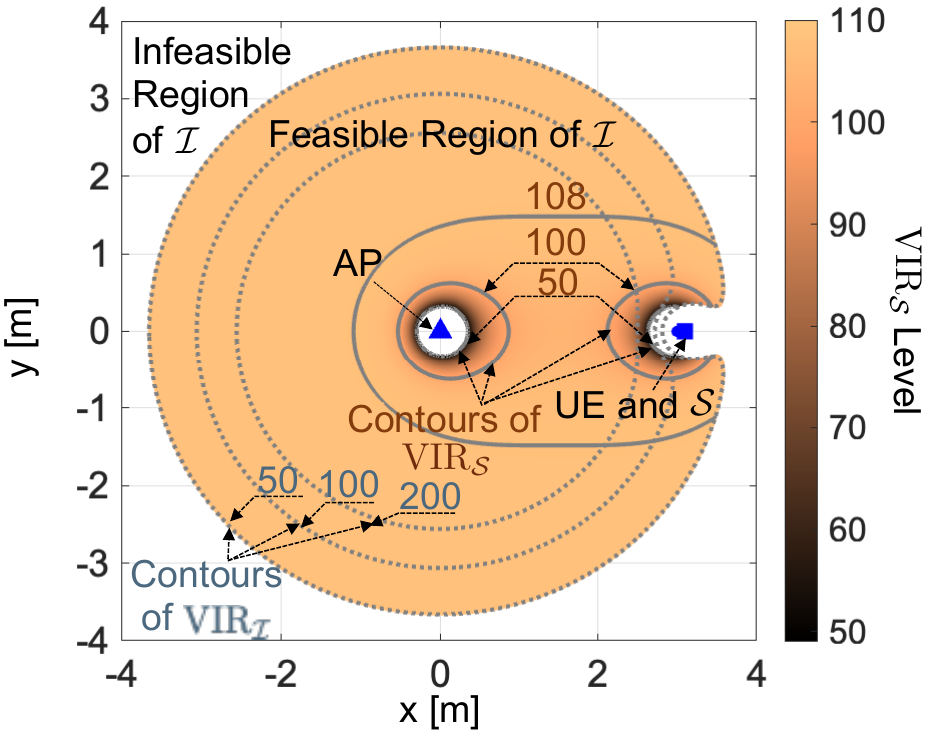} 
    \caption{The concept of Near-Field Domination. The system defines a mathematically derived \textbf{Feasible Region} (orange area) where the Variation-to-Interference Ratio (VIR) exceeds a critical threshold ($\gamma_{th} \approx 50$), ensuring the target signal physically overwhelms external interference~\cite{hu2023muse}.}
    \label{fig:musefi_vir}
\end{figure}
Traditional multi-user sensing relies on algorithmic separation (e.g., ICA), which often falters under strong interference. MUSE-Fi sidesteps this by enforcing a topological shift to a distributed \textit{Personal-AP Link} architecture, governed by the physics of wave propagation.

\paragraph{\textbf{The Variation-to-Interference Ratio (VIR)}}
To rigorously quantify separability, MUSE-Fi establishes the VIR metric. Based on the indoor path loss model, the signal variation power $P$ 
\revhu{follows a power-law decay}
with distance $d$ (i.e., $P \propto d^{-\alpha}$, where $\alpha \approx 4$). 
Consequently, the VIR between a target user (at distance $d_{tgt}$) and an interfering user (at distance $d_{int}$) can be approximated as:
\begin{equation}
    \label{eq:vir_model}
    \text{VIR} \approx \frac{P_{tgt}}{P_{int}} \propto \left( \frac{d_{int}}{d_{tgt}} \right)^{\alpha}.
\end{equation}
This fourth-power relationship implies that spatial proximity translates into exponential signal dominance. For instance, a mere $2\times$ distance difference yields a $16\times$ (approx. 12~dB) advantage in signal strength.

\paragraph{\textbf{Threshold and Feasible Region}}
MUSE-Fi defines a strict separability threshold $\gamma_{th}$ to ensure robust sensing. Theoretical models and empirical validations sets this threshold at $\gamma_{th} \approx 50$ (approx. 17~dB)~\cite{hu2023muse}.
The spatial area satisfying $\text{VIR} > \gamma_{th}$ constitutes the \textit{Feasible Region} (illustrated as the orange zone in Fig.~\ref{fig:musefi_vir}). Within this ``physical isolation bubble'' (typically $d_{tgt} < 0.5$~m), the target's signal is dominant so that external interference is effectively suppressed to noise. This mechanism allows the system to scale to multiple concurrent subjects by physically enforcing signal
\revhu{separability}
in the spatial domain.

\subsubsection{Digital Pillar---Overcoming Temporal Sparsity}
While physical 
\revhu{separability addresses}
the interference problem, the reliance on commodity traffic introduces a new challenge: \textbf{Data Sparsity}. Real-world applications (e.g., video streaming) generate bursty traffic, resulting in CSI/BFI streams that are effectively ``undersampled'' and discontinuous.

\begin{figure}[t]
    \centering
    \includegraphics[width=0.95\columnwidth]{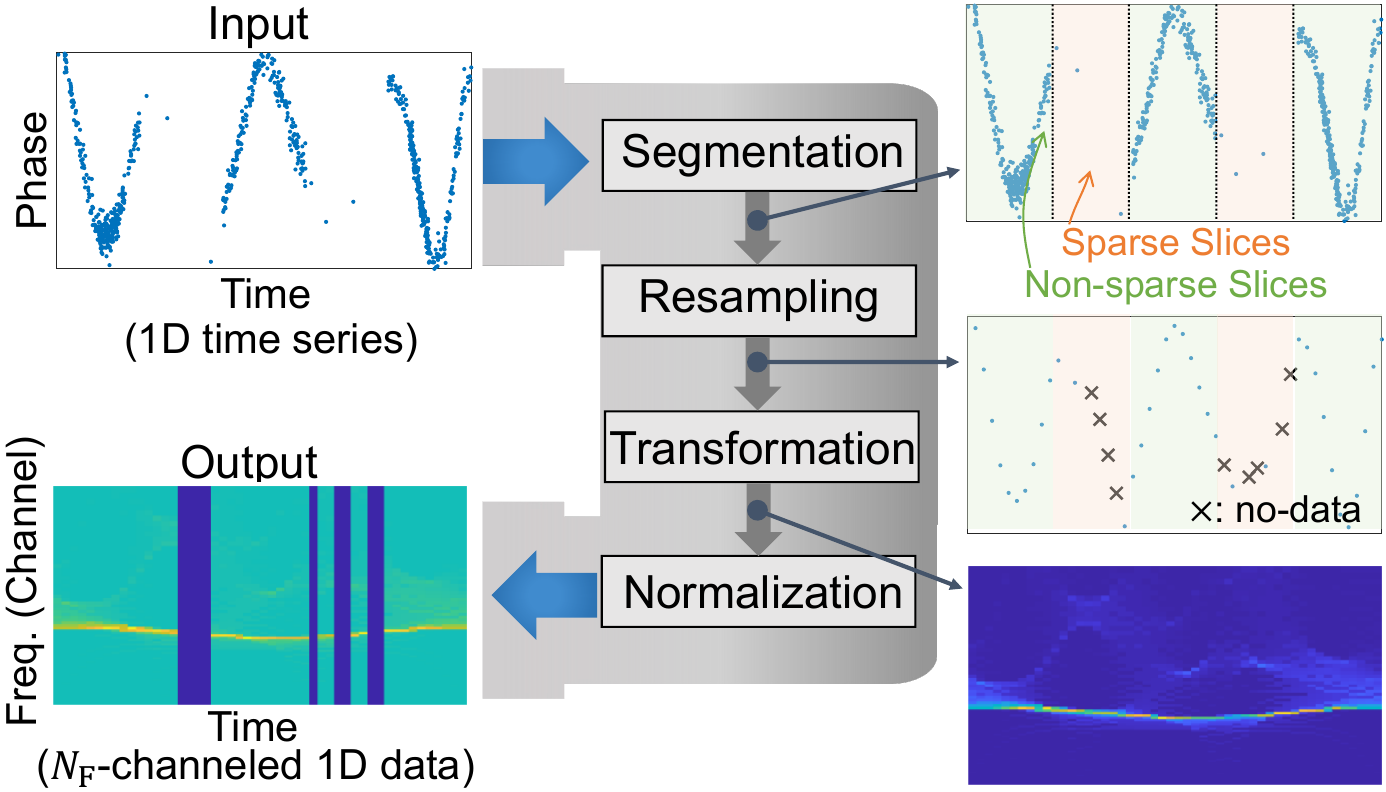} 
    \caption{The Data Transformation Pipeline for handling sparsity. As the first stage of the Sparse Recovery Algorithm, MUSE-Fi transforms discontinuous BFI streams into normalized spectrograms, formatting the bursty traffic into a structured input for the subsequent neural network~\cite{hu2023muse}.}
    \label{fig:musefi_sra}
\end{figure}

To address this, MUSE-Fi proposes a novel Sparse Recovery Algorithm (SRA). 
\revhu{Unlike simple interpolation, the SRA adopts a two-stage framework that explicitly handles data representation transformation and temporal reconstruction.}

\begin{itemize}
    \item \textbf{Stage I: Data Transformation Pipeline:} As illustrated in Fig.~\ref{fig:musefi_sra}, the system first pre-processes the raw, discontinuous BFI streams. It segments and resamples the bursty packets into a structured spectrogram, \revhu{explicitly} marking the missing temporal segments with ``no-data'' tags. This step formats irregular traffic into a tensor shape compatible with deep learning models.
    
    \item \textbf{Stage II: Deep Temporal Reconstruction:} The structured input is then fed into a Temporal Convolutional Network (TCN) autoencoder. This network learns the underlying physiological periodicity of human motion (e.g., respiration patterns) and utilizes contextual information to accurately 
    fill in the gaps left by the ``no-data'' tags.
    
    \item \textbf{Self-Supervised Strategy:} A 
    \revhu{key} innovation is the \revhu{self-supervised} training mechanism. Instead of requiring impossible-to-obtain ground truth for missing packets, SRA adopts a self-supervised approach. It masks non-sparse data segments to simulate sparsity, training the network to recover the original signal, thus adapting to diverse traffic patterns without manual labeling.
\end{itemize}


\subsubsection{Performance Evaluation and Capabilities}
To validate the efficacy of the proposed hybrid diversity, we examine MUSE-Fi's performance in challenging multi-user scenarios where signal superposition and traffic sparsity typically lead to sensing failure.

\paragraph{\textbf{Evaluation (Separability and Recovery)}}

\begin{figure}[t]
    \centering
    \setlength{\abovecaptionskip}{3pt} 
    
    \subfloat[Subject A.]{
        \includegraphics[height=2.5cm]{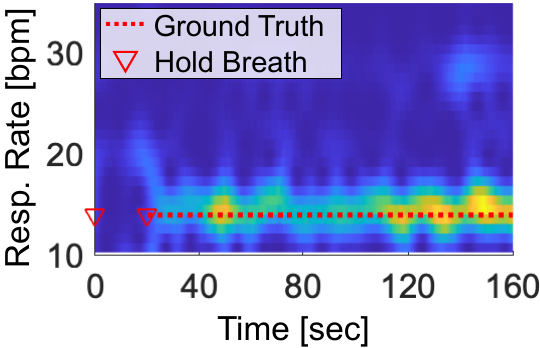}
        \label{fig:musefi_subj_a}
    }
    \subfloat[Subject B.]{
        \includegraphics[height=2.5cm]{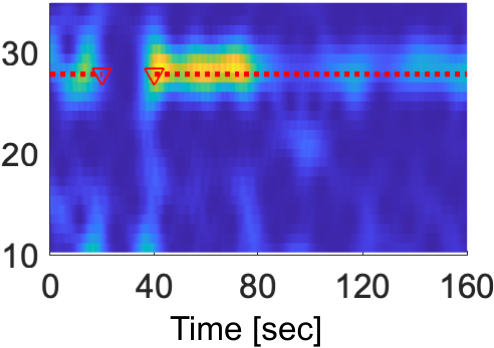}
        \label{fig:musefi_subj_b}
    }
    
    \vspace{2pt} 
    
    \subfloat[Subject C.]{
        \includegraphics[height=2.62cm]{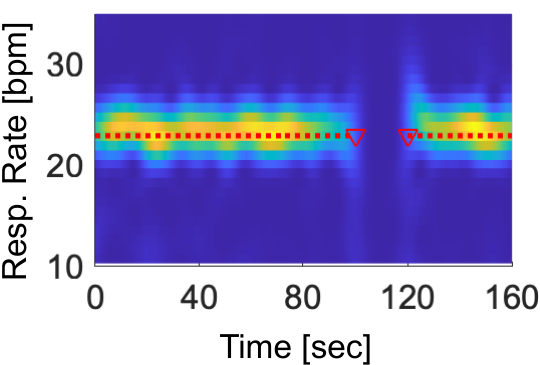}
        \label{fig:musefi_subj_c}
    }
    \subfloat[Subject D.]{
        \includegraphics[height=2.5cm]{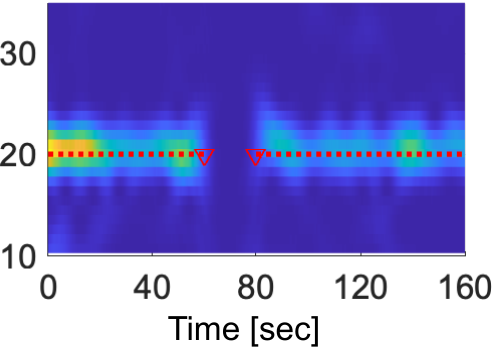}
        \label{fig:musefi_subj_d}
    }
    
    \caption{Performance evaluation of MUSE-Fi. The recovered respiration spectrograms for representative co-located subjects demonstrate distinct separation due to physical near-field isolation and temporal continuity due to digital sparse recovery~\cite{hu2023muse}.}
    \label{fig:musefi_eval}
\end{figure}
The fundamental premise of MUSE-Fi is that physical near-field isolates signals in space, while the digital sparse recovery of SRA restores them in time. 
Fig.~\ref{fig:musefi_eval} (adapted from~\cite{hu2023muse}) visualizes 
\revhu{the sensing outputs of the system for multiple simultaneously present subjects with typical spatial separations.}
The results confirm two critical mechanisms:
\begin{itemize}
    \item \textbf{Spatial Isolation:} Despite the crowded environment, the recovered spectrograms for different users are distinct and non-overlapping. Specific activity segments (e.g., breath holding) are clearly distinguishable for each individual, validating that the 
    \revhu{VIR-based near-field domination effectively prevents cross-user interference.}
    \item \textbf{Temporal Continuity:} Although the input BFI stream is inherently sparse due to bursty background traffic, the recovered spectrograms exhibit continuous, smooth spectral lines. 
    \revhu{The high fidelity of the reconstruction indicates that, by leveraging the inherent stability of BFI, the SRA is able to recover missing temporal segments, thereby preserving temporal continuity.}
\end{itemize}

\begin{table*}[t]
\centering
\small
\caption{Comparative Analysis of Link Diversity Methodologies}
\label{tab:link_comparison}
\renewcommand{\arraystretch}{1.2} 

\newcolumntype{L}{>{\raggedright\arraybackslash}X}

\begin{tabularx}{\textwidth}{@{} >{\raggedright\arraybackslash\bfseries}p{3.2cm} L L L @{}}
\toprule
\textbf{Methodology} & \textbf{Physical Link Diversity} & \textbf{Digital Link Diversity} & \textbf{Hybrid Diversity (Case Study)} \\
\midrule

\textbf{Representative Systems} & 
Widar3.0~\cite{zhang2021widar3}, LiFS~\cite{LiFS-MobiCom16}, MultiSense~\cite{zeng2020multisense} & 
BeamSense~\cite{wu2023beamsense}, BFMSense~\cite{yi2024bfmsense}, WiKI-Eve~\cite{hu2023wikieve} & 
MUSE-Fi~\cite{hu2023muse} \\
\addlinespace

\textbf{Core Goal} & 
Overcoming Geometric Blind Spots \& Occlusion & 
Overcoming Hardware Fragmentation \& Instability & 
Realizing Practical, Multi-User Sensing on Commodity Devices \\
\addlinespace

\textbf{Key Enabling Technology} & 
Distributed Topology \newline (Multi-static Tx-Rx pairs) & 
BFI \newline (Protocol-standardized) & 
Near-Field Domination + Sparse Recovery (SRA) \\
\addlinespace

\textbf{Primary Sensing Gain} & 
\textbf{Geometric Ubiquity}: Omnidirectional coverage and spatial disentanglement of multiple targets. & 
\textbf{Device Ubiquity}: Works on unmodified hardware; Stability via quantization filtering. & 
\textbf{Scalability}: Separating concurrent users physically and recovering continuous signals from bursty traffic. \\
\addlinespace

\textbf{Key Limitation} & 
High deployment cost; requires tight synchronization between distributed nodes. & 
Granularity Loss: Lacks absolute amplitude/phase; ill-suited for ranging. & 
Requires target proximity to the sensing device (Near-field constraint). \\

\bottomrule
\end{tabularx}
\end{table*}

\paragraph{\textbf{Unlocking Practical Sensing Capabilities}}
By establishing this robust physical-digital foundation, the system successfully unlocks a suite of capabilities essential for ubiquitous deployment:
\begin{itemize}
    \item \textbf{Scalable Multi-Subject Sensing:} It overcomes the interference bottleneck in crowded environments, enabling the simultaneous tracking of diverse fine-grained tasks (e.g., respiration monitoring, gesture interaction, and activity recognition) for multiple co-located users.

    \item \textbf{Generic Commodity Accessibility:} Unlike prior systems requiring specialized NICs (e.g., Intel 5300) to extract CSI, MUSE-Fi lowers the barrier to entry by utilizing \textit{standardized BFI}. This allows precision sensing to be deployed on unmodified commodity devices (e.g., smartphones) and operate purely on sparse background traffic (e.g., video streaming) without impacting network performance.
\end{itemize}


\subsection{Summary: Towards Ubiquitous ISAC}
\label{ssec:link_summary}

This chapter explored Link Diversity, moving beyond single-link observations to realize \textbf{Ubiquitous ISAC}. We navigated this landscape through two paradigms:
\begin{itemize}
    \item \textbf{Physical Link Diversity:} Densifies infrastructure with distributed transceivers to physically expand the observation aperture. While offering superior coverage and diversity, it faces challenges regarding deployment cost and hardware synchronization.
    \item \textbf{Digital Link Diversity:} Harnesses protocol-standardized feedback (BFI) to prioritize ubiquity. Despite sacrificing ranging fidelity due to quantization, BFI's inherent stability effectively filters hardware noise, making it promising for mobile sensing.
\end{itemize}

The MUSE-Fi case study demonstrates that practical ubiquity relies on the synergy of these paradigms. By integrating physical near-field isolation with digital sparse recovery, the system resolves complex multi-user interference on commodity devices.
Table~\ref{tab:link_comparison} summarizes these methodologies. Having established the potential of distributed topologies and digital feedback, we now shift to the antenna domain. The following chapter investigates \textbf{Spatial Diversity}, examining how MIMO arrays enable directional sensing for ISAC.

\section{Spatial Diversity: Empowering Directional ISAC via Active and Passive MIMO}     \label{sec:spatial}

This chapter explores the realization of \textit{Directional ISAC} through \textbf{Spatial Diversity}, which empowers the system to resolve angular geometry and actively steer electromagnetic energy. We begin by quantifying the fundamental resolution limit imposed by the finite physical antenna array of commodity hardware. To transcend this physical boundary, we investigate two complementary paradigms: \textit{Passive Rx-MIMO}, which leverages advanced signal processing to extract high-resolution spatial information from limited receiver antennas, and \textit{Active Tx-MIMO}, which introduces a paradigm shift by transforming the transmitter into a programmable directional probe. We then present a detailed case study of Beam-Fi, demonstrating how active beam steering eliminates sensing blind spots with almost not compromising communication throughput. Finally, we conclude with a comparative analysis of these paradigms.

\subsection{Theoretical Foundations: The Angular Resolution Limit}
\label{ssec:spatial_theory}

To understand the imperative for advanced spatial processing, we must first quantify the fundamental physical limits of commodity hardware. This Spatial Diversity challenge exhibits a perfect duality with the Frequency Diversity constraints discussed in Sec.~\ref{ssec:bandwidth_resolution}: just as temporal resolution is governed by the bandwidth $B$, angular resolution is fundamentally dictated by the number of antennas.

\subsubsection{The Constraint of Finite Antenna Count}
In a standard Uniform Linear Array (ULA) configuration with $J$ receive antennas spaced by a fixed distance $d$ (typically $d = \lambda/2$), the array's capability is determined strictly by $J$. Unlike specialized radar systems that deploy massive arrays, commodity Wi-Fi devices are constrained by form factor and cost, typically featuring a very limited \textit{antenna count} (e.g., $J=2$ to $4$). Consequently, improving physical resolution directly necessitates increasing the \textit{number of antennas}, which is often infeasible in consumer electronics.

\subsubsection{The Physical Resolution Limit}
Analogous to the temporal resolution limit ($\Delta \tau = 1/B$) detailed in Sec.~\ref{sec:frequency}, the angular resolution—defined as the minimum angular separation $\Delta \theta$ required to distinguish two subjects—is inversely proportional to the antenna count. According to classical array signal processing theory~\cite{vantrees2002optimum}, this limit is expressed as:
\begin{equation}
    \label{eq:spatial_resolution}
    \Delta \theta \approx \frac{\lambda}{(J-1)d \cos \theta},
\end{equation}
where $\theta$ is the AoA relative to the array's boresight. Since $d$ is fixed, the denominator is dominated by $J$. This relationship reveals that resolution improves linearly with more antennas.

\begin{figure}[t]
    \centering
    \setlength{\abovecaptionskip}{3pt}
    
    \subfloat[8 Antennas.]{
        \includegraphics[width=0.40\linewidth]{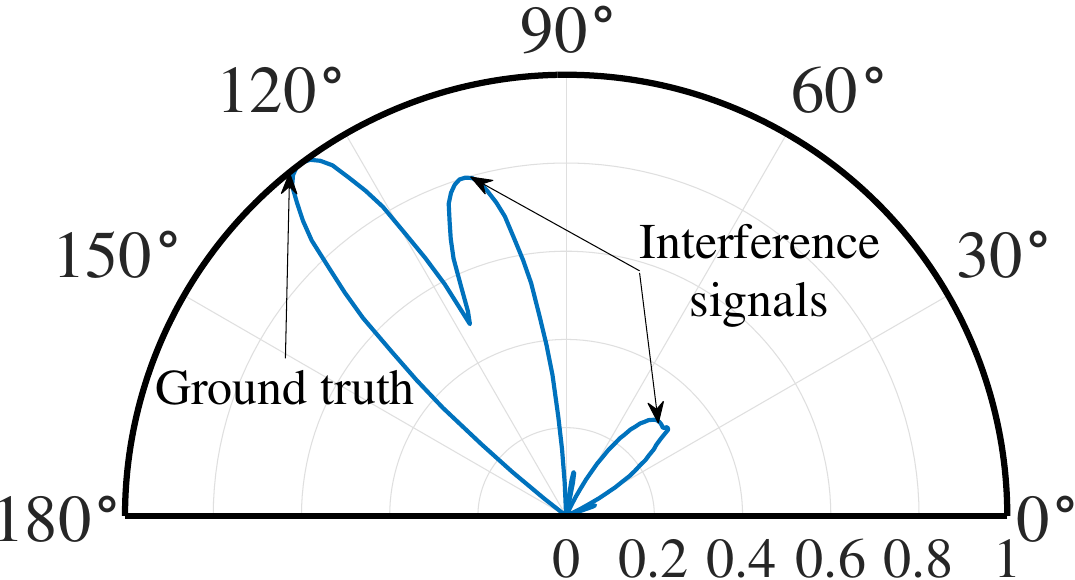}
        \label{sfig:8antennas}
    }
    \subfloat[4 Antennas.]{
        \includegraphics[width=0.40\linewidth]{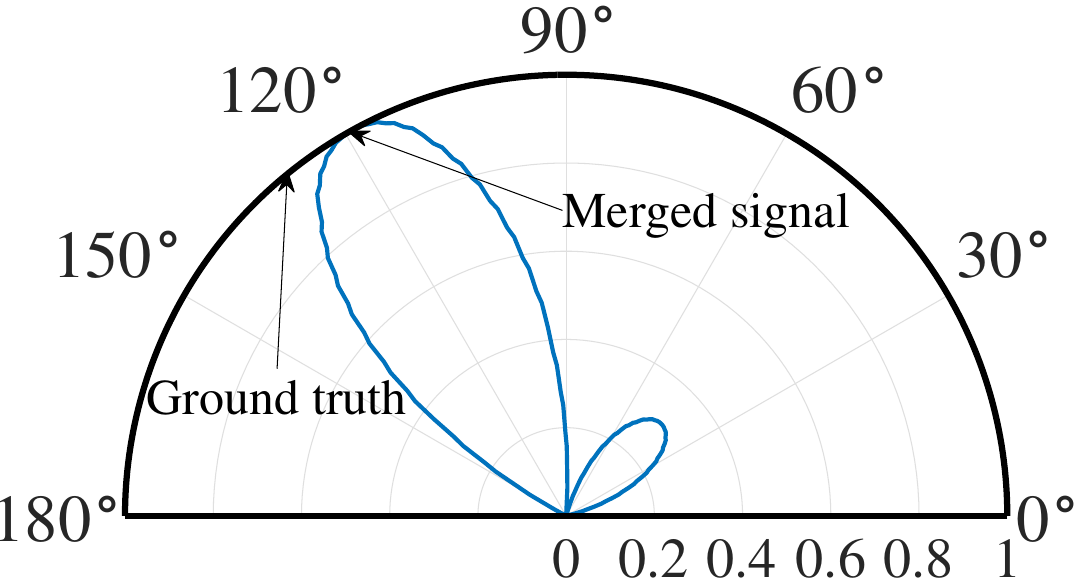}
        \label{sfig:4antennas}
    }
    
    \caption{Visualizing the Antenna-Resolution relationship. (a) With 8 antennas, the AoA spectrum exhibits a sharp peak, indicating high angular resolution. (b) Reducing to 4 antennas significantly widens the main lobe (blurring the direction), empirically verifying the theoretical resolution limit discussed in Eqn.~\eqref{eq:spatial_resolution} ~\cite{wang2025vrfi}.}
    \label{fig:AoA_spectrum}
\end{figure}

\subsubsection{The Challenge of Commodity Hardware}
Applying Eqn.~\eqref{eq:spatial_resolution} to even high-end commodity hardware reveals a severe deficit, a bottleneck widely recognized in foundational works like SpotFi~\cite{kotaru2015spotfi}. 
Consider a premium Wi-Fi AP equipped with $J = 4$ antennas. With standard half-wavelength spacing ($d = \lambda/2$), the angular resolution is severely restricted by the small $J$. Assuming the optimal viewing angle at boresight ($\theta = 0^{\circ}$):
\begin{equation}
    \Delta \theta_{J=4} \approx \frac{\lambda}{(4-1)\frac{\lambda}{2} \cdot 1} = \frac{2}{3} \text{ rad} \approx 38.2^{\circ}.
\end{equation}
This theoretical limit is visually corroborated by Fig.~\ref{sfig:4antennas}. With 4 antennas, the AoA spectrum exhibits a significantly widened main lobe compared to the idealized 8-antenna case. This implies that even a high-end AP acts as a coarse sensor with a blur radius of nearly $40^{\circ}$. The situation naturally deteriorates further for typical mobile clients (e.g., smartphones with 2 or 3 antennas). 
Consequently, relying solely on the \textit{limited antenna count} of a single device is insufficient for fine-grained sensing. This physical bottleneck necessitates the paradigm shift discussed in the remainder of this chapter: moving from hardware reliance to algorithmic spatial analysis (Passive Rx-MIMO) and active beam scanning (Active Tx-MIMO).

\begin{table*}[t]
\caption{Taxonomy and Comparison of Passive Rx-MIMO Sensing Paradigms}
\label{tab:rx_mimo_summary}
\centering
\small 
\renewcommand{\arraystretch}{1.2} 

\newcolumntype{L}{>{\raggedright\arraybackslash}X}

\begin{tabularx}{\textwidth}{@{} 
    >{\hsize=0.7\hsize}L   
    >{\hsize=1.5\hsize}L   
    >{\hsize=1.1\hsize}L 
    >{\hsize=0.7\hsize}L 
@{}}
\toprule
\textbf{Paradigm} & \textbf{Core Mechanism} & \textbf{Representative Works} & \textbf{Target Applications} \\ 
\midrule

\textbf{Stationary Physical Arrays} & 
\textbf{Frequency-Derived Virtual Array:} \newline 
Exploits OFDM subcarriers or sparse recovery on static antennas to resolve AoA. & 
ArrayTrack \cite{xiong2013arraytrack}
SpotFi \cite{kotaru2015spotfi}, 
\cite{gaber2014study},
RoArray \cite{gong2018roarray}, Phaser \cite{gjengset2014phaser}, S-Phaser \cite{han2018indoor}, M3 \cite{chen2019m3} & 
High-Accuracy Localization \\ 
\addlinespace 

\textbf{Joint Estimation} & 
\textbf{Multi-Dimensional Optimization:} \newline 
Jointly estimates AoA, ToF, and Doppler to resolve multipath in high-dimensional space. & 
Widar2.0 \cite{qian2018widar2}, mD-Track \cite{xie2019mdtrack}, MaTrack \cite{li2016matrack} & 
Device-Free Trajectory Tracking \\ 
\addlinespace 

\textbf{CSI Ratio} & 
\textbf{Differential Cancellation:} \newline 
Exploits amplitude/phase differences between antennas to cancel common RF noise. & 
FarSense \cite{zeng2019farsense}, FingerDraw \cite{wu2020fingerdraw}, RT-Fall \cite{wang2016rtfall} & 
Vital Signs \& Fine-grained Gestures \\ 
\addlinespace 

\textbf{Motion-Induced SAR} & 
\textbf{Trajectory-Derived Virtual Array:} \newline 
Synthesizes large apertures via explicit device motion (e.g., twisting) to enhance resolution. & 
Ubicarse \cite{kumar2014accurate}, Differential-MUSIC \cite{qian2017enabling} & 
Device-based Localization \\ 
\addlinespace 

\textbf{Representation Learning} & 
\textbf{Tensor Pattern Recognition:} \newline 
Learns mappings from spatial tensors to labels via Deep Learning. & 
BodyCompass \cite{yue2020bodycompass}, Wi-PIGR \cite{zhang2021wi}, WiWrite \cite{lin2020wiwrite}, R-TTWD~\cite{zhu2017rttwd}, AFall \cite{chen2022afall} & 
Activity Recognition \& User ID \\ 

\bottomrule
\end{tabularx}
\end{table*}
\subsection{The Passive Paradigm: Rx-MIMO for Spatial Analysis}
\label{ssec:passive_mimo}

The traditional approach to Spatial Diversity exploits the receiver's antenna array to decipher the spatial structure of incoming signals. Faced with the constrained finite antenna count of commodity hardware, the research community has moved beyond standard array processing, developing five distinct methodological paradigms, distinguished primarily by how they expand the spatial limit—either through signal processing algorithms, joint parameter estimation, or physical motion.

\subsubsection{Stationary Physical Array Processing}
\label{sssec:stationary_array}

This family of techniques focuses on maximizing the resolution of the static physical array. To overcome the hardware limit of few antennas, systems often construct ``virtual sensors'' by exploiting the frequency domain (OFDM subcarriers).
\begin{itemize}
    \item ArrayTrack~\cite{xiong2013arraytrack} employs spatial smoothing on the physical array to suppress multipath and accurately extract the direct path's AoA.
    
    \item SpotFi~\cite{kotaru2015spotfi} transcends the physical antenna count by mathematically constructing a virtual array with subcarriers, achieving super-resolution joint ToF-AoA estimation.
    
    \item The 2-D MP estimator~\cite{gaber2014study} applies a search-free Matrix Pencil algorithm to jointly resolve TDOA and DOA via generalized eigenvalue decomposition.
    
    \item RoArray~\cite{gong2018roarray} addresses robustness in cluttered environments. Instead of standard subspace methods, it employs sparse recovery theories to resolve correlated multipath signals that typically degrade estimation accuracy.
    
    \item Phaser~\cite{gjengset2014phaser} focuses on the hardware prerequisite, enabling phased array processing by rigorously calibrating phase offsets across RF chains. Subsequent systems like S-Phaser~\cite{han2018indoor} and M3~\cite{chen2019m3} further refine these geometric constraints to achieve robust localization using a single Access Point.
\end{itemize}

\subsubsection{Multi-Dimensional Joint Parameter Estimation}
A second line of research moves beyond AoA-only spatial processing by simultaneously exploiting multiple physical dimensions of CSI (AoA, ToF, Doppler) to separate multipath components.
\begin{itemize}
    \item Widar2.0~\cite{qian2018widar2} pioneers the \textit{single-link tracking} paradigm. It develops a maximum likelihood estimation (MLE) framework to jointly estimate AoA, ToF, and Doppler.
    \item mD-Track~\cite{xie2019mdtrack} introduces a unified optimization framework to resolve targets in the joint AoA-ToF-Doppler domain.
    \item MaTrack~\cite{li2016matrack} (Dynamic-MUSIC) intelligently merges static paths to isolate the weak dynamic reflection from the target, enabling AoA estimation with only 3 antennas.
\end{itemize}

\subsubsection{CSI Ratio and Antenna-Pair Differencing}
Another family of works circumvents geometric solving and instead exploits the relative variations between antennas to eliminate common RF impairments (CFO/SFO).
\begin{itemize}
    \item \textbf{Micro-Motion Sensing:} 
    FarSense~\cite{zeng2019farsense}, and FingerDraw~\cite{wu2020fingerdraw} propose the CSI-Ratio model (complex division of two antennas) to amplify micro-motions for vital signs and handwriting.
    \item \textbf{Robust Event Detection:} RT-Fall~\cite{wang2016rtfall} leverages the \textit{variance of phase differences} between antennas. Since falls induce rapid, non-coherent changes across the array, this differential metric serves as a robust segmenter.
\end{itemize}

\subsubsection{Motion-Induced Synthetic Aperture (SAR)}
\label{sssec:sar}
Unlike the stationary methods in Sec.~\ref{sssec:stationary_array}, this paradigm synthesizes a large aperture through explicit device motion. By moving a single antenna along a trajectory, it emulates a large physical array in the time domain.
\begin{itemize}
    \item Ubicarse~\cite{kumar2014accurate} allows a user to twist a handheld device, creating a circular Synthetic Aperture Radar (SAR) effect that enables accurate 3D self-localization.
    \item Differential-MUSIC~\cite{qian2017enabling} combines user motion with differential phase processing to enable phased-array signal processing on mobile devices with severe phase noise.
\end{itemize}

\subsubsection{Spatial Representation Learning}
Recent trends treat multi-antenna CSI as a high-dimensional feature tensor, learning the mapping between spatial patterns and semantic labels using data-driven models.
\begin{itemize}
    \item \textbf{Deep Spatial Learning:} BodyCompass~\cite{yue2020bodycompass} treats the MIMO channel as a spatial image, employing fully-connected networks to learn sleep postures. Wi-PIGR~\cite{zhang2021wi} and WiWrite~\cite{lin2020wiwrite} utilize CNN-based architectures to extract fine-grained spatial signatures for identification.
    
    \item \textbf{Feature-Based Recognition:} 
    R-TTWD~\cite{zhu2017rttwd} applies PCA to extract statistical features for robust through-the-wall detection.
    AFall~\cite{chen2022afall} extracts explicit Spatial AoA profiles to distinguish falls via physics-based models. FallDeFi~\cite{palipana2018falldefi} aggregates time-frequency features from multiple antennas and classifies them via Support Vector Machines (SVM).
\end{itemize}

\subsubsection*{\textbf{Summary and Limitations}}
Table~\ref{tab:rx_mimo_summary} summarizes these paradigms. While effective, the Passive Rx-MIMO approach is fundamentally constrained by its dependence on fortuitous illumination, leaving targets in spatial ``blind spots'' undetected. This necessitates a shift toward controlled \textit{active illumination}, motivating the \textit{Active Paradigm} (Tx-Beamforming) discussed next.


\subsection{The Active Paradigm: Tx-MIMO Directional Probing}
\label{ssec:active_tx}

The \textit{Active Paradigm} represents a fundamental shift in Wi-Fi ISAC, moving the focus from the receiver's passive listening to the transmitter's active illumination. Unlike Rx-MIMO which depends on fortuitous coverage, Tx-MIMO exploits the transmit antenna array to actively steer electromagnetic energy, effectively functioning as a programmable directional probe. This section traces the evolution of Tx-MIMO in Wi-Fi ISAC through three distinct phases, revealing how the perspective on beamforming has transformed from a source of interference to a powerful enabling resource.

\subsubsection{Phase I---Passive Calibration (Decoupling Channel from Precoding)}
In the nascent stage of ISAC, transmit beamforming was primarily viewed as a destructive factor for sensing stability. Modern Wi-Fi standards (e.g., 802.11ac/ax) employ closed-loop beamforming to maximize communication throughput, dynamically adjusting the precoding matrix (steering matrix) based on channel variations.

\begin{itemize}
    \item \textbf{The Conflict (Beamforming-Induced Impact):} As detailed in Sec.~\ref{ssec:bfi_protocol}, the measured CSI is a composite of the physical channel $\mathbf{H}_u$ and the beamforming matrix $\mathbf{V}_u$ associated with a specific user $u$, expressed as $\hat{\mathbf{H}}_u = \mathbf{H}_u \mathbf{V}_u$. This creates a fundamental conflict: communication-driven updates to $\mathbf{V}_u$ cause the measured $\hat{\mathbf{H}}_u$ to fluctuate even when the physical environment ($\mathbf{H}_u$) remains static. Such artificial variations can be mistakenly interpreted as human activity, severely degrading sensing stability.
    
    \item \textbf{Mitigation Strategy (SenCom):} The representative system, SenCom~\cite{he2023sencom,he2024forward}, treats this dynamic beamforming as ``noise'' to be neutralized. It proposes a passive calibration mechanism. By sniffing the sounding protocols, SenCom reconstructs the beamforming matrix $\mathbf{V}_u$ utilized by the AP and applies an inverse transformation to the received signal. The goal is to mathematically cancel out the effect of $\mathbf{V}_u$ and recover the underlying, stable physical channel $\mathbf{H}_u$. In this phase, Tx-MIMO is considered a nuisance to be removed rather than a tool to be utilized.
\end{itemize}

\subsubsection{Phase II---Feedback Manipulation (Seizing Control of the Beam)}
The second phase marks a pivotal turning point, where the \textit{controllability} of commercial Wi-Fi beams was revealed. This phase proved that the ``black box'' of AP beamforming could be influenced by client feedback.

\begin{itemize}
    \item \textbf{The Discovery (Feedback Dependency):} Beamforming decisions in commodity Wi-Fi are not completely implicit; they are deterministically driven by the \textit{BFI} explicitly fed back by clients. This reliance creates a control mechanism: if the feedback is altered, the beam direction changes accordingly.
    
    \item \textbf{Control Mechanism (BeamCraft):} BeamCraft~\cite{xu2024beamforming,he2025traffic} demonstrates this by injecting forged BFI packets. By crafting a feedback matrix orthogonal to the victim's channel, an attacker can trick the AP into steering its beam away from the victim (creating a null). While originally proposed as a security study, this work fundamentally proves a critical hypothesis for ISAC: the spatial energy distribution of a commodity Wi-Fi AP can be precisely controlled via client-side feedback without modifying the AP's firmware. This capability unlocks the potential for active spatial steering.
\end{itemize}

\subsubsection{Phase III---\hbrev{Joint Optimization (Beamforming Design for ISAC)}}
The final phase embraces Tx-MIMO as an \hbox{optimization-driven mechanism}, serving as the cornerstone for \textit{Directional ISAC}. By shifting from passive reception to active illumination, systems can now \textit{design} the feedback to simultaneously serve sensing and communication.

\begin{itemize}
    \item \textbf{From Avoidance to Optimization:} Unlike Phase II which focused on destructive nulling, this phase pursues beamforming design. The core innovation lies in optimizing the trade-off between communication reliability and sensing coverage. The system formulates a constrained optimization problem: guaranteeing sufficient main lobe gain to sustain the communication link's Modulation and Coding Scheme (MCS), while actively shaping the radiation pattern to form directional sensing beams that  ``illuminate'' targets in blind spots.
    
    \item \textbf{Active Design Implementation:} This philosophy is materialized in VersaBeam~\cite{he2025versabeam} and \textit{Beam-Fi}~\cite{he2025beamfi}, catering to different network complexities. Specifically, VersaBeam is tailored for basic SU-MIMO scenarios, while Beam-Fi addresses the more complex MU-MIMO environments. Given the higher complexity and representativeness of multi-user scenarios, Beam-Fi will be detailed as the Case Study in the following section.
\end{itemize}

\subsection{Case Study: Realizing Programmable Directionality via Beamforming Feedback Control}
\label{ssec:case_study_realization}

To demonstrate the ``Active Paradigm'' on commodity hardware, we introduce Beam-Fi~\cite{he2025beamfi} as a representative case study. Beam-Fi addresses the rigorous multi-user environment, where the system must manage complex inter-user interference constraints. Note that VersaBeam~\cite{he2025versabeam} represents a simplified special form of this framework, tailored for single-user settings without inter-user interference.
\begin{figure}[t]
    \centering
    \setlength{\abovecaptionskip}{3pt} 
    \subfloat[Legacy ISAC (Dilemma).]{
        \includegraphics[width=0.45\linewidth]{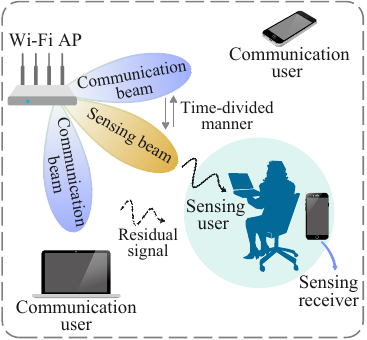}
        \label{fig:beamfi_strategy_a}
    }
    \hfill 
    \subfloat[Active Coordination (Strategy).]{
        \includegraphics[width=0.45\linewidth]{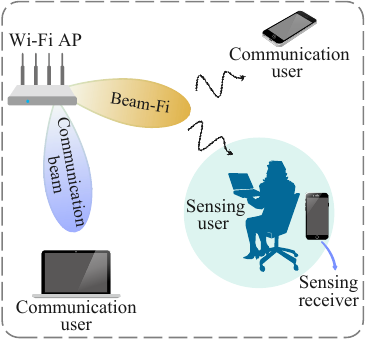}
        \label{fig:beamfi_strategy_b}
    }
    \caption{The paradigm shift in MU-MIMO sensing. (a) Legacy systems struggle with the trade-off between throughput and sensing coverage. (b) The active paradigm (Beam-Fi) adopts a coordination strategy to simultaneously serve both by actively scheduling the beam direction~\cite{he2025beamfi}.}
    \label{fig:beamfi_strategy}
\end{figure}
\subsubsection{The Conflict (Blind Spots in Legacy MU-MIMO)}
In standard MU-MIMO, the AP employs precoding algorithms to maximize communication throughput. This strategy concentrates energy on \hbrev{communication users (C-Users)} while actively suppressing signals in other directions to prevent interference. Consequently, \hbrev{sensing users (S-Users)} located in these ``null spaces'' or interference-dominated zones suffer from extremely low SNR, creating effective blind spots.

As illustrated in Fig.~\ref{fig:beamfi_strategy_a}, legacy systems face a dilemma: serving sensing users in a time-divided manner sacrifices throughput, while passively utilizing residual signals results in poor sensing coverage. The active paradigm (Fig.~\ref{fig:beamfi_strategy_b}) resolves this by treating the communication beam as a programmable resource, strategically coordinating it to cover sensing targets without disrupting data transmission.

\begin{figure}[b]
    \centering
    \setlength{\abovecaptionskip}{3pt}
    \subfloat[MCS-margin $\rightarrow$ admissible loss.]{
        \includegraphics[width=0.45\linewidth]{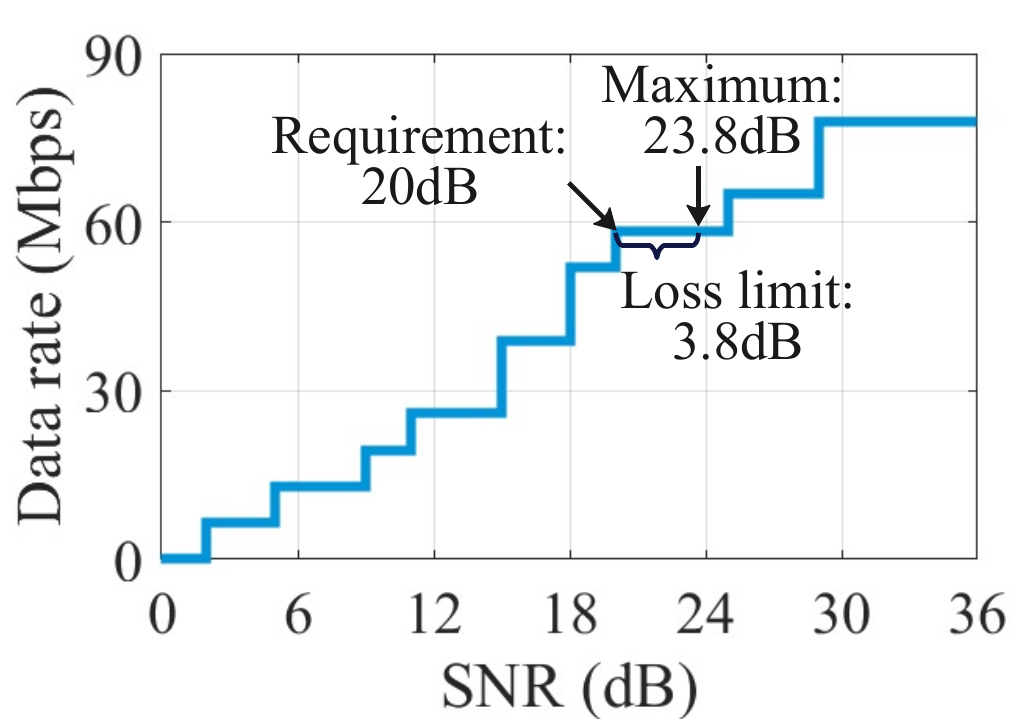}
        \label{sfig:mcs_snr}
    }
    \hfill
    \subfloat[Bounded search for $\mathbf{w}^\mathrm{opt}$.]{
        \includegraphics[width=0.45\linewidth]{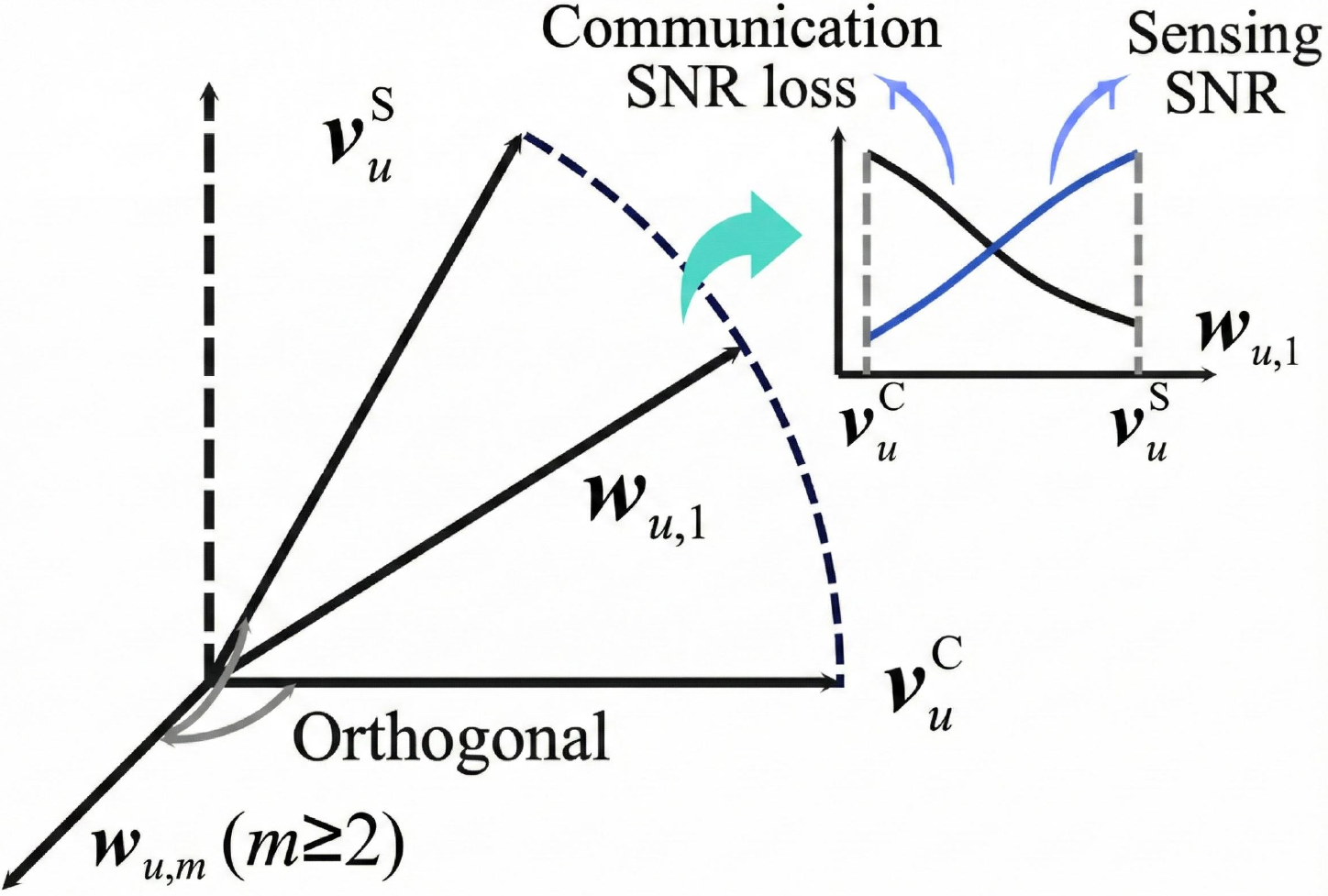}
        \label{sfig:search}
    }
    \caption{Beam design logic. (a) Determine the admissible SNR loss using the \textbf{MCS Margin}. (b) To reduce complexity, the matrix optimization is simplified to a vector search. The algorithm searches for the optimal weight vector $\mathbf{w}^\mathrm{opt}$ between the sensing direction $\mathbf{v}_u^\mathrm{S}$ and communication direction $\mathbf{v}_u^\mathrm{C}$, ensuring the user's SNR stays above the MCS threshold~\cite{he2025beamfi}.}
    \label{fig:beam_scheduling}
\end{figure}

\begin{table*}[b]
\caption{Taxonomy and Comparison of Spatial Diversity Methodologies: Passive vs. Active}
\label{tab:spatial_diversity_summary}
\centering
\small 
\renewcommand{\arraystretch}{1.2} 

\newcolumntype{L}{>{\raggedright\arraybackslash}X}

\begin{tabularx}{\textwidth}{@{} 
    >{\hsize=0.5\hsize}L 
    >{\hsize=1.3\hsize}L 
    >{\hsize=0.9\hsize}L 
    >{\hsize=1.3\hsize}L 
@{}}
\toprule
\textbf{Paradigm} & \textbf{Core Mechanism} & \textbf{Representative Works} & \textbf{Key Gain \& Limitation} \\ 
\midrule

\textbf{Passive Rx-MIMO} & 
\textbf{Spatial Spectrum Estimation:} \newline 
Exploits signal processing (MUSIC, SAR, Learning) to resolve AoA from received signals. & 
SpotFi \cite{kotaru2015spotfi}, Widar2.0 \cite{qian2018widar2}, Ubicarse \cite{kumar2014accurate}, BodyCompass \cite{yue2020bodycompass} & 
\textbf{Gain:} High Angular Resolution (Decimeter Localization). \par\smallskip
\textbf{Limit:} Passive Blindness (Dependent on fortuitous illumination). \\ 
\addlinespace 

\textbf{Active Tx-MIMO} & 
\textbf{Programmable Directionality:} \newline 
Exploits \textbf{BFI Forging} to steer transmit beams via feedback control. & 
BeamCraft \cite{xu2024beamforming}, VersaBeam \cite{he2025versabeam}, Beam-Fi \cite{he2025beamfi} & 
\textbf{Gain:} SNR Enhancement \& Coverage (No Blind Spots). \par\smallskip
\textbf{Limit:} Implementation Overhead (Requires feedback manipulation). \\ 

\bottomrule
\end{tabularx}
\end{table*}

\subsubsection{Methodology (SNR-Constrained Beamforming Design)}
To implement this coordination on commercial devices with low latency, the framework employs a two-step active design. Note that while the system state is defined by matrices $\mathbf{W}$ and $\mathbf{V}$, the optimization step simplifies these into specific beam vectors (denoted by lowercase $\mathbf{w}$ and $\mathbf{v}$) to enable real-time geometric search.

\paragraph{\textbf{Spatial Resource Scheduling (User Grouping)}}
Since the AP must prioritize communication quality, it cannot arbitrarily steer beams. Instead, the system employs a greedy strategy to match the \hbrev{S-User with a paired C-User}. Specifically, the system calculates the \textit{spatial correlation} between the target's physical direction and the beamforming feedback matrices of available C-Users~\cite{he2025beamfi}. This ensures that the C-User's natural beam pattern already covers the S-User's vicinity.

\paragraph{\textbf{Optimization via MCS Margin}}
Once grouped, the system must fine-tune the weights. As shown in Fig.~\ref{sfig:mcs_snr}, Wi-Fi data rates follow a discrete ``staircase'' function based on the Modulation and Coding Scheme (MCS)~\cite{he2025beamfi}. Any SNR exceeding the current MCS requirement constitutes a communication margin (or admissible SNR loss) that can be traded for sensing.

To solve this efficiently, Beam-Fi reduces the high-dimensional matrix optimization to a lightweight vector search. It focuses on the dominant beam vector $\mathbf{w}_{u,1}$ (the first column of $\mathbf{W}_u$) and performs a bounded search within the subspace spanned by the communication channel vector $\mathbf{v}^\mathrm{C}$ and the sensing channel vector $\mathbf{v}^\mathrm{S}$ (Fig.~\ref{sfig:search}). By applying a binary search algorithm, the system identifies the optimal vector $\mathbf{w}^\mathrm{opt}$ that rotates the beam towards the sensing target as much as possible, strictly stopping at the point where the communication SNR hits the MCS threshold limit.

\subsubsection{Performance Evaluation and Capabilities}
To validate the active programmable directionality enabled by Spatial Diversity, Beam-Fi is implemented on commodity Wi-Fi to assess the trade-off between sensing gain and communication overhead.

\begin{figure}[t]
    \centering
    \setlength{\abovecaptionskip}{3pt}
    
    \subfloat[Communication Throughput.]{
        \includegraphics[width=0.45\linewidth]{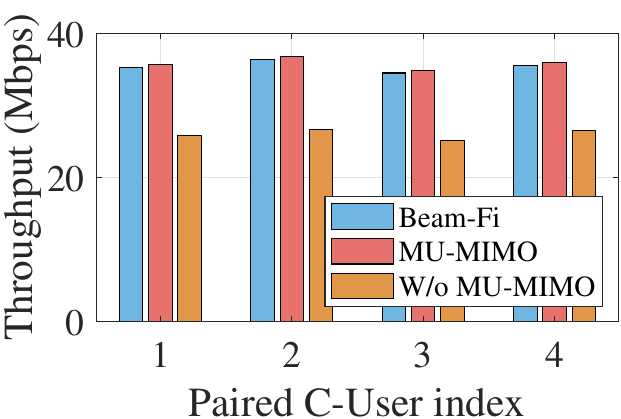}
        \label{sfig:eval_comm}
    }
    \hfill 
    \subfloat[Sensing SNR.]{
        \includegraphics[width=0.45\linewidth]{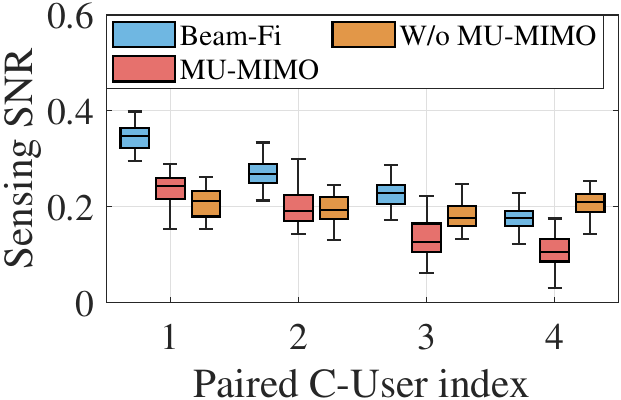}
        \label{sfig:eval_sensing}
    }
    
    \caption{The Win-Win performance of the Active Paradigm. The x-axis ``Paired C-User index'' denotes candidate users sorted by their spatial correlation to the sensing target (Index 1 is the most aligned ``Host User''). (a) Communication throughput remains robust regardless of the pairing choice, whereas (b) sensing quality is highly sensitive to this alignment, peaking when paired with the optimal host~\cite{he2025beamfi}.}
    \label{fig:beamfi_eval}
\end{figure}

\paragraph{\textbf{Evaluation (Performance Decoupling)}}
The Active Paradigm aims to decouple sensing performance from communication reliability. Evaluated across varying spatial alignments (Fig.~\ref{fig:beamfi_eval}), the results reveal a distinct ``Win-Win'' characteristic:

\begin{itemize}
    \item \textbf{Robust Communication:} Throughput remains robust (loss $<2\%$) regardless of the pairing choice~\cite{he2025beamfi}. This stability validates that the SNR-constrained optimization effectively acts as a protective buffer, preventing sensing tasks from degrading network service.
    
    \item \textbf{Significant Sensing Gain:} Conversely, Sensing SNR is highly sensitive to spatial scheduling, boosting to nearly 0.4 only with the optimal C-User (Index 1). This confirms that active coordination turns potential ``blind spots'' into ``sweet spots'' without compromising the primary link.
\end{itemize}

\paragraph{\textbf{Unlocking Programmable Sensing Capabilities}}
This methodology transforms the communication beam from a static broadcast into a steerable spotlight, unlocking two critical capabilities:
\begin{itemize}
    \item \textbf{Omnidirectional Active Coverage:} Unlike passive Rx-MIMO relying on fortuitous coverage, the active Tx-MIMO mechanism eliminates blind spots. By steering energy towards the target via the \hbrev{paired communication user}, the system guarantees high-quality illumination regardless of the target's angular position, overcoming the coverage irregularity of legacy systems.
    
    \item \textbf{Scalable Multi-User Sensing:} Exploiting MU-MIMO spatial multiplexing, the system serves as a parallel sensing engine. By pairing distinct targets with spatially distributed \hbrev{communication users}, the AP generates multiple simultaneous high-gain beams. This enables concurrent tracking of fine-grained activities (e.g., respiration, gestures) for multiple users~\cite{he2025beamfi}, a feat unattainable by single-stream sensing.
\end{itemize}

\subsection{Summary: Towards Directional ISAC}
This chapter explored Spatial Diversity, the dimension governing angular resolution and energy steering. We mapped the evolution of this domain through two distinct paradigms:

\begin{itemize}
    \item \textbf{Passive Rx-MIMO (Receiver-Side Analysis):} Focuses on extracting spatial geometry via advanced signal processing. While effective for tracking, it is constrained by \textit{passive blindness}---relying entirely on fortuitous illumination and remaining vulnerable to energy nulls.
    
    \item \textbf{Active Tx-MIMO (Programmable Directionality):} Transforms the AP from a passive radiator into an active probe. By manipulating feedback to steer beams, it converts interference management into a controllable resource, ensuring targets are actively highlighted rather than neglected.
\end{itemize}

Table \ref{tab:spatial_diversity_summary} summarizes these paradigms. As demonstrated by the Case Study, achieving true \textbf{Directional ISAC} requires transcending passive observation to embrace active control, realizing a high-fidelity sensing paradigm that coexists symbiotically with high-speed data transmission.
\section{Open Challenges and Future Directions}
\label{sec:future_work}

This tutorial has established the first bottom-up framework for Wi-Fi ISAC, systematically decoding how the four dimensions of physical-layer diversity—Temporal, Frequency, Link, and Spatial—fundamentally empower sensing capabilities on commodity hardware. By shifting the perspective from high-level applications to underlying signal characteristics, we 
\revhu{reveal} that the potential for high-precision ISAC is already inherent in the evolving Wi-Fi standards. 

With these physical foundations now firmly laid, the research frontier shifts from \textit{capability exploration} to \textit{systemic integration}. 
\revhu{In other words, t}he path toward ubiquitous ISAC lies not in discovering new diversities, but in harmonizing 
\revhu{existing ones}. The grand challenge ahead is to bridge the gap between \textit{physical possibility} and \textit{practical ubiquity}—evolving from isolated high-performance prototypes to a standardized, full-stack, 
\revhu{widely deployable, and robust} ecosystem.

\subsection{Standardization and Native Protocol Support}
Current Wi-Fi sensing research, including many systems discussed in this tutorial, predominantly relies on ``parasitic'' sensing modes. Researchers often exploit firmware hacks or driver modifications on specific network interface cards (e.g., Intel 5300, Qualcomm AX210) to extract CSI \cite{yi2024enabling}. This dependency on hardware-specific workarounds severely restricts large-scale deployment.

\revhu{A promising direction} lies in Native ISAC Support. The emerging IEEE 802.11bf (WLAN Sensing) standard marks a pivotal transition from communicating and sensing to communicating for sensing. Future research 
\revhu{should} focus on designing native sensing waveforms and measurement setup frames that can coexist efficiently with data transmission.
Furthermore, to democratize development, there is an urgent need for Generalized Sensing APIs. Similar to the Wi-Fi RTT API on Android, future operating systems should expose ``CSI-as-a-Service,'' allowing upper-layer applications to request specific diversity resources—such as steering a spatial beam or scanning a frequency band—without managing the underlying chipset intricacies.

\subsection{Orchestration of Multi-Dimensional Diversity}
While individual case studies 
\revhu{may represent local optima, the next frontier} is the Holistic Multi-Dimensional Fusion. However, simply stacking these technologies triggers Joint Cross-Diversity Optimization challenges, where optimizing one dimension inadvertently degrades another:

\begin{itemize}
    \item \textbf{Space vs. Time:} High-power active beamforming (Spatial) maximizes range but drastically increases self-interference.
    This can saturate analog cancellation circuits, disabling monostatic full-duplex \revhu{operation and thereby forcing time-division sensing with longer measurement intervals (Temporal).}
    
    \item \textbf{Link vs. Frequency:} Wideband synthesis (Frequency) requires aggregating feedback across bands. 
    \revhu{However, each band switching triggers a link re-establishment, disrupting the continuity of real-time BFI sensing (Link).}
\end{itemize}

\revhu{Future architectures should explicitly address these systemic couplings, so that different diversity dimensions can reinforce rather than undermine one another. Moreover, this orchestration can be extended beyond the transceiver to the environment itself. By integrating Reconfigurable Intelligent Surfaces (RIS)~\cite{zhang2023multi, hu2024cross, hu2023holofed}, future systems can artificially create additional link and spatial diversity, actively mitigating blockage and enhancing sensing resolution in complex scenarios.}

\subsection{Resource Efficiency and Collaborative Coexistence}
Sensing is not free. As highlighted by Ma et al. \cite{ma2020wifi}, exploiting higher diversity—such as occupying 320 MHz bandwidth or performing frequent beam scans—incurs significant costs in energy consumption and spectrum utility.
\revhu{Therefore, practical ISAC deployments must explicitly account for the following two aspects.}

\textbf{Sensing-Communication Trade-off:} 
\revhu{In ISAC systems, sensing and communication inherently compete for limited spectral, temporal, and energy resources.}
Future research must \revhu{therefore} quantify the theoretical bounds of ISAC efficiency. Key questions include: What is the minimum spectral resource required to achieve a target sensing resolution? How can we design ``Green ISAC'' protocols that piggyback sensing onto necessary communication traffic to realize low-overhead sensing?

\textbf{Collaborative Coexistence:} As active sensing becomes prevalent, the network will face the severe mutual interference and spectral congestion. Uncoordinated active probing from multiple APs will lead to severe co-channel interference. We envision the development of 
\revhu{collaborative ISAC protocols}, where neighboring APs coordinate their sensing slots and spatial beams, akin to interference management in cellular networks, to ensure 
\revhu{collaborative}
coexistence~\cite{he2022collaborative}.

\subsection{Universal Generalization and Semantic Intelligence}
While deep learning has transitioned Wi-Fi sensing from manual feature engineering to data-driven representation learning~\cite{nirmal2021deep}, a persistent bottleneck in Wi-Fi sensing is domain dependency—models trained in one environment often fail in another due to multipath variations~\cite{ chen2023crossdomain, hu2026crossdomain}. While utilizing diversity mitigates this, a \revhu{potentially} transformative direction is to move beyond environment-specific training towards 
AI-native ISAC\revhu{, which can be enabled by the following two promising approaches.}

\paragraph{\textbf{Wireless Foundation Models}}
\revhu{Inspired by the success of foundation models such as large language models},
future research should leverage the abundance of unlabeled CSI/BFI data to pre-train wireless foundation models~\cite{ahmad2024wifibased}. By learning general representations of radio wave propagation from massive, diverse datasets, these models could achieve ``zero-shot'' transferability across different environments and hardware platforms. 
\revhu{This represents a shift from small-scale supervised training toward large-scale self-supervised representation learning for ISAC.}

\paragraph{\textbf{From Trajectory to Semantics}}
Building upon these general representations, 
the sensing targets of ISAC will evolve from
\revhu{raw physical properties}
to 
\revhu{higher-level} semantic understanding. Future systems 
\revhu{are expected to}
integrate RF sensing with Large Multimodal Models (LMMs), using Wi-Fi as a 
\revhu{complementary}
context provider. This convergence will power intelligent environments capable of not just detecting motions, but understanding complex human activities, interactions, and intent, fundamentally bridging the gap between raw RF signals and human-centric semantics.

\subsection{The Privacy-Utility Paradox and Physical Security}
As Wi-Fi evolves from a communication standard to a sensing modality, it 
\revhu{inevitably raises new privacy challenges}
that must be addressed alongside its utility~\cite{geng2025survey}.
\revhu{In particular, these privacy challenges manifest in the following two aspects:}
\begin{itemize}
    \item \textbf{Plaintext Vulnerability:} Standardized feedback features (e.g., BFI) are currently transmitted as plaintext management frames. This allows attackers to 
    \revhu{infer sensitive information}
    without authentication, such as occupancy detection~\cite{xiao2025leakybeam} or keystroke inference~\cite{ali2017recognizing, hu2023wikieve}.
    \item \textbf{Active Sensing Risks:} The transition to active modes transforms APs into programmable 
    \revhu{sensors}.
    Without safeguards, 
    \revhu{attackers}
    \revhu{can}
    exploit directional beams to ``see'' through walls~\cite{wang2024throughwall}, turning infrastructure into \revhu{malicious} surveillance tools.
\end{itemize}

Future work must develop {physical layer privacy} mechanisms, such as beamforming perturbation or friendly jamming~\cite{hu2024wishield}. 
The goal is 
degrade sensing quality for unauthorized eavesdroppers while maintaining high fidelity for legitimate communication and authorized sensing 
\revhu{applications}.
\section{Conclusion}
\label{sec:conclusion}

In this tutorial, we have charted the technological trajectory of ``ISAC-izing'' commodity Wi-Fi, presenting a systematic primer for transforming standard communication devices into high-precision sensing nodes. By dissecting the complex physical layer into four distinct dimensions of diversity—\textit{temporal, frequency, link, and spatial}—we have established a bottom-up framework that bridges the gap between underlying signal characteristics and upper-layer sensing applications.

\revhu{Our analysis has shown that throughput-driven physical-layer upgrades inadvertently endow Wi-Fi with radar-like sensing potentials.}
We \revhu{have} concretely validated \revhu{these potentials} through four distinct paradigms that form the pillars of modern Wi-Fi sensing:
\begin{itemize}
    \item \textbf{Synchronized ISAC:} Validated by ISAC-Fi, which exploits \textit{temporal diversity} via active self-interference cancellation to enable absolute ranging;
    \item \textbf{High-Resolution ISAC:} Exemplified by UWB-Fi, which harnesses \textit{frequency diversity} through discrete channel sampling to achieve centimeter-level resolution;
    \item \textbf{Ubiquitous ISAC:} Demonstrated by MUSE-Fi, which utilizes \textit{link diversity} to separate multi-user signals for robust concurrent sensing;
    \item \textbf{Directional ISAC:} Realized by Beam-Fi, which leverages \textit{spatial diversity} through programmable active beamforming for precise directional scanning.
\end{itemize}

These 
\revhu{paradigms, validated through representative case studies,} demonstrate that high-fidelity sensing is not an extrinsic add-on, but an intrinsic capability waiting to be unlocked 
\revhu{through the careful orchestration of physical-layer diversities.}

\revhu{As the IEEE 802.11bf era approaches,}
the boundary between communication and sensing is rapidly dissolving. We envision a future where Wi-Fi ISAC becomes as ubiquitous as the connectivity it currently provides—evolving from a passive data pipe into an active and intelligent medium that perceives the physical world. This transformation will not only redefine wireless networks but also unlock a new dimension of context-awareness for the Internet of Things, making our environments smarter, safer, and more responsive.

\bibliographystyle{IEEEtran}
\balance   
\bibliography{ref}

@inproceedings{xie2015precise,
  title={{Precise Power Delay Profiling with Commodity Wi-Fi}},
  author={Xie, Yaxiong and Li, Zhenjiang and Li, Mo},
  booktitle={Proc. of the 21st ACM MobiCom},
  pages={53--64},
  year={2015}
}

@inproceedings{xiong2015tonetrack,
	title	=	{{ToneTrack: Leveraging Frequency-Agile Radios for Time-Based Indoor Wireless Localization}},
	author	=	{Xiong, Jie and Sundaresan, Karthikeyan and Jamieson, Kyle},
	booktitle	=	{Proc. of the 21st ACM MobiCom},
	pages	=	{537-549},
	year	=	{2015}
}

@inproceedings{vasisht2016decimeter,
  author    =   {Vasisht, Deepak and Kumar, Swarun and Katabi, Dina},
  booktitle =   {Proc. of the 13th USENIX NSDI},
  title     =   {{Decimeter-Level Localization with a Single WiFi Access Point}},
  year      =   {2016},
  pages     =   {165–178}
}

@InProceedings{Xu2020low_light,
author = {Xu, Ke and Yang, Xin and Yin, Baocai and Lau, Rynson W.H.},
title = {{Learning to Restore Low-Light Images via Decomposition-and-Enhancement}},
booktitle = {Proc. of the IEEE/CVF CVPR},
month = {June},
year = {2020}
}

@inproceedings{kumar2014accurate,
  title={{Accurate Indoor Localization with Zero Start-up Cost}},
  author={Kumar, Swarun and Gil, Stephanie and Katabi, Dina and Rus, Daniela},
  booktitle={Proc. of the 20th ACM MobiCom},
  pages={483--494},
  year={2014}
}

@misc{weisstein_crt,
    author = {Weisstein, Eric W.},
    title = {{Chinese Remainder Theorem}},
    howpublished = {\url{http://mathworld.wolfram.com/ChineseRemainderTheorem.html}},
    note = {Accessed: 2025-12-19}
}

@book{soumekh1999synthetic,
  title={{Synthetic Aperture Radar Signal Processing}},
  author={Soumekh, Mehrdad},
  volume={7},
  number={1999},
  year={1999},
  publisher={New York: Wiley}
}

@inproceedings{van1995channel,
  title={{On Channel Estimation in OFDM systems}},
  author={Van De Beek, J-J and Edfors, Ove and Sandell, Magnus and Wilson, Sarah Kate and Borjesson, P Ola},
  booktitle={1995 IEEE 45th Vehicular Technology Conference. Countdown to the Wireless Twenty-First Century},
  volume={2},
  pages={815--819},
  year={1995},
  organization={IEEE}
}

@misc{sclrr,
	title = {{Security Camera Laws, Rights, and Rules}},
	howpublished = {\url{https://www.safewise.com/security-camera-laws/}},
	note = {Accessed: 2022-05-19}
}

@inproceedings{LiFS-MobiCom16,
  author    =   {Wang, Ju and Jiang, Hongbo and Xiong, Jie and Jamieson, Kyle and Chen, Xiaojiang and Fang, Dingyi and Xie, Binbin},
  booktitle =   {Proc. of the 22nd ACM MobiCom},
  title     =   {{LiFS: Low Human-Effort, Device-Free Localization with Fine-Grained Subcarrier Information}},
  year      =   {2016},
  pages     =   {243–256}
}

@inproceedings{liu2015tracking,
  author    =   {Liu, Jian and Wang, Yan and Chen, Yingying and Yang, Jie and Chen, Xu and Cheng, Jerry},
  booktitle =   {Proc. of the 16th ACM MobiHoc},
  title     =   {{Tracking Vital Signs During Sleep Leveraging Off-the-Shelf WiFi}},
  year      =   {2015},
  pages     =   {267–276}
}

@inproceedings{wang2014eyes,
  title={{E-Eyes: Device-Free Location-Oriented Activity Identification Using Fine-Grained WiFi Signatures}},
  author={Wang, Yan and Liu, Jian and Chen, Yingying and Gruteser, Marco and Yang, Jie and Liu, Hongbo},
  booktitle={Proc. of the 20th ACM MobiCom},
  pages={617--628},
  year={2014}
}

@article{wang2017device,
  title={{Device-Free Human Activity Recognition Using Commercial WiFi Devices}},
  author={Wang, Wei and Liu, Alex X and Shahzad, Muhammad and Ling, Kang and Lu, Sanglu},
  journal={IEEE Journal on Selected Areas in Communications},
  volume={35},
  number={5},
  pages={1118--1131},
  year={2017},
  publisher={IEEE}
}

@inproceedings{kotaru2015spotfi,
  author    =   {{Kotaru, Manikanta and Joshi, Kiran and Bharadia, Dinesh and Katti, Sachin}},
  booktitle =   {Proc. of 29th ACM SIGCOMM},
  title     =   {{SpotFi: Decimeter Level Localization Using WiFi}},
  year      =   {2015},
  pages     =   {269–282}
}

@article{halperin2011tool,
  title={{Tool Release: Gathering 802.11n Traces with Channel State Information}},
  author={Halperin, Daniel and Hu, Wenjun and Sheth, Anmol and Wetherall, David},
  journal={ACM SIGCOMM computer communication review},
  volume={41},
  number={1},
  pages={53--53},
  year={2011},
  publisher={ACM New York, NY, USA}
}

@article{li2017indotrack,
  title={{Indotrack: Device-Free Indoor Human Tracking With Commodity Wi-Fi}},
  author={Li, Xiang and Zhang, Daqing and Lv, Qin and Xiong, Jie and Li, Shengjie and Zhang, Yue and Mei, Hong},
  journal={Proc. of the 19th UbiComp},
  volume={1},
  number={3},
  pages={1--22},
  year={2017},
  publisher={ACM New York, NY, USA}
}

@misc{schulz2017nexmon,
  title={{Nexmon: The C-Based Firmware Patching Framework}},
  author={Schulz, Matthias and Wegemer, Daniel and Hollick, Matthias},
  howpublished={\url{https://github.com/seemoo-lab/nexmon}},
  year={2017},
  note={Online; accessed 25 Dec 2025}
}

@misc{espressif_esp_csi,
  title={{ESP-CSI}},
  author={{Espressif Systems}},
  howpublished={\url{https://github.com/espressif/esp-csi}},
  year={2020},
  note={Online; accessed 25 Dec 2025}
}

@inproceedings{hu2023muse,
  title={{MUSE-Fi: Contactless MUti-person SEnsing Exploiting Near-field Wi-Fi Channel Variation}},
  author={Hu, Jingzhi and Zheng, Tianyue and Chen, Zhe and Wang, Hongbo and Luo, Jun},
  booktitle={Proc. of the 29th ACM MobiCom},
  pages={1--15},
  year={2023}
}

@inproceedings{li2024uwb,
  title={{UWB-Fi: Pushing Wi-Fi towards Ultra-wideband for Fine-Granularity Sensing}},
  author={Li, Xin and Wang, Hongbo and Chen, Zhe and Jiang, Zhiping and Luo, Jun},
  booktitle={Proc. of the 22nd ACM MobiSys},
  pages={42--55},
  year={2024}
}

@inproceedings{li2025ccs,
title = {{CCS-Fi: Widening Wi-Fi Sensing Bandwidth via Compressive Channel Sampling}},
author = {Li, Xin and Wang, Hongbo and Hu, Jingzhi and Chen, Zhe and Jiang, Zhiping and Luo, Jun},
year = {2025},
month = {05},
booktitle={Proc. of the 44th IEEE INFOCOM}
}

@inproceedings{tan2019multitrack,
  title={{MultiTrack: Multi-user Tracking and Activity Recognition using Commodity WiFi}},
  author={Tan, Sheng and Zhang, Linghan and Wang, Zi and Yang, Jie},
  booktitle={Proc. of the 37th ACM CHI},
  pages={1--12},
  year={2019}
}

@article{zeng2020multisense,
  title={{MultiSense: Enabling Multi-person Respiration Sensing with Commodity WiFi}},
  author={Zeng, Youwei and Wu, Dan and Xiong, Jie and Liu, Jinyi and Liu, Zhaopeng and Zhang, Daqing},
  journal={Proc. of the 22nd ACM UbiComp},
  volume={4},
  number={3},
  pages={1--29},
  year={2020}
}

@inproceedings{song2024siwis,
  title={{SiWiS: Fine-grained Human Detection Using Single WiFi Device}},
  author={Song, Kunzhe and Wang, Qijun and Zhang, Shichen and Zeng, Huacheng},
  booktitle={Proc. of the 30th ACM MobiCom},
  pages={1439--1454},
  year={2024}
}

@article{chen2024isac,
  title={{ISAC-Fi: Enabling Full-Fledged Monostatic Sensing Over Wi-Fi Communication}},
  author={Chen, Zhe and Hu, Chao and Zheng, Tianyue and Cao, Hangcheng and Yang, Yanbing and Chu, Yen and Jiang, Hongbo and Luo, Jun},
  journal={IEEE Journal of Selected Areas in Sensors},
  year={2024},
  publisher={IEEE}
}

@inproceedings{he2025versabeam,
title = {{VersaBeam: Versatile Beamforming for Integrated Sensing and Communication over Commodity Wi-Fi}},
author = {He, Yinghui and Xu, Mingming and Xiao, Fu and Luo, Jun},
year = {2025},
month = {05},
booktitle={Proc. of the 44th IEEE INFOCOM}
}

@inproceedings{he2023sencom,
  title={{SenCom: Integrated Sensing and Communication With Practical WiFi}},
  author={He, Yinghui and Liu, Jianwei and Li, Mo and Yu, Guanding and Han, Jinsong and Ren, Kui},
  booktitle={Proc. of the 29th ACM MobiCom},
  pages={1--16},
  year={2023}
}

@inproceedings{xu2024beamforming,
  title={{Beamforming made Malicious: Manipulating Wi-Fi Traffic via Beamforming Feedback Forgery}},
  author={Xu, Mingming and He, Yinghui and Li, Xin and Hu, Jingzhi and Chen, Zhe and Xiao, Fu and Luo, Jun},
  booktitle={Proc. of the 30th ACM MobiCom},
  pages={908--922},
  year={2024}
}

@article{xiao2023surveya,
  title = {{A Survey on Wireless Device-Free Human Sensing: Application Scenarios, Current Solutions, and Open Issues}},
  author = {Xiao, Jiang and Li, Huichuwu and Wu, Minrui and Jin, Hai and Deen, M. Jamal and Cao, Jiannong},
  year = {2023},
  month = may,
  journal = {ACM Computing Surveys},
  volume = {55},
  number = {5},
  pages = {1--35}
}

@article{chen2023crossdomain,
  title = {{Cross-Domain WiFi Sensing with Channel State Information: A Survey}},
  author = {Chen, Chen and Zhou, Gang and Lin, Youfang},
  year = {2023},
  month = nov,
  journal = {ACM Computing Surveys},
  volume = {55},
  number = {11},
  pages = {1--37}
}

@article{he2020wifi,
  title = {{WiFi Vision: Sensing, Recognition, and Detection With Commodity MIMO-OFDM WiFi}},
  author = {He, Ying and Chen, Yan and Hu, Yang and Zeng, Bing},
  year = {2020},
  month = sep,
  journal = {IEEE Internet of Things Journal},
  volume = {7},
  number = {9},
  pages = {8296--8317}
}

@article{hernandez2023wifia,
  title = {{WiFi Sensing on the Edge: Signal Processing Techniques and Challenges for Real-World Systems}},
  author = {Hernandez, Steven M. and Bulut, Eyuphan},
  year = 2023,
  journal = {IEEE Communications Surveys \& Tutorials},
  volume = {25},
  number = {1},
  pages = {46--76}
}

@article{liu2020wireless,
  title = {{Wireless Sensing for Human Activity: A Survey}},
  author = {Liu, Jian and Liu, Hongbo and Chen, Yingying and Wang, Yan and Wang, Chen},
  year = 2020,
  journal = {IEEE Communications Surveys \& Tutorials},
  volume = {22},
  number = {3},
  pages = {1629--1645}
}

@article{ma2020wifi,
  title = {{WiFi Sensing with Channel State Information: A Survey}},
  author = {Ma, Yongsen and Zhou, Gang and Wang, Shuangquan},
  year = {2020},
  month = may,
  journal = {ACM Computing Surveys},
  volume = {52},
  number = {3},
  pages = {1--36}
}

@article{yang2023sensefi,
  title = {{SenseFi: A Library and Benchmark on Deep-Learning-Empowered WiFi Human Sensing}},
  author={Yang, Jianfei and Chen, Xinyan and Zou, Han and Lu, Chris Xiaoxuan and Wang, Dazhuo and Sun, Sumei and Xie, Lihua},
  journal={Patterns},
  volume={4},
  number={3},
  year={2023},
  publisher={Elsevier}
}

@article{ahmad2024wifibased,
  title = {WiFi-Based Human Sensing With Deep Learning: Recent Advances, Challenges, and Opportunities},
  author = {Ahmad, Iftikhar and Ullah, Arif and Choi, Wooyeol},
  year = {2024},
  journal = {IEEE Open Journal of the Communications Society},
  volume = {5},
  pages = {3595--3623},
}

@article{xue2020deepmv,
  title={{DeepMV: Multi-View Deep Learning for Device-Free Human Activity Recognition}},
  author={Xue, Hongfei and Jiang, Wenjun and Miao, Chenglin and Ma, Fenglong and Wang, Shiyang and Yuan, Ye and Yao, Shuochao and Zhang, Aidong and Su, Lu},
  journal={Proc. of the ACM IMWUT},
  volume={4},
  number={1},
  pages={1--26},
  year={2020}
}

@inproceedings{wang2022wimesh,
  title={{Wi-Mesh: A WiFi Vision-Based Approach for 3D Human Mesh Construction}},
  author={Wang, Yichao and Ren, Yili and Chen, Yingying and Yang, Jie},
  booktitle={Proc. of the 20th ACM SenSys},
  pages={362--376},
  year={2022}
}

@article{wang2024multi,
  title={Multi-Subject 3D Human Mesh Construction Using Commodity WiFi},
  author={Wang, Yichao and Ren, Yili and Yang, Jie},
  journal={Proc. of the ACM IMWUT},
  volume={8},
  number={1},
  pages={1--25},
  year={2024},
  publisher={ACM New York, NY, USA}
}

@inproceedings{xiong2013arraytrack,
  title={{ArrayTrack: A Fine-Grained Indoor Location System}},
  author={Xiong, Jie and Jamieson, Kyle},
  booktitle={Proc. of the 10th USENIX NSDI},
  pages={71--84},
  year={2013}
}

@article{zhang2021widar3,
  title={{Widar3.0: Zero-Effort Cross-Domain Gesture Recognition With Wi-Fi}},
  author={Zhang, Yi and Zheng, Yue and Qian, Kun and Zhang, Guidong and Liu, Yunhao and Wu, Chenshu and Yang, Zheng},
  journal={IEEE Transactions on Pattern Analysis and Machine Intelligence},
  volume={44},
  number={11},
  pages={8671--8688},
  year={2022}
}

@article{yi2024enabling,
  title={{Enabling WiFi Sensing on New-Generation WiFi Cards}},
  author={Yi, Enze and Zhang, Fusang and Xiong, Jie and Niu, Kai and Yao, Zhiyun and Zhang, Daqing},
  journal={Proc. of the ACM IMWUT},
  volume={7},
  number={4},
  pages={1--26},
  year={2024}
}

@inproceedings{hu2024m2fi,
  title={{M2-Fi: Multi-Person Respiration Monitoring via Handheld WiFi Devices}},
  author={Hu, Jingzhi and Jiang, Hongbo and Zheng, Tianyue and Hu, Jingyang and Wang, Hongbo and Cao, Hangcheng and Chen, Zhe and Luo, Jun},
  booktitle={Proc. of the 43rd IEEE INFOCOM},
  pages={1221--1230},
  year={2024}
}

@inproceedings{hu2023wikieve,
  title={{Password-Stealing Without Hacking: Wi-Fi Enabled Practical Keystroke Eavesdropping}},
  author={Hu, Jingyang and Wang, Hongbo and Zheng, Tianyue and Hu, Jingzhi and Chen, Zhe and Jiang, Hongbo and Luo, Jun},
  booktitle={Proc. of the 30th ACM CCS},
  pages={239--252},
  year={2023}
}

@article{chen2024beamthief,
  title={{Echoes of Fingertip: Unveiling POS Terminal Passwords Through Wi-Fi Beamforming Feedback}},
  author={Chen, Siyu and Jiang, Hongbo and Hu, Jingyang and Zheng, Tianyue and Wang, Mengyuan and Xiao, Zhu and Liu, Daibo and Luo, Jun},
  journal={IEEE Transactions on Mobile Computing},
  year={2024},
  note={Early Access}
}

@inproceedings{xiao2025leakybeam,
  title={{Lend Me Your Beam: Privacy Implications of Plaintext Beamforming Feedback in WiFi}},
  author={Xiao, Rui and Chen, Xiankai and He, Yinghui and Han, Jun and Han, Jinsong},
  booktitle={Proc. of the 32nd NDSS},
  year={2025}
}

@inproceedings{todt2025bfid,
  title={{BFId: Identity Inference Attacks Utilizing Beamforming Feedback Information}},
  author={Todt, Julian and Morsbach, Felix and Strufe, Thorsten},
  booktitle={Proc. of the 32nd ACM CCS},
  year={2025}
}

@inproceedings{chen2024beamcount,
  title={{BeamCount: Indoor Crowd Counting Using Wi-Fi Beamforming Feedback Information}},
  author={Chen, Siyu and Jiang, Hongbo and Xiong, Jie and Hu, Jingyang and Wang, Penghao and Liu, Chao and Xiao, Zhu and Li, Bo},
  booktitle={Proc. of the 25th ACM MobiHoc},
  pages={1--10},
  year={2024}
}

@inproceedings{wu2023beamsense,
  title={{Enabling Ubiquitous WiFi Sensing With Beamforming Reports}},
  author={Wu, Chenhao and Huang, Xuan and Huang, Jun and Xing, Guoliang},
  booktitle={Proc. of the 37th ACM SIGCOMM},
  pages={20--32},
  year={2023}
}

@inproceedings{yi2024bfmsense,
  title={{BFMSense: WiFi Sensing Using Beamforming Feedback Matrix}},
  author={Yi, Enze and Wu, Dan and Xiong, Jie and Zhang, Fusang and Niu, Kai and Li, Wenwei and Zhang, Daqing},
  booktitle={Proc. of the 21st USENIX NSDI},
  pages={1123--1138},
  year={2024}
}

@article{wang2024mukifi,
  title={{MuKI-Fi: Multi-Person Keystroke Inference with BFI-Enabled Wi-Fi Sensing}},
  author={Wang, Hongbo and Hu, Jingyang and Zheng, Tianyue and Hu, Jingzhi and Chen, Zhe and Jiang, Hongbo and Zheng, Yuanjin and Luo, Jun},
  journal={IEEE Transactions on Mobile Computing},
  volume={23},
  number={10},
  pages={9835--9850},
  year={2024}
}

@article{wang2025vrfi,
  title={{{VR-Fi}: Positioning and Recognizing Hand Gestures via {VR}-Embedded {Wi-Fi} Sensing}},
  author={Wang, Hongbo and Hu, Jingzhi and Zheng, Tianyue and Chen, Zhe and Luo, Jun},
  journal={IEEE Transactions on Mobile Computing},
  year={2025},
  publisher={IEEE},
  doi={10.1109/TMC.2025.3562383}
}

@book{vantrees2002optimum,
  title={{Optimum Array Processing: Part IV of Detection, Estimation, and Modulation Theory}},
  author={Van Trees, Harry L},
  year={2002},
  publisher={John Wiley \& Sons}
}

@article{ren2022gopose,
  title={{GoPose: 3D Human Pose Estimation Using WiFi}},
  author={Ren, Yili and Wang, Zi and Wang, Yichao and Tan, Sheng and Chen, Yingying and Yang, Jie},
  journal={Proc. of the ACM IMWUT},
  volume={6},
  number={2},
  pages={1--25},
  year={2022},
  publisher={ACM New York, NY, USA}
}

@article{ren2021winect,
  title={{Winect: 3D Human Pose Tracking for Free-Form Activity Using Commodity WiFi}},
  author={Ren, Yili and Wang, Zi and Tan, Sheng and Chen, Yingying and Yang, Jie},
  journal={Proc. of the ACM IMWUT},
  volume={5},
  number={4},
  pages={1--29},
  year={2021},
  publisher={ACM New York, NY, USA}
}

@inproceedings{yan2024person,
  title={{Person-In-Wifi 3D: End-To-End Multi-Person 3D Pose Estimation With Wi-Fi}},
  author={Yan, Kangwei and Wang, Fei and Qian, Bo and Ding, Han and Han, Jinsong and Wei, Xing},
  booktitle={Proc. of the 37th IEEE/CVF CVPR},
  pages={969--978},
  year={2024}
}

@article{han2018indoor,
  title={{Indoor Localization With a Single Wi-Fi Access Point Based on OFDM-MIMO}},
  author={Han, Shuai and Li, Yi and Meng, Weixiao and Li, Cheng and Liu, Tianqi and Zhang, Yanbo},
  journal={IEEE systems journal},
  volume={13},
  number={1},
  pages={964--972},
  year={2018},
  publisher={IEEE}
}

@article{li2016matrack,
  title     = {{Dynamic-MUSIC: Accurate Device-Free Indoor Localization}},
  author    = {Li, Xiang and Li, Shengjie and Zhang, Daqing and Xiong, Jie and Wang, Yasha and Mei, Hong},
  journal   = {Proc. of the 18th ACM UbiComp},
  volume    = {2016},
  number    = {UbiComp},
  pages     = {196--207},
  year      = {2016}
}

@article{chen2019m3,
  title={{M3: Multipath Assisted Wi-Fi Localization with a Single Access Point}},
  author={Chen, Zhe and Zhu, Guorong and Wang, Sulei and Xu, Yuedong and Xiong, Jie and Zhao, Jin and Luo, Jun and Wang, Xin},
  journal={IEEE Transactions on Mobile Computing},
  volume={20},
  number={2},
  pages={588--602},
  year={2019},
  publisher={IEEE}
}

@inproceedings{xie2019mdtrack,
  title={{Md-Track: Leveraging Multi-Dimensionality for Passive Indoor Wi-Fi Tracking}},
  author={Xie, Yaxiong and Xiong, Jie and Li, Mo and Jamieson, Kyle},
  booktitle={Proc. of the 25th ACM MobiCom},
  pages={1--16},
  year={2019}
}

@article{zhang2021wi,
  title={{Wi-PIGR: Path Independent Gait Recognition With Commodity Wi-Fi}},
  author={Zhang, Lei and Wang, Cong and Zhang, Daqing},
  journal={IEEE Transactions on Mobile Computing},
  volume={21},
  number={9},
  pages={3414--3427},
  year={2021},
  publisher={IEEE}
}

@article{yue2020bodycompass,
  title={{BodyCompass: Monitoring Sleep Posture with Wireless Signals}},
  author={Yue, Shichao and Yang, Yuzhe and Wang, Hao and Rahul, Hariharan and Katabi, Dina},
  journal={Proc. of the ACM IMWUT},
  volume={4},
  number={2},
  pages={1--25},
  year={2020},
  publisher={ACM New York, NY, USA}
}

@article{zeng2019farsense,
  title     = {{FarSense: Pushing the Range Limit of WiFi-based Respiration Sensing with CSI Ratio of Two Antennas}},
  author    = {Zeng, Youwei and Wu, Dan and Xiong, Jie and Yi, Enze and Gao, Ruiyi and Zhang, Daqing},
  journal   = {Proc. of the ACM IMWUT},
  volume    = {3},
  number    = {3},
  pages     = {1--26},
  year      = {2019}
}

@article{wu2020fingerdraw,
  title={{FingerDraw: Sub-Wavelength Level Finger Motion Tracking with WiFi Signals}},
  author={Wu, Dan and Gao, Ruiyang and Zeng, Youwei and Liu, Jinyi and Wang, Leye and Gu, Tao and Zhang, Daqing},
  journal={Proc. of the ACM IMWUT},
  volume={4},
  number={1},
  pages={1--27},
  year={2020},
  publisher={ACM New York, NY, USA}
}

@article{wu2021witraj,
  title={{WiTraj: Robust Indoor Motion Tracking with WiFi Signals}},
  author={Wu, Dan and Zeng, Youwei and Gao, Ruiyang and Li, Shenjie and Li, Yang and Shah, Rahul C and Lu, Hong and Zhang, Daqing},
  journal={IEEE Transactions on Mobile Computing},
  volume={22},
  number={5},
  pages={3062--3078},
  year={2021},
  publisher={IEEE}
}

@ARTICLE{he2025traffic,
  author={He, Yinghui and Xu, Mingming and Li, Xin and Hu, Jingzhi and Chen, Zhe and Xiao, Fu and Luo, Jun},
  journal={IEEE Transactions on Mobile Computing}, 
  title={{Traffic Manipulation Via Beamforming Feedback Forgery in Practical Wi-Fi Systems}}, 
  year={2025},
  pages={1-16},
  doi={10.1109/TMC.2025.3649818}}

@article{qian2017enabling,
  title={{Enabling Phased Array Signal Processing for Mobile WiFi Devices}},
  author={Qian, Kun and Wu, Chenshu and Yang, Zheng and Zhou, Zimu and Wang, Xu and Liu, Yunhao},
  journal={IEEE Transactions on Mobile Computing},
  volume={17},
  number={8},
  pages={1820--1833},
  year={2017},
  publisher={IEEE}
}

@inproceedings{jiang2020wipose,
  title     = {{Towards 3D Human Pose Construction Using WiFi}},
  author    = {Jiang, Wenjun and Xue, Hongfei and Miao, Chenglin and Wang, Shiyang and Lin, Sen and Tian, Chong and Murali, Srinivasan and Hu, Haochen and Sun, Zhi and Su, Lu},
  booktitle = {Proc. of the 26th ACM MobiCom},
  pages     = {1--14},
  year      = {2020}
}

@inproceedings{li2025uceiverfi,
  title     = {{$\mu$Ceiver-Fi: Exploiting Spectrum Resources of Multi-Link Receiver for Fine-Granularity Wi-Fi Sensing}},
  author    = {Li, Xin and He, Yinghui and Luo, Jun},
  booktitle = {Proc. of the 31st ACM MobiCom},
  pages     = {1045--1059},
  year      = {2025}
}

@article{wang2016rtfall,
  title={{RT-Fall: A Real-Time and Contactless Fall Detection System with Commodity WiFi Devices}},
  author={Wang, Hao and Zhang, Daqing and Wang, Yasha and Ma, Junyi and Wang, Yuxiang and Li, Shengjie},
  journal={IEEE Transactions on Mobile Computing},
  volume={16},
  number={2},
  pages={511--526},
  year={2016},
  publisher={IEEE}
}

@inproceedings{lin2020wiwrite,
  title={{WiWrite: An Accurate Device-Free Handwriting Recognition System with COTS WiFi}},
  author={Lin, Chi and Xu, Tingting and Xiong, Jie and Ma, Fenglong and Wang, Lei and Wu, Guowei},
  booktitle={Proc. of 40th IEEE ICDCS},
  pages={700--709},
  year={2020},
  organization={IEEE}
}

@article{chen2022afall,
  title={{AFall: Wi-Fi-Based Device-Free Fall Detection System Using Spatial Angle of Arrival}},
  author={Chen, Sheng and Yang, Wei and Xu, Yang and Geng, Yangyang and Xin, Bangzhou and Huang, Liusheng},
  journal={IEEE Transactions on Mobile Computing},
  volume={22},
  number={8},
  pages={4471--4484},
  year={2022},
  publisher={IEEE}
}

@article{palipana2018falldefi,
  title={{FallDeFi: Ubiquitous Fall Detection using Commodity Wi-Fi Devices}},
  author={Palipana, Sameera and Rojas, David and Agrawal, Piyush and Pesch, Dirk},
  journal={Proc. of the ACM IMWUT},
  volume={1},
  number={4},
  pages={1--25},
  year={2018},
  publisher={ACM New York, NY, USA}
}

@inproceedings{youssef2007challenges,
  title={{Challenges: Device-Free Passive Localization for Wireless Environments}},
  author={Youssef, Moustafa and Mah, Matthew and Agrawala, Ashok},
  booktitle={Proc. of the 13th ACM MobiCom},
  pages={222--229},
  year={2007}
}

@inproceedings{joshi2015wideo,
  title={{$\{$Wideo$\}$: Fine-Grained Device-Free Motion Tracing Using $\{$Rf$\}$ Backscatter}},
  author={Joshi, Kiran and Bharadia, Dinesh and Kotaru, Manikanta and Katti, Sachin},
  booktitle={Proc. of the 12th NSDI},
  pages={189--204},
  year={2015}
}

@inproceedings{gjengset2014phaser,
  title     = {{Phaser: Enabling Phased Array Signal Processing on Commodity WiFi Access Points}},
  author    = {Gjengset, Jon and Xiong, Jie and McPhillips, Graeme and Jamieson, Kyle},
  booktitle = {Proc. of the 20th ACM MobiCom},
  pages     = {153--164},
  year      = {2014}
}

@article{gong2018roarray,
  title={{RoArray: Towards More Robust Indoor Localization Using Sparse Recovery with Commodity WiFi}},
  author={Gong, Wei and Liu, Jiangchuan},
  journal={IEEE Transactions on Mobile Computing},
  volume={18},
  number={6},
  pages={1380--1392},
  year={2018},
  publisher={IEEE}
}

@inproceedings{qian2018widar2,
  title     = {{Widar2.0: Passive Human Tracking with a Single Wi-Fi Link}},
  author    = {Qian, Kun and Wu, Chenshu and Yang, Zheng and Liu, Yunhao and He, Fugui and Xing, Tianzhang},
  booktitle = {Proc. of the 16th ACM MobiSys},
  pages     = {81--93},
  year      = {2018}
}

@article{sun2015widraw,
  title={{WiDraw: Enabling Hands-free Drawing in the Air on Commodity WiFi Devices}},
  author={Sun, Li and Sen, Souvik and Koutsonikolas, Dimitrios and Kim, Kyu-Han},
  journal={Proc. of the 21st ACM MobiCom},
  pages={77--89},
  year={2015}
}

@inproceedings{pu2013whole,
  title={{Whole-Home Gesture Recognition Using Wireless Signals}},
  author={Pu, Qifan and Gupta, Sidhant and Gollakota, Shyamnath and Patel, Shwetak},
  booktitle={Proc. of the 19th ACM MobiCom},
  pages={27--38},
  year={2013}
}

@inproceedings{fan2019rf,
  title={{RF-based Inertial Measurement}},
  author={Fan, Yusen and Zhang, Feng and Wu, Chenshu and Liu, K. J. Ray},
  booktitle={Proc. of the ACM SIGCOMM},
  pages={1--14},
  year={2019}
}

@inproceedings{qian2017inferring,
  title={{Inferring Motion Direction using Commodity Wi-Fi for Interactive Exergames}},
  author={Qian, Kun and Wu, Chenshu and Zhou, Zimu and Zheng, Yue and Yang, Zheng and Liu, Yunhao},
  booktitle={Proc. of the 2017 ACM CHI},
  pages={1961--1972},
  year={2017}
}

@article{he2025beamfi,
  title={{Beam-Fi: Integrated Sensing and Communication via MU-MIMO upon Commodity Wi-Fi}},
  author={He, Yinghui and Xu, Mingming and Chen, Zhe and Xiao, Fu and Luo, Jun},
  journal={Proc. of the 27th UbiComp},
  volume={9},
  number={3},
  pages={84:1--84:22},
  year={2025},
  publisher={ACM New York, NY, USA}
}

@inproceedings{tan2021integrated,
  title={{Integrated Sensing and Communication in 6G: Motivations, Use Cases, Requirements, Challenges and Future Directions}},
  author={Tan, Danny Kai Pin and He, Jia and Li, Yanchun and Bayesteh, Alireza and Chen, Yan and Zhu, Peiying and Tong, Wen},
  booktitle={Proc. of the 1st IEEE JC\&S},
  pages={1--6},
  year={2021},
  organization={IEEE}
}

@inproceedings{wang2016human,
  author    = {Wang, Hao and Zhang, Daqing and Ma, Junyi and Wang, Yasha and Wang, Yuxiang and Wu, Dan and Gu, Tao and Xie, Bing},
  booktitle = {Proc. of the 2016 ACM UbiComp},
  title     = {{Human Respiration Detection with Commodity WiFi Devices: Do User Location and Body Orientation Matter?}},
  pages     = {25--36},
  year      = {2016}
}

@inproceedings{gringoli2019free,
  author    = {Gringoli, Francesco and Schulz, Matthias and Link, Jakob and Hollick, Matthias},
  booktitle = {Proc. of the 13th ACM WiNTECH},
  title     = {{Free Your CSI: A Channel State Information Extraction Platform for Modern Wi-Fi Chipsets}},
  pages     = {21--28},
  year      = {2019}
}

@article{garcia2021ieee,
  author    = {Garcia-Rodriguez, Adrian and Bellalta, Boris and Carrascosa, M and Lopez-Perez, D and others},
  title     = {{IEEE 802.11be: Wi-Fi 7 Strikes Back}},
  journal   = {IEEE Communications Magazine},
  volume    = {59},
  number    = {4},
  pages     = {102--108},
  year      = {2021}
}

@article{jiang2021picoscenes,
  author    = {Jiang, Zhiping and Luan, Tom H. and Ren, Xincheng and Lv, Dongtao and Hao, Han and Wang, Jing and Zhao, Kun and Xi, Wei and Xu, Yueshen and Li, Rui},
  title     = {{Eliminating the Barriers: Demystifying Wi-Fi Baseband Design and Introducing the PicoScenes Wi-Fi Sensing Platform}},
  journal   = {IEEE Internet of Things Journal},
  volume    = {8},
  number    = {16},
  pages     = {13488--13510},
  year      = {2021}
}

@misc{ieee80211ax,  

	author	=	{{IEEE Std 802.11ax-2021 (Amendment to IEEE Std 802.11-2020)}},   

	title	=	{{IEEE Standard for Information Technology--Telecommunications and Information Exchange between Systems Local and Metropolitan Area Networks--Specific Requirements Part 11: Wireless LAN Medium Access Control (MAC) and Physical Layer (PHY) Specifications Amendment 1: Enhancements for High-Efficiency WLAN}},   

	year	=	{2021}

}

@article{dardari2009ranging,
  title={{Ranging with Ultrawide Bandwidth Signals in Multipath Environments}},
  author={Dardari, Davide and Conti, Andrea and Ferner, Ulric and Giorgetti, Andrea and Win, Moe Z},
  journal={Proc. of the IEEE},
  volume={97},
  number={2},
  pages={404--426},
  year={2009},
  publisher={IEEE}
}

@inproceedings{bharadia2013full,
  author    = {Bharadia, Dinesh and McMilin, Emily and Katti, Sachin},
  booktitle = {Proc. of the 27th ACM SIGCOMM},
  title     = {{Full Duplex Radios}},
  year      = {2013},
  pages     = {375--386}
}

@inproceedings{wang2024wi2dmeasure,
  title={{Wi2DMeasure: WiFi-based 2D Object Size Measurement}},
  author={Wang, Xuanzhi and Wang, Junzhe and Niu, Kai and Xiong, Jie and Zhang, Fusang and Yi, Enze and Yu, Anlan and Yao, Zhiyun and Zhang, Daqing},
  booktitle={Proc. of the 22nd ACM Sensys},
  pages={253--266},
  year={2024}
}

@article{wang2025freebfi,
  title={{FreeBFI: Enabling Fine-grained BFI Sensing with an Arbitrary Number of Antennas}},
  author={Wang, Junzhe and Li, Wenwei and Zhou, Jiarun and Xiong, Jie and Wang, Xuanzhi and Wang, Qiwei and Yao, Zhiyun and Zhang, Xusheng and Zhang, Duo and Zhang, Daqing},
  journal={Proc. of the ACM IMWUT},
  volume={9},
  number={4},
  pages={1--32},
  year={2025},
  publisher={ACM New York, NY, USA}
}

@article{li2024efficient,
  title={{Efficient Beamforming Feedback Information-Based Wi-Fi Sensing by Feature Selection}},
  author={Li, Xin and Hu, Jingzhi and Luo, Jun},
  journal={IEEE Wireless Communications Letters},
  volume={13},
  number={9},
  pages={2347--2351},
  year={2024},
  publisher={IEEE}
}

@article{li2025enabling,
  title={{Enabling Ultra-Wideband Wi-Fi Sensing via Sparse Channel Sampling}},
  author={Li, Xin and Hu, Jingzhi and Wang, Hongbo and Chen, Zhe and Luo, Jun},
  journal={IEEE Journal on Selected Areas in Communications},
  volume={43},
  number={11},
  pages={3782--3795},
  year={2025},
  publisher={IEEE}
}

@article{ratnam2024optimal,
  title   = {{Optimal Preprocessing of WiFi CSI for Sensing Applications}},
  author  = {Ratnam, Vishnu V. and Chen, Hao and Chang, Hao-Hsuan and Sehgal, Abhishek and Zhang, Jianzhong},
  journal = {IEEE Transactions on Wireless Communications},
  volume  = {23},
  number  = {9},
  pages   = {10820--10833},
  year    = {2024},
  publisher = {IEEE}
}

@article{liu2022integrated,
  title     = {{Integrated Sensing and Communications: Toward Dual-Functional Wireless Networks for 6G and Beyond}},
  author    = {Liu, Fan and Cui, Yuanhao and Masouros, Christos and Xu, Jie and Han, Tony Xiao and Eldar, Yonina C. and Buzzi, Stefano},
  journal   = {IEEE Journal on Selected Areas in Communications},
  volume    = {40},
  number    = {6},
  pages     = {1728--1767},
  year      = {2022},
  publisher = {IEEE}
}

@article{radwan2025tutorial,
  title     = {{A Tutorial-cum-Survey on Self-Supervised Learning for Wi-Fi Sensing: Trends, Challenges, and Outlook}},
  author    = {Radwan, Ahmed Y. and Yildirim, Mustafa and Hasanzadeh, Navid and Tabassum, Hina and Valaee, Shahrokh},
  journal   = {IEEE Communications Surveys and Tutorials},
  year      = {2025},
  publisher = {IEEE}
}

@article{du2024overview,
  title     = {{An Overview on IEEE 802.11bf: WLAN Sensing}},
  author    = {Du, Rui and Hua, Haocheng and Xie, Hailiang and Song, Xianxin and Lyu, Zhonghao and Hu, Mengshi and Xin, Yan and McCann, Stephen and Montemurro, Michael and Han, Tony Xiao and others},
  journal   = {IEEE Communications Surveys and Tutorials},
  volume    = {27},
  number    = {1},
  pages     = {184--217},
  year      = {2024},
  publisher = {IEEE}
}

@article{tan2022commodity,
  title     = {{Commodity WiFi Sensing in Ten Years: Status, Challenges, and Opportunities}},
  author    = {Tan, Sheng and Ren, Yili and Yang, Jie and Chen, Yingying},
  journal   = {IEEE Internet of Things Journal},
  volume    = {9},
  number    = {18},
  pages     = {17832--17843},
  year      = {2022},
  publisher = {IEEE}
}

@article{liu2022survey,
  title={{A Survey on Fundamental Limits of Integrated Sensing and Communication}},
  author={Liu, An and Huang, Zhe and Li, Min and Wan, Yubo and Li, Wenrui and Han, Tony Xiao and Liu, Chenchen and Du, Rui and Tan, Danny Kai Pin and Lu, Jianmin and others},
  journal={IEEE Communications Surveys \& Tutorials},
  volume={24},
  number={2},
  pages={994--1034},
  year={2022},
  publisher={IEEE}
}

@article{hu2026crossdomain,
  title     = {{Cross-Domain Continual Learning for Edge Intelligence in Wireless ISAC Networks}},
  author    = {Hu, Jingzhi and Li, Xin and Su, Zhou and Luo, Jun},
  journal   = {IEEE Transactions on Wireless Communications},
  volume    = {25},
  pages     = {1109--1122},
  year      = {2026},
  doi       = {10.1109/TWC.2025.3588914}
}

@article{gaber2014study,
  title     = {{A Study of Wireless Indoor Positioning Based on Joint TDOA and DOA Estimation Using 2-D Matrix Pencil Algorithms and IEEE 802.11ac}},
  author    = {Gaber, Abdo and Omar, Abbas},
  journal   = {IEEE Transactions on Wireless Communications},
  volume    = {14},
  number    = {5},
  pages     = {2440--2454},
  year      = {2014},
  publisher = {IEEE}
}

@article{choi2022sensor,
  title     = {{Sensor-Aided Learning for Wi-Fi Positioning with Beacon Channel State Information}},
  author    = {Choi, Jeongsik},
  journal   = {IEEE Transactions on Wireless Communications},
  volume    = {21},
  number    = {7},
  pages     = {5251--5264},
  year      = {2022},
  publisher = {IEEE}
}

@article{zhang2023multi,
  title     = {{Multi-Person Passive WiFi Indoor Localization with Intelligent Reflecting Surface}},
  author    = {Zhang, Ganlin and Zhang, Dongheng and He, Ying and Chen, Jinbo and Zhou, Fang and Chen, Yan},
  journal   = {IEEE Transactions on Wireless Communications},
  volume    = {22},
  number    = {10},
  pages     = {6534--6546},
  year      = {2023},
  publisher = {IEEE}
}

@article{hu2024cross,
  title     = {{Cross-Domain Learning Framework for Tracking Users in RIS-Aided Multi-Band ISAC Systems with Sparse Labeled Data}},
  author    = {Hu, Jingzhi and Niyato, Dusit and Luo, Jun},
  journal   = {IEEE Journal on Selected Areas in Communications},
  volume    = {42},
  number    = {10},
  pages     = {2754--2768},
  year      = {2024},
  publisher = {IEEE}
}

@article{hu2023holofed,
  title     = {{HoloFed: Environment-Adaptive Positioning via Multi-Band Reconfigurable Holographic Surfaces and Federated Learning}},
  author    = {Hu, Jingzhi and Chen, Zhe and Zheng, Tianyue and Schober, Robert and Luo, Jun},
  journal   = {IEEE Journal on Selected Areas in Communications},
  volume    = {41},
  number    = {12},
  pages     = {3736--3751},
  year      = {2023},
  publisher = {IEEE}
}

@article{he2024forward,
  title={{Forward-Compatible Integrated Sensing and Communication for WiFi}},
  author={He, Yinghui and Liu, Jianwei and Li, Mo and Yu, Guanding and Han, Jinsong},
  journal={IEEE Journal on Selected Areas in Communications},
  volume={42},
  number={9},
  pages={2440--2456},
  year={2024},
  publisher={IEEE}
}

@article{hu2024wishield,
  title     = {{WiShield: Privacy Against Wi-Fi Human Tracking}},
  author    = {Hu, Jingyang and Jiang, Hongbo and Chen, Siyu and Zhang, Qibo and Xiao, Zhu and Liu, Daibo and Liu, Jiangchuan and Li, Bo},
  journal   = {IEEE Journal on Selected Areas in Communications},
  volume    = {42},
  number    = {10},
  pages     = {2970--2984},
  year      = {2024},
  publisher = {IEEE}
}

@article{jiang2023design,
  title     = {{On the Design and Performance of QRD-Based Beamforming Feedback for Wi-Fi Sensing}},
  author    = {Jiang, Yihang and Gong, Yi and Zeng, Yuan and Lv, Yi and Han, Tony Xiao and Ding, Rentian and Du, Rui},
  journal   = {IEEE Transactions on Wireless Communications},
  volume    = {23},
  number    = {5},
  pages     = {5261--5271},
  year      = {2023},
  publisher = {IEEE}
}

@article{zhang2022wi,
  title     = {{Wi-Fi Sensing for Joint Gesture Recognition and Human Identification from Few Samples in Human-Computer Interaction}},
  author    = {Zhang, Ronghui and Jiang, Chunxiao and Wu, Sheng and Zhou, Quan and Jing, Xiaojun and Mu, Junsheng},
  journal   = {IEEE Journal on Selected Areas in Communications},
  volume    = {40},
  number    = {7},
  pages     = {2193--2205},
  year      = {2022},
  publisher = {IEEE}
}

@article{chi2024xfall,
  title     = {{XFall: Domain-Adaptive Wi-Fi-Based Fall Detection with Cross-Modal Supervision}},
  author    = {Chi, Guoxuan and Zhang, Guidong and Ding, Xuan and Ma, Qiang and Yang, Zheng and Du, Zhenguo and Xiao, Houfei and Liu, Zhuang},
  journal   = {IEEE Journal on Selected Areas in Communications},
  volume    = {42},
  number    = {9},
  pages     = {2457--2471},
  year      = {2024},
  publisher = {IEEE}
}

@article{wang2025wical,
  title     = {{WiCAL: Accurate Wi-Fi-Based 3D Localization Enabled by Collaborative Antenna Arrays}},
  author    = {Wang, Fuhai and Li, Zhe and Xiong, Rujing and Mi, Tiebin and Qiu, Robert Caiming},
  journal   = {IEEE Journal on Selected Areas in Communications},
  volume    = {43},
  number    = {11},
  pages     = {3752--3765},
  year      = {2025},
  doi       = {10.1109/JSAC.2025.3584540}
}

@article{zhu2017rttwd,
  title     = {{R-TTWD: Robust Device-Free Through-The-Wall Detection of Moving Human with WiFi}},
  author    = {Zhu, Hai and Xiao, Fu and Sun, Lijuan and Wang, Ruchuan and Yang, Panlong},
  journal   = {IEEE Journal on Selected Areas in Communications},
  volume    = {35},
  number    = {5},
  pages     = {1090--1103},
  year      = {2017},
  doi       = {10.1109/JSAC.2017.2679578}
}

@article{ali2017recognizing,
  title     = {{Recognizing Keystrokes Using WiFi Devices}},
  author    = {Ali, Kamran and Liu, Alex X. and Wang, Wei and Shahzad, Muhammad},
  journal   = {IEEE Journal on Selected Areas in Communications},
  volume    = {35},
  number    = {5},
  pages     = {1175--1190},
  year      = {2017},
  publisher = {IEEE}
}

@article{meng2024graphar,
  title     = {{GrapHAR: A Lightweight Human Activity Recognition Model by Exploring the Sub-Carrier Correlations}},
  author    = {Meng, Wei and Liu, Zhicong and Li, Bing and Cui, Wei and Zhou, Joey Tianyi and Zhang, Le},
  journal   = {IEEE Transactions on Wireless Communications},
  volume    = {23},
  number    = {4},
  pages     = {2755--2770},
  year      = {2024},
  doi       = {10.1109/TWC.2023.3302861}
}

@article{su2023realtime,
  title     = {{A Real-Time Cross-Domain Wi-Fi-Based Gesture Recognition System for Digital Twins}},
  author    = {Su, Jian and Mao, Qiankun and Liao, Zhenlong and Sheng, Zhengguo and Huang, Chenxi and Zhang, Xuedong},
  journal   = {IEEE Journal on Selected Areas in Communications},
  volume    = {41},
  number    = {11},
  pages     = {3690--3701},
  year      = {2023},
  doi       = {10.1109/JSAC.2023.3310073}
}

@article{wang2024throughwall,
  title     = {{Through-the-Wall Detection and Localization of Autonomous Mobile Device in Indoor Scenario}},
  author    = {Wang, Jiacheng and Du, Hongyang and Niyato, Dusit and Zhou, Mu and Kang, Jiawen and Xiong, Zehui and Jamalipour, Abbas},
  journal   = {IEEE Journal on Selected Areas in Communications},
  volume    = {42},
  number    = {1},
  pages     = {161--176},
  year      = {2024},
  doi       = {10.1109/JSAC.2023.3322819}
}

@article{nirmal2021deep,
  title     = {{Deep Learning for Radio-Based Human Sensing: Recent Advances and Future Directions}},
  author    = {Nirmal, Isura and Khamis, Abdelwahed and Hassan, Mahbub and Hu, Wen and Zhu, Xiaoqing},
  journal   = {IEEE Communications Surveys and Tutorials},
  volume    = {23},
  number    = {2},
  pages     = {995--1019},
  year      = {2021},
  publisher = {IEEE}
}

@article{he2016wifi,
  title     = {{Wi-Fi Fingerprint-Based Indoor Positioning: Recent Advances and Comparisons}},
  author    = {He, Suining and Chan, S.-H. Gary},
  journal   = {IEEE Communications Surveys and Tutorials},
  volume    = {18},
  number    = {1},
  pages     = {466--490},
  year      = {2016},
  doi       = {10.1109/COMST.2015.2464084}
}

@article{he2022collaborative,
  title     = {{Collaborative Sensing in Internet of Things: A Comprehensive Survey}},
  author    = {He, Shibo and Shi, Kun and Liu, Chen and Guo, Bicheng and Chen, Jiming and Shi, Zhiguo},
  journal   = {IEEE Communications Surveys and Tutorials},
  volume    = {24},
  number    = {3},
  pages     = {1435--1474},
  year      = {2022},
  publisher = {IEEE}
}

@article{geng2025survey,
  title     = {{A Survey of Wireless Sensing Security from a Role-Based View}},
  author    = {Geng, Ruixu and Wang, Jianyang and Yuan, Yuqin and Zhan, Fengquan and Zhang, Tianyu and Zhang, Rui and Huang, Pengcheng and Zhang, Dongheng and Chen, Jinbo and Hu, Yang and others},
  journal   = {IEEE Communications Surveys and Tutorials},
  year      = {2025},
  publisher = {IEEE}
}

\end{document}